\documentclass[
superscriptaddress,
 amsmath,amssymb,
 aps,
prstab,
]{revtex4-1}

\usepackage{graphicx}
\usepackage{dcolumn}
\usepackage{bm}

\usepackage{subfigure}
\begin{document}

\preprint{Arxiv}

\title{Self-Diffusiophoretic Colloidal Propulsion Near a Solid Boundary}


\author{Ali Mozaffari}\thanks{These two authors contributed equally to this work.}
\affiliation{Benjamin Levich Institute and Department of Chemical Engineering, City College of the City University of New York, New York, NY 10031, USA}
\author{Nima Sharifi-Mood}\thanks{These two authors contributed equally to this work.}
\affiliation{Department of Chemical and Biomolecular Engineering, University of Pennsylvania, Philadelphia, PA 19104, USA}
\author{Joel Koplik}
\affiliation{Benjamin Levich Institute and Department of Physics, City College of the City University of New York, NY 10031, USA}
\author{Charles Maldarelli}
\email{cmaldarelli@ccny.cuny.edu.}
\affiliation{Benjamin Levich Institute and Department of Chemical Engineering, City College of the City University of New York, New York, NY 10031, USA}
\begin{abstract}
We study the diffusiophoretic self-propulsion of a colloidal catalytic particle due to a surface chemical reaction in a vicinity of a  solid wall.  Diffusiophoresis is a chemico-mechanical transduction mechanism in which a concentration gradient of an interacting solute produces an unbalanced force on a colloidal particle and drives it along the gradient. We consider a spherical particle with an axisymmetric reacting cap covering the polar angle range $0\le \theta\le \theta_{cap}$ in the presence of a repulsive solute, near an infinite planar wall, and solve the coupled solute concentration and Stokes equations, using a mixture of numerical and analytic arguments. The resulting  particle trajectory is determined by $\theta_{cap}$ and the initial orientation of the symmetry axis with respect to the plane. At normal incidence the particle initially moves away from or towards the wall, depending on whether the cap faces towards or away, respectively, but even in the latter case the particle never reaches the wall due to hydrodynamic lubrication resistance.  For other initial orientations, when  $\theta_{cap}\le 115^{\circ}$ the particle either moves away immediately or else rotates along its trajectory so as to cause the active side to face the wall and the particle to rebound.  For higher coverage we find trajectories where the particle skims along the wall at constant separation or else comes to rest. We provide a phase diagram giving the nature of the trajectory (repulsion, rebound, skimming or stationary) as  a function of $\theta_{cap}$ and the initial orientation. 
\end{abstract}

\maketitle 

\section{Introduction}
 The newest generation of ultra-miniaturized engines now under development are constructed using colloidal sized objects, which are engineered to react with solute in a solution in which the colloids are immersed, and convert the chemical reaction energy  into mechanical self-propulsion without moving parts (for reviews see Howse \cite{eh10}, Sen and Velegol \cite{Paxton:2006gf,ms2009,sIS2012}, Wang  \cite{Wang2009,WM2009,CKOW20011, GSW2012,Wang2012a}, Schmidt \cite{doi:10.1146/annurev.matsci.34.040203.115827}, Pumera \cite{SM2009,PUM2010}, Kapral \cite{kap2013}, and Ozin \cite{oz2005}).   These autonomous molecular locomotors or artificial swimmers represent a bottom-up approach to mechanical design made possible by advances in micro and nano-fabrication, and are at the center of a wide range of potential applications  at the micro- and nano-scale. The locomoters themselves can be used as micropumps \cite{KarPump}, or as roving sensors if their chemical propulsion is affected by the presence of solute \cite{Wang2009a}. The locomoters can be configured with chemical receptors to function as  transporters  for capturing, towing and delivering  molecular cargo \cite{doi:10.1021/nl072275j, SSIS2010, Wang2012a} as for example drugs to selected targets \cite{wang2010aa,wang2012cc,wang2012cd}, molecular building blocks to  assemble supramolecular structures \cite{Hamley}, or analytes in microfluidic networks for lab on a chip operations \cite{doi:10.1021/ja803529u, schmidt2011, schmidt2012aa, wang2013a}. Colloidal motors with chemical receptors  have also shown promise as the engines of  shuttles for the capture and movement of biological cells \cite{bkhcw2011, ssss2011, wang2012c} .

The operating principle for the self-propulsion of  molecular motors is  the chemico-mechanical transduction mechanism which converts the reaction energy into the propulsive motion.  The first set of synthetic locomotors were bimetallic rods made of platinum and gold (a few microns in length and a few hundred nanometers in diameter) prepared by sequential electrodeposition in a membrane template. A catalytic  oxidation of a solute, hydrogen   peroxide takes place on the platinum (anode) side of the rod to produce protons in solution, and electrons. The protons migrate to the opposite gold (cathode) side of the rod through the solution, while the electrons conduct through the rod to the gold side. At this cathode end, hydrogen peroxide combines with the released protons and the electrons conducted through the metal to produce water \cite{Paxton:2004rt,Kline:2005ly,doi:10.1021/ja056069u,doi:10.1021/ja0643164,doi:10.1021/la0615950,PhysRevE.81.065302,posner2011}. The accumulation of protons at one end of the bisegmented rod, relative to the opposite end creates an electric dipole field around the rod directed from the anode to the cathode.  The field acts on the diffuse layer of positive charge in the water  around the cylindrical sides of the particle (which balances the negative charge of the metal) driving fluid from the anode to the cathode and propelling the rod in the opposite direction.

More recent research has adopted a tubular configuration for the engine, rather than a nano-rod structure, in order to increase  the  efficiency of the hydrodynamic propulsion. Using either thin-film roll-up technology \cite{ADMA:ADMA200801589, SMLL:SMLL200900021, C0CS00078G, doi:10.1146/annurev.matsci.34.040203.115827}, or template membrane electrodeposition, conically shaped hollow cylinders are formed  tens of microns in length and a few microns in diameter with an inner catalytic platinum surface which catalyzes the reaction of hydrogen peroxide to water and oxygen on the inside of the cylinder. The buildup of oxygen creates bubbles which are expelled through the wider opening of the cylinder, creating a recoil which propels the object.

\begin{figure} [h]
\centering
	     \subfigure[]{\label{diffusiophoresis1a}\includegraphics[width=0.45\textwidth]{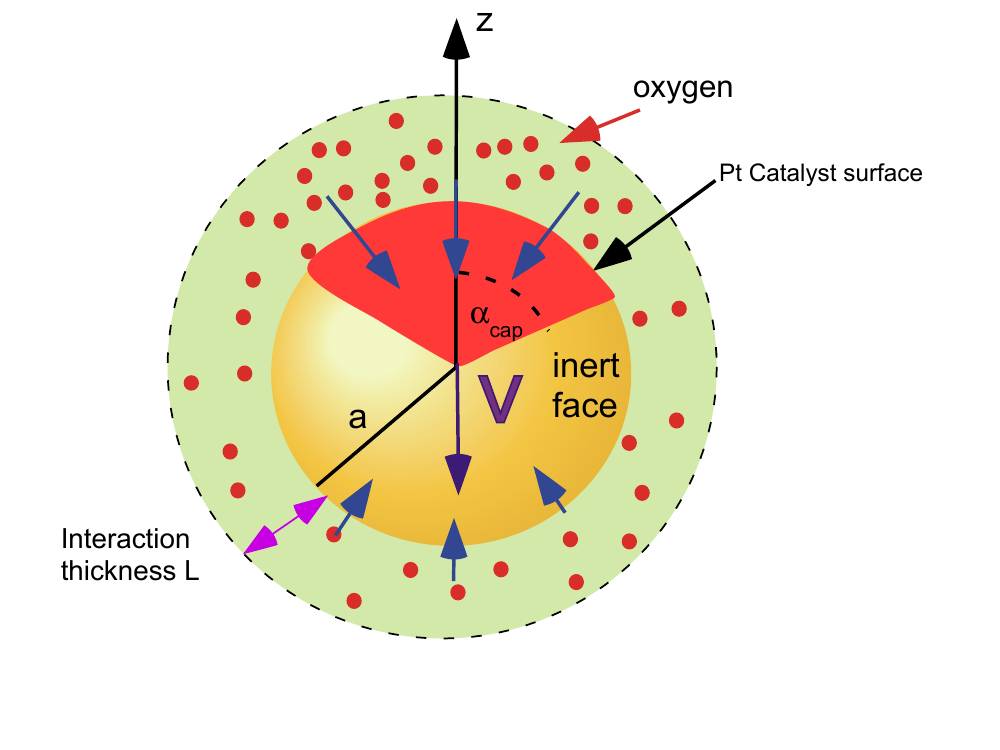}}
	     \subfigure[] {\label{diffusiophoresis2a}\includegraphics[width=0.45\textwidth]{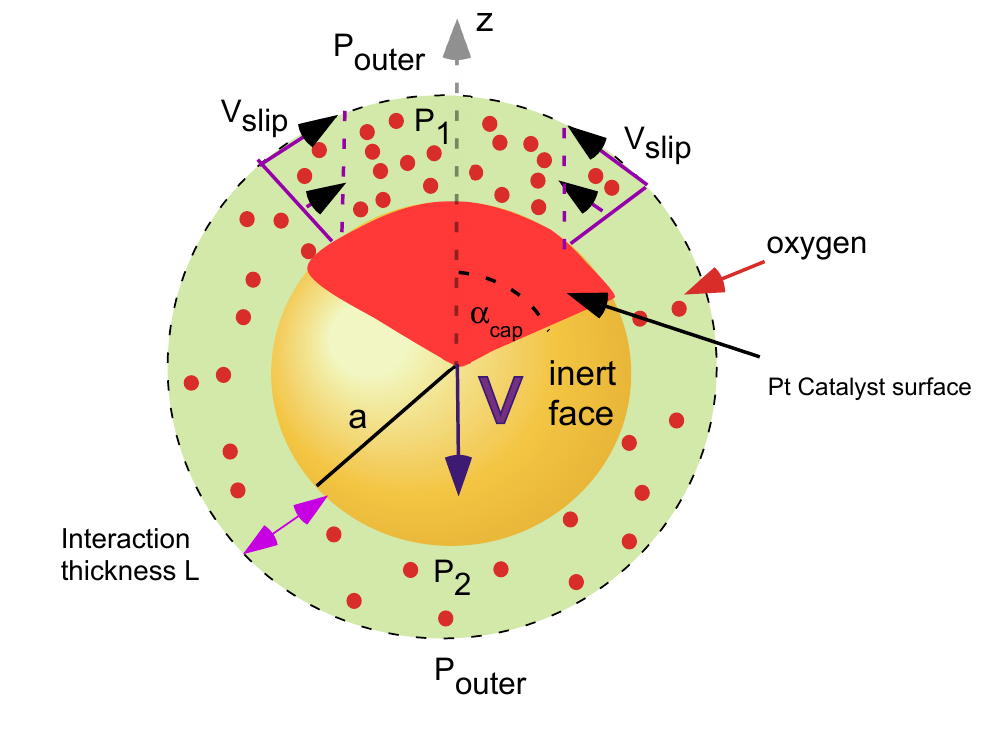}}
	      \caption{\footnotesize{A Janus colloid molecular motor:  (a) A difference in concentration of product (oxygen) created by the reaction of hydrogen peroxide on the  catalyst half of a Janus particle  generates an unbalanced intermolecular force on the particle (shown here as repulsive and acting over a length scale $L$) which propels the particle towards the lower  concentration, shown by the particle velocity vector $V$. (b) The  thin film hydrodynamic analysis of Anderson \textit{et al} \cite{alp82,anderson83, AP84,a89} for $a \gg L$: Greater net repulsion of the solute against the particle on the high concentration side in the intermolecular interaction layer generates a lower pressure $P_{1}$ relative to outside the intermolecular layer ($P_{outer}$) to balance this repulsion compared to the pressure $P_{2}$ at the low concentration side. This causes a streaming flow towards the high concentration side (a slip velocity) forcing the particle to  moves in the opposite direction or leading with the passive side as in (a).}} 
	      \label{diffusiophoresisschematic}
\end{figure}

A third molecular engine configuration, which we will focus on in this study, is constructed around spherical colloids, and has the advantage that it is the easiest of the engines to fabricate. In this design, as shown in Fig. \ref{diffusiophoresisschematic}, nonconducting (usually polystyrene) spheres a few microns in diameter are first coated on one side with platinum to form a Janus colloid (see Howse, Golestanian and coworkers and references \cite{PhysRevE.82.015304,  PhysRevLett.99.048102, KYCS2010, eh10}). The functionalized colloids are then immersed in a hydrogen peroxide solution; because the underlying polystyrene is nonconducting, there is no electrochemical driving force for oxidation, and the platinum catalyzes the hydrogen peroxide in solution directly to water and oxygen. The experiments are usually undertaken in two regimes:  The first is at a high hydrogen peroxide concentration, in which oxygen production  exceeds the equilibrium solubility in aqueous solution, and microbubble production is observed on the active side of the colloid. As in the tubular jet configuration, the recoil force on the colloid projects the particle in the direction of the passive side.
The second case is one in which the hydrogen peroxide concentration is low, and oxygen bubbles are not evident, but the colloid is still observed to move in the same direction (leading with the passive side). Howse, Golestanian and collaborators \cite{PhysRevE.82.015304,  PhysRevLett.99.048102} have argued that unbalanced intermolecular interactions between the hydrogen peroxide reactant and the oxygen product create the propulsion. This type of motion can be explained by the diagram in Fig. \ref{diffusiophoresis1a}. Assuming that the (oxygen) product species produced by the reaction repels the colloid particle,  the excess of this product on one side of the colloid due to the asymmetric distribution driven by the catalytic conversion leads to a net repulsive force projecting the particle in the direction of the passive side. The intermolecular interaction occurs over a small length scale, denoted by $L$ in the figure, and  is of the order of a few nanometers. This propulsion force is termed diffusiophoresis after Derjaguin \cite{Derjaguin,Derjaguin1993138} and later Anderson \emph{et~al.} \cite{alp82,anderson83, AP84,a89} who studied this problem for the case of a gradient in solute applied directly across the particle (rectified diffusiophoresis). Although, the neutral solute gradient driven self-diffusiophoresis has been widely considered as an accepted mechanism for the self-propulsion of Janus catalytic swimmers due to hydrogen peroxide decomposition on the platinum surface, recent experimental studies reveal that the electrochemical effect due to ion diffusion can not be entirely neglected and further investigation should be undertaken to elucidate and disentangle these effects \cite{poon}.\\ 
\indent Unbalanced intermolecular interactions which are caused by an asymmetric reaction around a particle, and which in turn drive a propulsive motion, represents a general type of  mechanism  by which a molecular engine  can operate, and is adopted in this study. For particles in an infinite liquid phase hydrodynamic models have been developed to predict the diffusiophoretic velocity \cite{PhysRevLett.94.220801, PhysRevLett.99.048102,1367-2630-9-5-126, pdo09,refId}. The hydrodynamics was first studied theoretically in a general way by Derjaguin \cite{Derjaguin,Derjaguin1993138} and later Anderson {\em{et al}}~~\cite{alp82,anderson83, AP84,a89} in the case in which gradients of solute are applied directly across the colloid.  These authors studied the problem in the limit in which the interaction length $L$ is much smaller than the particle radius $a$, i.e. $\lambda=L/a \ll 1$, and the particle moves through a continuous Newtonian phase in Stokes flow. Within a continuum hydrodynamic framework, the effect of the  intermolecular interaction of the particle with the solute is accounted for by adding a body force to the Stokes equations, $-C\nabla \phi$, where $C$ is the solute concentration and $\phi({r})$ is the potential energy of interaction of the solute with the colloid \cite{alp82,Nima_1}.  In the limit $\lambda=L/a \ll 1$,  the flow field divides into a thin inner region (within a distance $L$ of the particle much smaller than the radius $a$ ) where the interaction force is effective, and an outer region where it is relatively unimportant (Fig. \ref{diffusiophoresis2a}).  The inward repulsion of  the solute with the particle  creates a lower pressure in the thin layer (relative to outside the layer) which is necessary to  balance this outward force. Since the repulsion  of the solute  to the particle is greater along the side of the particle where the solute concentration is larger, the pressure is  lowest there, and this generates a streaming flow to the higher concentration side as the particle moves in the direction of lower concentration (towards the passive side).  This streaming flow creates a  tangential velocity $v_{slip}$ at the outer edge of the inner region which is a function of the concentration gradient along the surface at this edge, $v_{slip}  =  - b \nabla _s C_s$ where $s$ measures distance along the surface, $C_s$ is the concentration of solute along the outer edge and $b$ is a slip coefficient which is a function of the intermolecular forces and the concentration distribution of solute  in the inner region.  From the perspective of the  outer region, the flow at the outer edge of  the inner region represents a ``slip" velocity on the particle surface. To leading order, the flow in the outer region and the translational  diffusiophoretic velocity $V$ of the particle are solved  in terms of this slip velocity. The solution of the mass conservation equation for the concentration field in the outer region allows for the calculation of the gradient in the  solute concentration at the outer edge of the inner region from which the particle velocity obtains. Anderson {\em{et al}}~~\cite{alp82,anderson83, AP84,a89} found for the slip coefficient 
\begin{equation}
b=  - \frac{{k_{B}TL^{2}}}{\mu }\int\limits_0^\infty  {y\left[ {\exp \left( { - \phi(y) /k_BT} \right) - 1} \right]dy},
\label{anderson2}
\end{equation}
where $\phi(r)$ is the interaction potential energy, $k_{B}T$ is the thermal energy,  $\mu$ is the solution viscosity and $y=(R-1)/\lambda$ is the rescaled radius with $\lambda = L/a$. 
\begin{figure} [b]
\centering
\includegraphics[width=0.4 \textwidth]{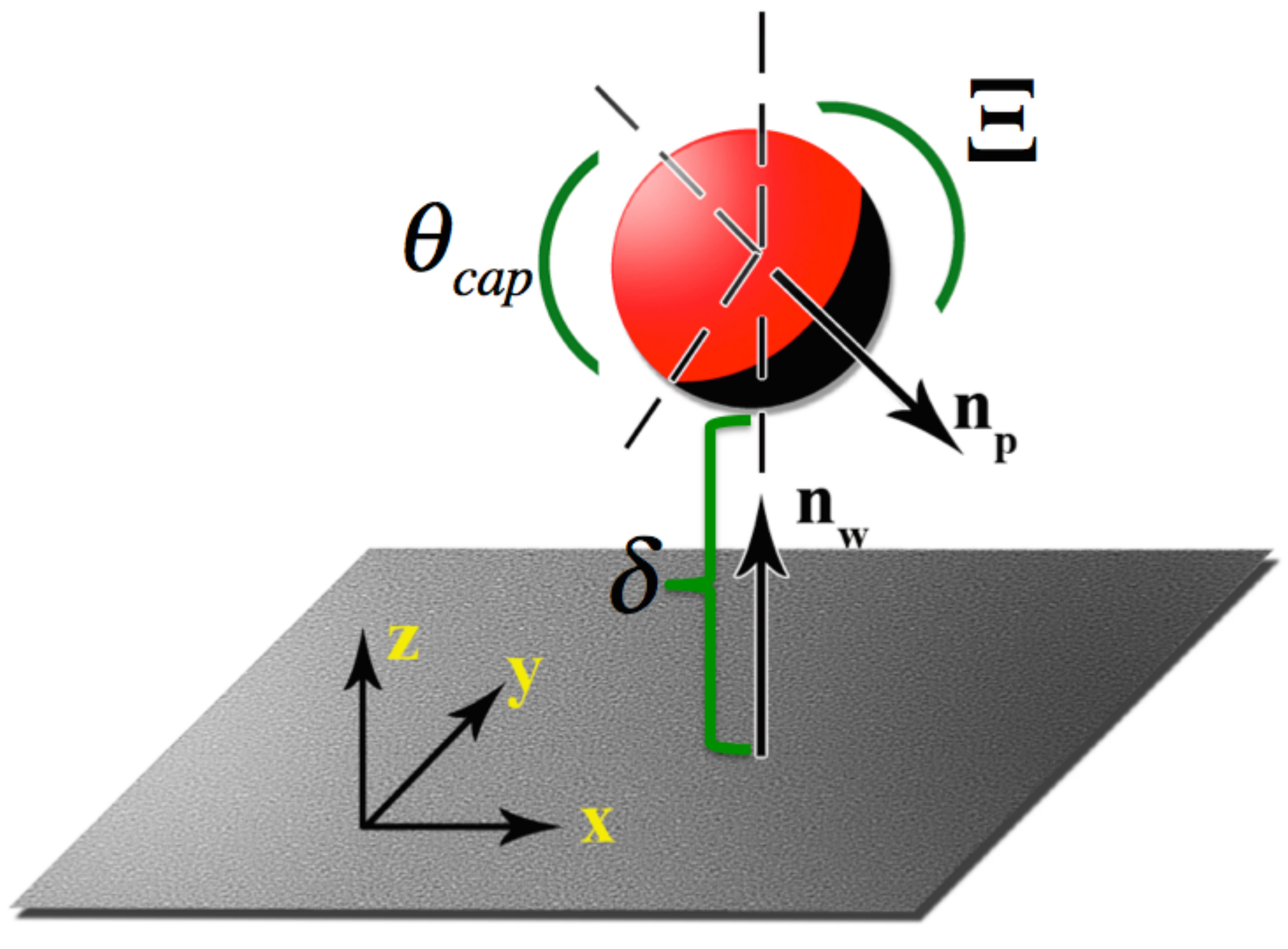}	     	    
 \caption{\footnotesize{A Janus swimmer with active area of size $\theta_{cap}$ near a solid boundary, with active area (in red) tilted at an angle $\Xi$ from the perpendicular to the boundary surface and with the colloid a distance $\delta$ from the wall. $\Xi=0$} represents a colloid with its active area directly facing the wall. } 
\label{Janusnearwalls}
 \end{figure}
More recently, the validation of this continuum framework for sub-micron size colloidal particle has been tested against molecular dynamics simulation study \cite{Nima_2} and a micro-mechanical colloidal perspective \cite{Brady}. \\
The self-diffusiophoresis of a non-Brownian Janus colloid with thin interaction layer and a well-defined net potential interaction between  solute and the colloid in an infinite medium has been now well-established and it was shown that the swimming velocity of the colloid can be described via two non-dimensional numbers so called Damk\"{o}hler $Da$, and P\'eclet $Pe$ numbers (see below for their definitions).
In order to describe the  hydrodynamics of  self-diffusiophoresis, in which the gradient responsible for driving the phoretic motion is generated by a surface reaction, the solutal concentration fields have to be resolved for the reaction-diffusion problem. The hydrodynamic studies assume an irreversible surface reaction of a reactant to products which occurs only along the reactive cap (Fig. \ref{diffusiophoresis2a}, $0 \le \theta  \le {\theta _{cap}}$),  and produces a constant flux (dimensional)  $N_{0}$ of product.   Along the rest of the colloid surface, the solute fluxes are zero since the solutes can not penetrate into the particle. This assumption of constant flux along the reactive part is equivalent to assuming the reaction rate is slow relative to diffusion of the reactants \cite{Nima_1} and product (small Damk\"{o}hler  ($Da$) number  where $Da$ is defined generally as  $Da=\frac{ka}{D}$ where $k$ is the surface rate of reaction of the reactant $A$ and $D$ in the diffusion coefficient of the reactant.  In the limit  $Da \to 0$, to leading order, the concentration of the solutes is spherically symmetric,  and to first order, the flux off of the cap, due to the reaction, is a constant and of order $Da$.  To calculate the self-diffusiophoretic velocity ${\bf{V}}$, the hydrodynamic studies assume Anderson's expression for the slip velocity (Eq. \ref{anderson2}). To compute the concentration field at the outer edge of the inner region, two approximations are made. First, only the concentration field in the outer region is solved, with the constants in the outer solution evaluated by applying  the boundary conditions directly at $R = 1$ (the colloid surface). This computation then provides the solution for $C^{\,outer}(R \to 1,\theta)=C_{s}(\theta)$. As a second approximation, the Anderson expression for the slip velocity coefficient (Eq. \ref{anderson2}) is used to calculate the velocity.  With these assumptions, the studies indicate that for the case in which only the product ``B'' interacts with the colloid,  the self-diffusiophoretic velocity  is to leading order in $\lambda$ given by:
\begin{equation}
V_{\infty}  =b \frac{N_{0}}{D}\left[\frac{1-\text{cos}^{2}\theta_{cap}}{4} \right],
\label{popescu}
\end{equation}
where $V_{\infty}$ is the dimensional diffusiophoretic velocity. For a fixed value of the dimensional flux, $N_0$, the velocity is again found to be independent of the particle radius $a$. The above expression has been used by Golestanian \textit{et al}\cite{PhysRevLett.94.220801, PhysRevLett.99.048102,1367-2630-9-5-126} to model the observed motion of Janus colloids driven by the asymmetric decomposition of hydrogen peroxide to oxygen on the platinum functionalized side of the particles.  Note that the above relationship assume the colloid motion is non-Brownian, i.e. the time scale for rotational diffusion of the colloid is much longer than the time scale for the directed motion of the particle $\frac{{{\tau _{\rm{Brownian}}}}}{{\frac{a}{{{U_c}}}}} \gg 1$, the typical value of this ratio for micron-size Janus colloid which swim with the speed of $5-50 \frac{\mu m}{s}$ is about $30-6000$ and therefore it is a legitimate assumption to ignore Brownian rotation of the colloid in the calculation for the short time $t \ll \tau _{\rm{Brownian}}$, however the role of Brownian rotational diffusion will be important in long time and consequently results in a diffusive behavior \cite{PhysRevLett.94.220801}. Later on, using the same Anderson mobility framework,  Popescu \emph{et~al.}\cite{pdo09,refId} extended the above expression for colloid particles with axisymmetric, ellipsoidal shapes. For intermediate to high values of $Da$, it was shown that the swimming velocity of the colloid increases monotonically until it reaches a plateau where the system is diffusion limited \cite{golestanian_size}.
\indent The influence of solute advection for rectified diffusiophoresis was studied by Keh and Weng \cite {Keh} and recently Khair \cite{Khair_Peclet}. It was found that for a thin interaction layer compared to colloid size, the effect of solute advection to the diffusiophoretic velocity is $O(Pe^2)$ and particle velocity persistently decreases as P\'eclet number increases (P\'eclet number is the ratio of solute advection to solute diffusion which defined as $\sim \frac{U_ca}{D}$ where $U_c$ is the characteristic diffusiophoretic velocity). In self-diffusiophoresis on the other hand, Milchen and Lauga \cite{Lauga_Peclet} found that the effect of solute advection on diffusiophoretic velocity is $O(Pe)$ and hence depending on attractive or repulsive net interaction of the solute with the colloid, the solute advection can increase or decrease the self-diffusiophoretic velocity respectively. More recently, the effect of external shear flow in self-diffusiophoresis has been examined by Frankel and Khair \cite{Frankel}. The thin interaction layer assumption and the slip velocity argument can be relaxed by introducing a body force term in Stokes equation which is a consequence of net interaction between the solutes and also a new term would appear in mass balance conservation which is a contribution to the flux of solute due to the net interactions with the colloid. In this case, it was shown that the swimming velocity of the colloid decreases as the size of interaction layer becomes larger \cite{Nima_1,Sabbas}.
An alternative approach to study the colloidal particle motion in the presence of gradient of another species is the micro-mechanical perspective in which the particle and solutes are Brownian particles in continuum solvent and in this treatment the constraint of small and low volume fraction of solutes can be relaxed \cite{Brady}. Moreover it was shown the micro-mechanical perspective can successfully recapitulate the continuum treatment of diffusiophoresis in the region of overlap \cite{Brady,Nima_1}. 
The shape of the particle itself can also brings about propulsion of the particle, in this case it was shown that a non-spherical particle is able to move even with uniform catalytic activity \cite{Wei,Chemical_Sailing}. What remains still as a potential challenge is a controlled motion at micro-nanoscale and it's been suggested that by applying an external field, the Janus swimmers can be steered towards a specific location \cite {ssss2011}.\\
 \indent The investigations discussed  above are an accounting of  the self-propulsion of a reactive Janus swimmer in an infinite medium. Since a lot of anticipated applications of colloidal motors involve in the motion of these swimmers in the vicinity of an interface or a solid wall, it's crucial to understand the effect of interface on the dynamics of colloidal particle. Moreover, we want to respond the question weather it is feasible to steer these artificial microswimmers via an interface similar to the observation for biological swimmers, e.g. spermatozoid \cite{Sperm_navigation} or bacteria \cite {Mehdi} . Despite swimming in the vicinity of a boundary can display a strikingly more complex set of trajectories than the rectilinear motion the swimmer executes in an infinite fluid space. The richer dynamics of colloidal propulsion near a boundary derives from the diffusive interaction of the solute concentration field generated at the surface of the  active area of the colloid with the boundary  which changes the diffusiophoretic propulsive force. In addition, there is related work on electrophoretic and dipolophoretic motion of colloids \cite{Keh1, Brenner,Bazant,Miloh} and biological swimmers such as bacteria \cite{Bacteria_Surface,Lauga_Spagnolie} near walls and finally the motion of biological swimmers, e.g. bacteria close to interfaces is crucial for understanding in broad range of problems involving biofilm formation on surfaces \cite{biofilm,Lauga_Stone,Goldstein,Liana}.\\
In this study we will consider an  infinite solid surface as the boundary, which we assume cannot be penetrated by the solute (see Fig. \ref{Janusnearwalls}). For a colloidal particle which is self-diffusiophoretically swims in a proximity of a solid wall, the role of the wall is not only important because of the hydrodynamic interactions, but also it distorts the concentration field around the particle and consequently alters the dynamics of colloidal motors. We denote by $\Xi $ the tilt angle of the outward normal of the center of the spherical cap of the active area of the swimmer with the perpendicular  to the boundary (defining $\Xi=0$ to be the orientation in which the active area normal points towards the boundary), and $\delta$ the edge-to-edge perpendicular distance of the sphere to the wall.  We will assume Stokes flow,  and a slow reaction so that the flux of solute  on the active area is specified.  The mass transfer is regarded as quasi-static, i.e. the time scale for diffusive equilibration $\frac{a^{2}}{D}$ is much shorter than the time scale for movement of the particle over its radius $a$, with the velocity of the particle as estimated by the velocity far from the boundary, $V_{\infty}$, Eq. \ref{popescu}. This corresponds to P\'eclet numbers $Pe=\frac{U_{\infty}a}{D} \ll 1$, where $D$ is solute diffusion coefficient. In addition, we will also assume that the intermolecular interaction length scale $L$ is much smaller than the colloid radius $a$ so that the simplifying framework in which propulsion is driven by a slip velocity on the surface dictated by the gradient in the surface concentration of the solute around the colloid. As in our discussion above, we will assume that the catalytic reaction on the surface produces a solute which acts repulsively with the colloid so that the accumulation of solute on the active side of the swimmer pushes the swimmer in the opposite direction, or equivalently creates a slip on the active area in the direction away from the passive side.  Hence the propulsion velocity $V$ (relative to the value far from the wall, $V_{\infty}$, eq. \ref{popescu}) becomes  an instantaneous function of the wall-colloid separation distance, $\delta$, the tilt $\Xi$ and the size of the catalytic cap $\theta_{cap}$. 
Our aim is to obtain solutions for the velocities and rotation for different inclination angles of a particle with respect to the wall as a function of $\delta$ and $\theta_{cap}$, and from these solutions identify trajectories that can correspond to swimmers being repelled form the surface, skimming over surfaces and stationary in the vicinity of the surface.  These regimes of motion are especially important in applications, where swimmers are usually in the vicinity of boundaries, and engineered motions such as skimming can be useful to precisely direct the swimming motion. Tasinkevych \emph{et~al.} \cite{Uspal} first examined diffusiophoretic-driven motion of a catalytically-active Janus colloid near a wall using the Anderson ``slip velocity" framework and constant production of solute for arbitrary active areas, wall separation distances and inclination angles. They used  boundary element methods and the Lorentz reciprocal theorem  to obtain particle trajectories, and  find that diffusiophoretic-driven Janus colloids can skim along the surface for large active areas, and become stationary at very large areas. Our aim in this study is to develope a more complete detailed analytical solution using bispherical coordinates, and obtain master curves  for the  translational and angular velocities as a function of separation distance, inclination angle and size of the active area so that  trajectories from any initial configuration can be computed. Furthermore, we construct a phase diagram to detail the different trajectories as a function of the initial orientation and size of the active surface area. This paper is organized as the following, in Sec.~\ref{formulation} we formulate the mass balance conservation for solute and the hydrodynamics problems with a detail analytical solution and in Sec.~\ref{R&D} we provide the discussion of the concentration field, swimming velocity and trajectories of catalytic colloidal particle close to the wall.
\section{Formulation and Analytical Solution}\label{formulation} 
\begin{figure} 
\centering
\includegraphics[width=0.5 \textwidth]{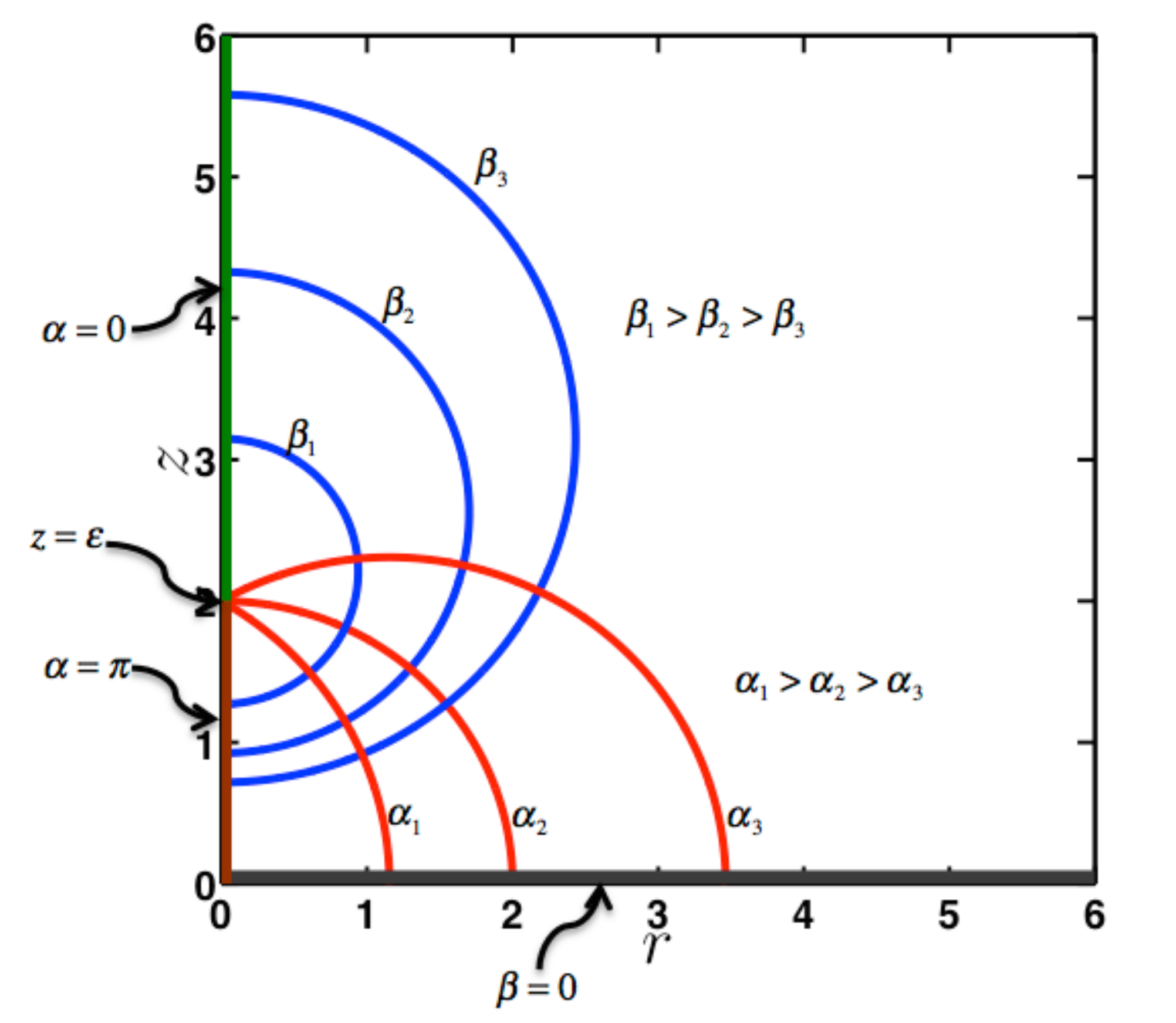}	     	    
 \caption{\footnotesize{Sketch of a bispherical coordinates $(\alpha ,\beta ,\phi)$.} } 
\label{bisf}
 \end{figure}

\indent To formulate the problem we begin with the non-dimensional Stokes and continuity equations,
\begin{align}
&\nabla  \cdot {\bf{v}} = 0, \label{hyd2}\\
&\nabla ^2 {\bf{v}} - \nabla p = 0, \label{hyd1}
\end{align}
where velocity and pressure are non-dimensionalized with $V_\infty$ and $\displaystyle{\frac{{\mu \;V_\infty  }}{R}}$ respectively, where $R$ is particle radius and  $V_\infty$ is defined in Eq.~\ref{popescu} . The sphere and planar wall geometries can be easily described by utilizing a set of orthogonal curvilinear bispherical coordinates $(\alpha ,\beta ,\phi)$. Bispherical coordinates are related to cylindrical coordinates through the following transformations
\begin{eqnarray}
\displaystyle{r = \frac{{\epsilon \;\sin \alpha \;}}{{\cosh \beta  - \cos \alpha }}},\label{p14}\\
\displaystyle{z = \frac{{\epsilon \;\sinh \beta }}{{\cosh \beta  - \cos \alpha }}},\label{p15}
\end{eqnarray} 
 The range of coordinates are limited to $0 \leqslant \alpha  \leqslant  \pi $, $-\infty < \beta < \infty $, $0\leqslant \phi  \leqslant2\pi $,  and $\epsilon $ is a positive scale factor. $\beta  = \beta _0 $ represents a sphere whose center is located on the $z$-axis at $\displaystyle{z = \epsilon \;\coth \beta _0} $, with radius $\displaystyle{\frac{\epsilon }{{\sinh \beta _0 }}}$. $\beta  = 0$ is a sphere of infinite radius and also the plane $z = 0$, which is a wall in the present problem. 
By combining Eq. \ref{p14} and \ref{p15} and eliminating $\beta$, it follows that $\alpha=$ constant describes  circular arcs which pass through the points $z =  \pm \epsilon $ and are symmetric about $z$-axis.  The rotation of the surface of $\displaystyle{\alpha  = \frac{\pi }{2}}$ around the $z$-axis produces a sphere with radius $\epsilon$, and for the cases of $\alpha  > \frac{\pi }{2}$ and $\alpha  < \frac{\pi }{2}$ spindle-like and apple-like shapes are obtained, respectively.  The value $\alpha  = \pi $ corresponds to the line on the $z$-axis from $- \epsilon $ to $ + \epsilon $, and the value $\alpha  = 0$ corresponds to the rest of the $z$-axis (Fig.\ref{bisf}). Since $z <0$ is irrelevant in the physical problem, $\beta$ just takes non-negative values. 

The non-dimensional boundary conditions are:
\begin{align}
&\left. {\bf{v}} \right|_{\beta  = 0}  = 0,\label{bond1}\\ 
&\left. {\bf{v}} \right|_{\beta  = \beta _0 }  ={\bf{v}}_s  + {\bf{U}} + {\bf{\Omega }} \times ({\bf{r}}-{\bf{r}}_p),\label{bond2}\\ 
&{\bf{v}} \to 0,\;\;\ as \;\;\ \sqrt {r^2  + z^2 }  \to \infty.
\end{align}
where ${\bf{r}}$ is position vector on the particle surface and ${\bf{r}}_p$ is the location of particle center of mass which one can show $\displaystyle{\beta _0  = \cosh ^{ - 1} (\frac{{\delta  + R}}{R})}$ . The main objective is to find translational and angular velocity vector $({\bf{U}},{\bf{\Omega }})$ from the fact that the particle is freely suspended in a viscous medium. In order to obtain these unknown velocities we need to find the distribution of solute around the particle, since the slip velocity at the particle surface is ${\bf{v}}_s  =  - \nabla_s C $ where $C$ is the solute distribution at the particle surface, non-dimensionalized with $\displaystyle{\frac{N_0 R}{D}}$.

\subsection{Concentration Field} 
The mass transfer relation which governs the solute distribution in steady state around the colloid is the Laplace equation
 \begin{eqnarray}
\nabla ^2 C = 0 \label{p13}
\end{eqnarray}
We can easily satisfy the boundary conditions at the wall as well as the particle surface in bispherical coordinates $(\alpha ,\beta ,\phi)$,  
 where a general 3-D solution of Laplace's equation can be expressed as \cite{JFM2}: 
\begin{equation}
\begin{split}
&C(\alpha ,\beta ,\phi ) = \sqrt {\cosh \beta  - \cos \alpha }  \\ 
&\times \sum\limits_{m = 0}^\infty  {\sum\limits_{n = m}^\infty  [{\widetilde{A}_{n,m} \sinh (n + \frac{1}{2})\beta  + \widetilde{B}_{n,m} \cosh (n + \frac{1}{2})\beta }] } \;P_n^m (\cos \alpha )\;\cos (m\phi  + \gamma _m ).
 \label{p22} \\ 
\end{split}
\end{equation}
Here, $P_n^m (\cos \alpha)$ is an associated Legendre polynomial of the first kind, and the constants ${\widetilde{A}_{n,m} }$, ${\widetilde{B}_{n,m} }$ and $\gamma _m $ can be obtained upon imposing appropriate boundary conditions at solid surfaces. Using constant flux production of solute at the active side of the particle and zero flux at the passive part of the particle as well as the wall surface, the boundary conditions can be written in nondimensional form as:
\begin{eqnarray}
\left. {\frac{{\partial C}}{{\partial \beta }}} \right|_{\beta  = 0}  = 0 ,\label{p23}
\end{eqnarray}
\begin{eqnarray}
\frac{{(\cosh \beta _0  - \cos \alpha )}}{\epsilon }\left. {\frac{{\partial C}}{{\partial \beta }}} \right|_{\beta  = \beta _0 }  = f(\alpha ,\phi ) ,\label{p24}
\end{eqnarray}
and $f(\alpha ,\phi )$ is the coverage function defined as:
\begin{equation}
\;f(\alpha ,\phi ) = \left\{ {\begin{array}{*{20}{l}}
{1\;\;\;\;{\bf{r}} \in {\rm{Active}}\;{\rm{side}}}\\
{0\;\;\;\;{\bf{r}} \in {\rm{Passive}}\;{\rm{side}}}
\end{array}} \right..\label{p25}
\end{equation}
\indent The relevant recursive relations to find concentration field can be found in the Appendix A, where the conditions (\ref{p23}-\ref{p24}) are used to find the unknown coefficients. \\

\subsection{Stokes Flow  Solution for the Velocity Field}
\begin{figure}
\centering
\subfigure[]{\label{sch1}\includegraphics[width=0.24\textwidth]{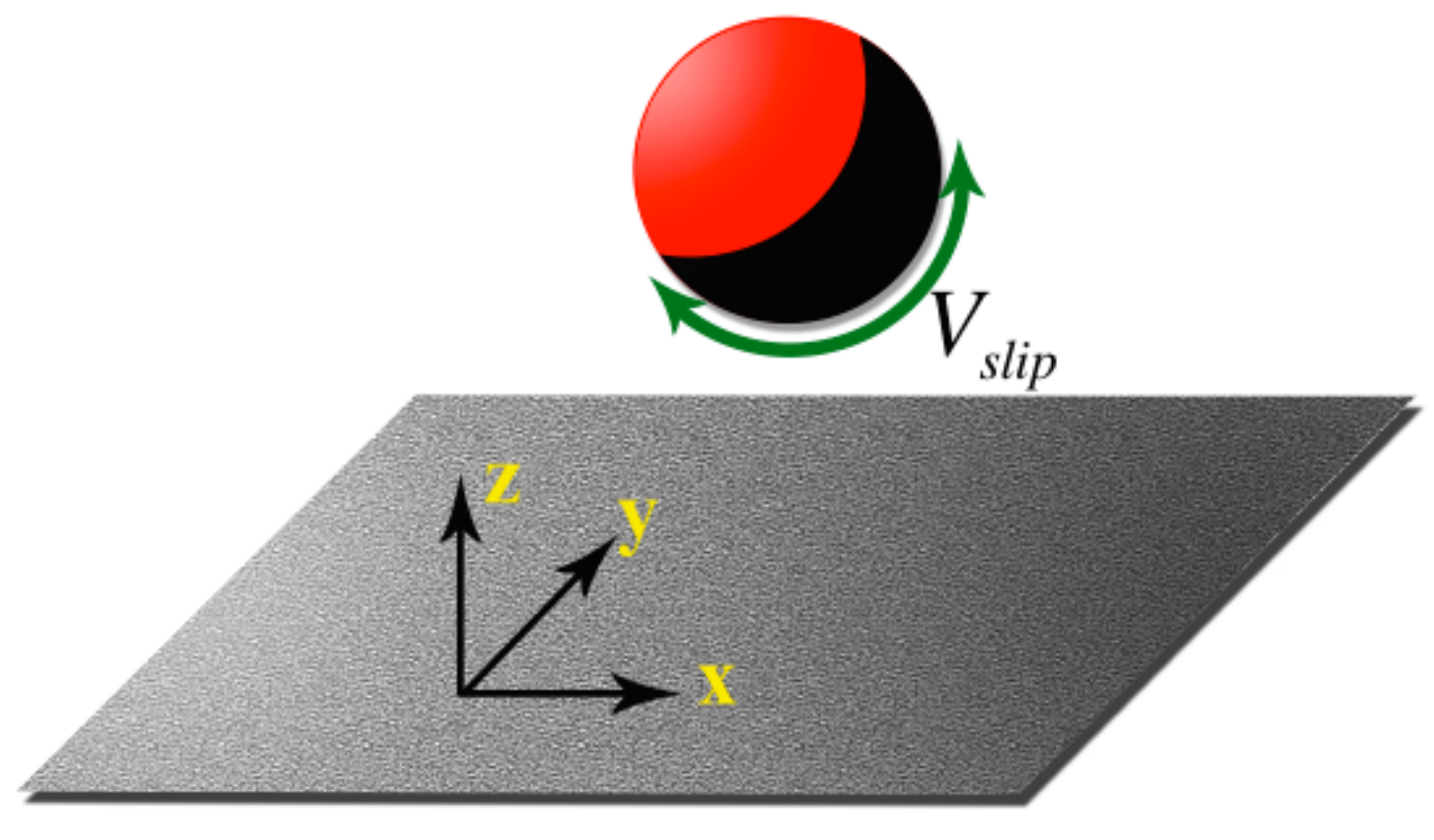}}
\subfigure[]{\label{sch2}\includegraphics[width=0.24\textwidth]{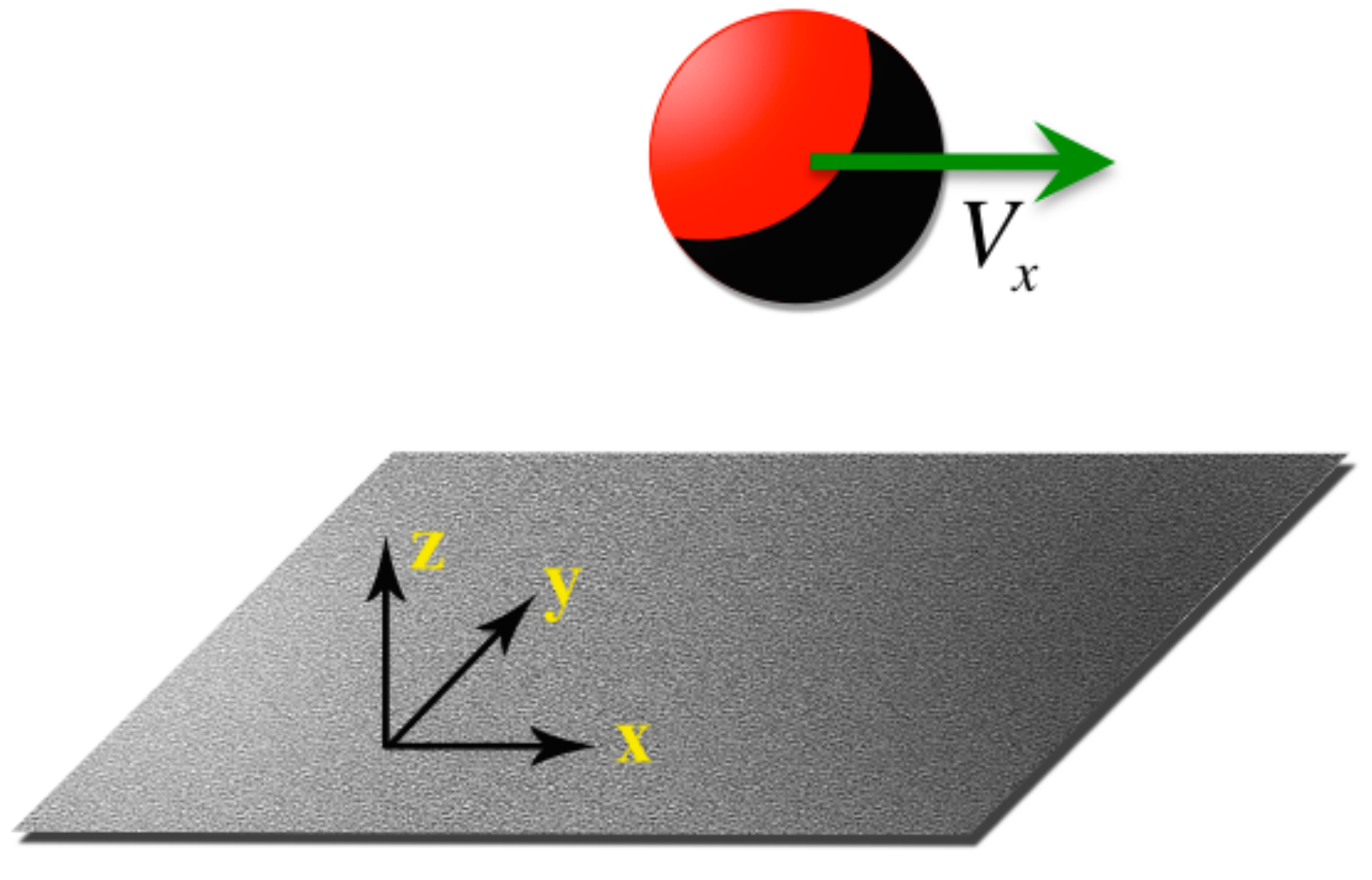}}
\subfigure[]{\label{sch3}\includegraphics[width=0.24\textwidth]{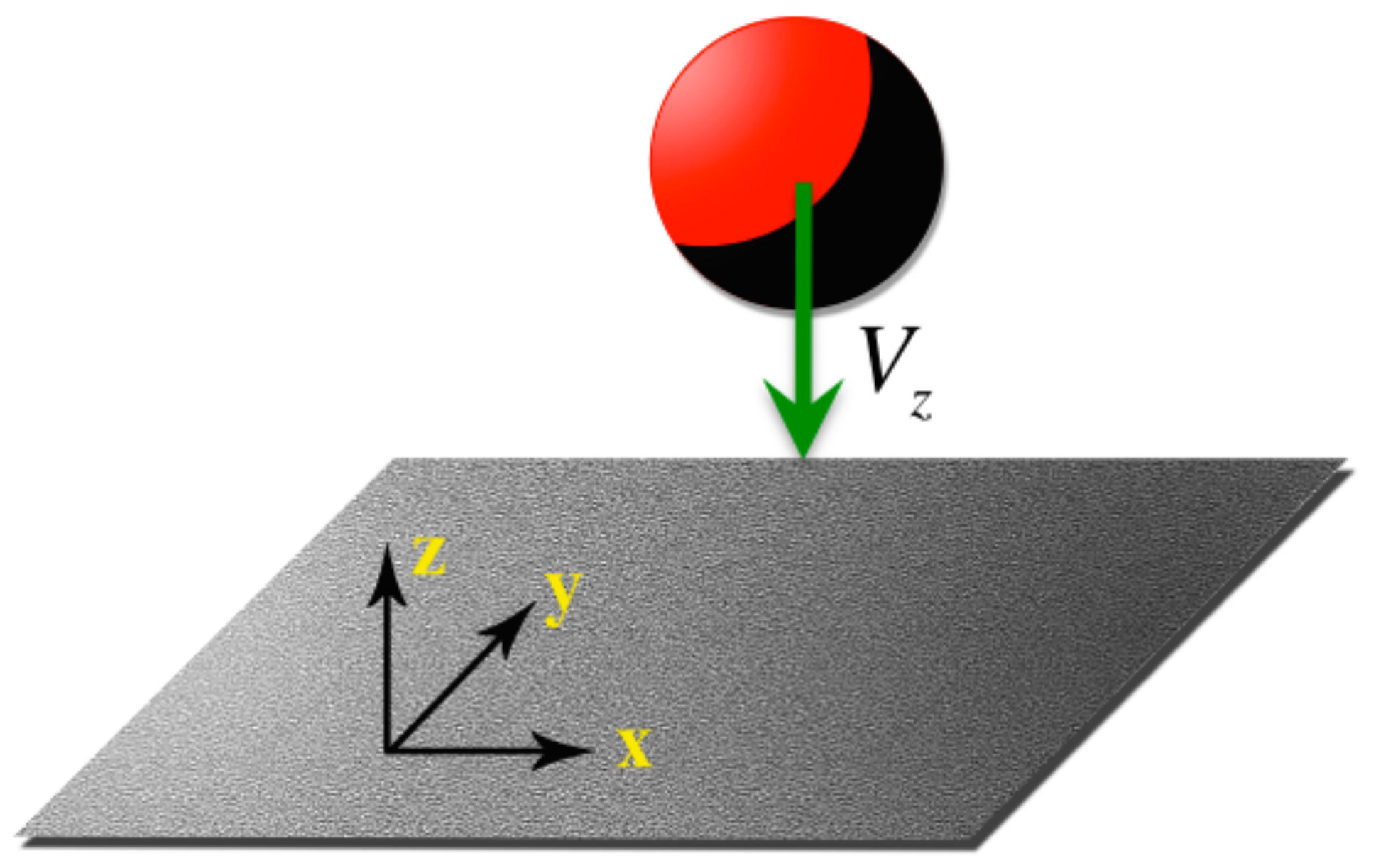}}
\subfigure[]{\label{sch4}\includegraphics[width=0.24\textwidth]{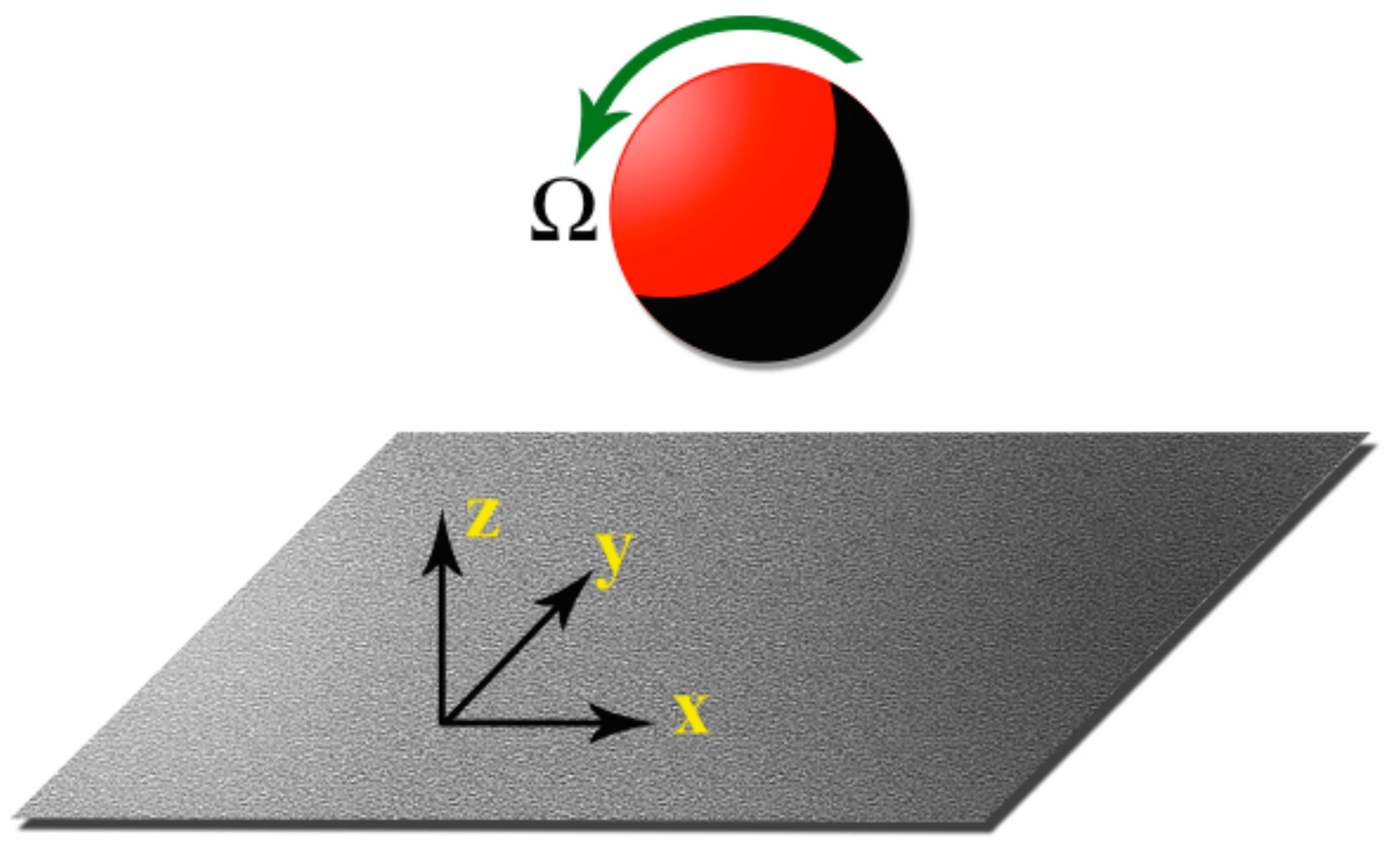}}
\caption{Decomposition of the velocity field into (a) surface diffusiophoretic slip with no net movement (translation or rotation) of the particle, (b) translation in x direction with no slip or rotation or motion in the z direction, (c) translation in z direction with no slip and rotation and motion in the x direction and (d) rotation with no slip or translation.}
\label{sch} 
\end{figure}
The number of independent components of the force and torque on the particle is reduced by symmetry to just three. The particle has a symmetry axis, $\hat{\bf{n}}$, running through the center of the cap towards the center of the particle, which by an appropriate choice of coordinate system can be chosen to lie in the $x$-$z$ plane, where the $\hat{\bf{z}}$ axis is normal to the solid boundary plane. In this coordinate system, the particle and concentration fields are symmetric about the $x$-$z$ plane, and there is no force in the $y$-direction. Likewise, in the absence of any tendency to rotate out of this plane, the angular velocity and torque must be in the $y$-direction. Thus
\begin{equation}
F_y  = T_x  = T_z  = 0.
\end{equation}
An alternative argument is to note that the only vectors in the problem are $\hat{\bf{n}}$ and $\hat{\bf{z}}$, so the force must be a linear combination of these, and the only pseudo vector available is $\hat{\bf{n}} \times \hat{\bf{z}}$, so the torque must lie in that direction. Choosing $\hat{\bf{n}}$ as above, the same non-zero force and torque components are obtained.
 
The fluid velocity field outside the thin interaction layer is governed by the Stokes flow and continuity equations. Due to the linearity of Stokes equation and the boundary conditions associate with it, the hydrodynamic problem can be decomposed into four separate problems as illustrated in Fig.~\ref{sch}
\begin{enumerate}
  \item 
  ``a'': The colloidal particle is stationary, i.e. ($V_x=0$ and $V_z=0$ and $\Omega_y$=0), but the fluid slips with the diffusiophoretic slip velocity ${\bf{v}}_s$.
  \item 
  ``b'': The colloidal particle is translating with velocity $V_x$ in the parallel direction along the wall with  no rotation ($\Omega_y=0$) or translation in the $z$ direction ($V_z=0$). In this case, there are force and torque shown as $ F^{T} _x$ and ${\rm T}_y ^{T} $ exerted on the particle.
  \item 
   ``c'': The colloidal particle is translating with velocity $V_z$ in the normal direction to the wall with no rotation ($\Omega_y=0$) or translational motion along the wall ($V_x=0$). In this case, the motion is axisymmetric and there is a force $ F^{N} _z$ on the particle in the $z$ direction.
    \item 
   ``d'': The colloidal particle is rotating around the $y$ axis with angular velocity $\Omega_y$ and there is no translational motion in any direction ($V_z=0$ and $V_x=0$). In this case, there are force and torque as $F^{R} _x $ and ${\rm T}^{R} _y $ on the particle.
\end{enumerate}
\indent The general procedure for solving each one of these problems is the same except for the boundary conditions on the colloid surface. In addition, the solutions to problems ($b$) - ($d$) have already been examined in the literature, and we use the previous results on these problems to validate our calculations \cite{Maude,Brennerwall,DO,ONeal,JFM2}.\\

\indent It was shown previously the slow motion of a sphere in a vicinity of a planar wall can be solved analytically in cylindrical coordinate where the components of the velocity can be written as \cite{JFM2}:
\begin{align}
 &p( {\alpha ,\beta ,\phi } ) = \frac{{\sqrt {\cosh \beta  - \cos \alpha } }}{\epsilon } \nonumber\\ 
 &\times \sum\limits_{m = 0}^\infty  {\sum\limits_{n = m}^\infty  {(A_{n,m} \sinh (n + \frac{1}{2})\beta  + B_{n,m} \cosh (n + \frac{1}{2})\beta )} } \;P_n^m (\cos \alpha )\;\cos (m\phi  + \gamma _m ),\label{aux1} 
\end{align} 
\begin{eqnarray}
\begin{split}
 &v_r  = \frac{{\sin \alpha }}{{2\sqrt {\cosh \beta  - \cos \alpha } }} \\ 
 & \times \sum\limits_{m = 0}^\infty  {\sum\limits_{n = m}^\infty  {(A_{n,m} \sinh (n + \frac{1}{2})\beta  + B_{n,m} \cosh (n + \frac{1}{2})\beta )} } \;P_n^m (\cos \alpha )\;\cos (m\phi  + \gamma _m ) \\ 
&  + \sqrt {\cosh \beta  - \cos \alpha } \;[\sum\limits_{n = 1}^\infty  {(E_{n,0} \sinh (n + \frac{1}{2})\beta  + F_{n,0} \cosh (n + \frac{1}{2})\beta )} \;P_n^1 (\cos \alpha )\;\cos (\gamma _0 ) \\ 
  &+ \frac{1}{2}\sum\limits_{m = 1}^\infty  {\sum\limits_{n = m + 1}^\infty  {(E_{n,m} \sinh (n + \frac{1}{2})\beta  + F_{n,m} \cosh (n + \frac{1}{2})\beta )} } \;P_n^{m + 1} (\cos \alpha )\;\cos (m\phi  + \gamma _m ) \\ 
 & + \frac{1}{2}\sum\limits_{m = 1}^\infty  {\sum\limits_{n = m - 1}^\infty  {(G_{n,m} \sinh (n + \frac{1}{2})\beta  + H_{n,m} \cosh (n + \frac{1}{2})\beta )} } \;P_n^{m - 1} (\cos \alpha )\;\cos (m\phi  + \gamma _m )],
\label{vel1}  \\ 
\end{split}
\end{eqnarray}

\begin{eqnarray}
\begin{split}
 &v_\phi   = \sqrt {\cosh \beta  - \cos \alpha } \;[\sum\limits_{n = 1}^\infty  {(G_{n,0} \sinh (n + \frac{1}{2})\beta  + H_{n,0} \cosh (n + \frac{1}{2})\beta )} \;P_n^1 (\cos \alpha )\;\sin (\gamma _0 ) \\ 
 & + \frac{1}{2}\sum\limits_{m = 1}^\infty  {\sum\limits_{n = m + 1}^\infty  {(E_{n,m} \sinh (n + \frac{1}{2})\beta  + F_{n,m} \cosh (n + \frac{1}{2})\beta )} } \;P_n^{m + 1} (\cos \alpha )\;\sin (m\phi  + \gamma _m ) \\ 
 & - \frac{1}{2}\sum\limits_{m = 1}^\infty  {\sum\limits_{n = m - 1}^\infty  {(G_{n,m} \sinh (n + \frac{1}{2})\beta  + H_{n,m} \cosh (n + \frac{1}{2})\beta )\;} } P_n^{m - 1} (\cos \alpha )\;\sin (m\phi  + \gamma _m )],  
\label{vel2}  \\ 
\end{split}
\end{eqnarray}
\begin{eqnarray}
\begin{split}
 &v_z  = \frac{{\sinh \beta }}{{2\sqrt {\cosh \beta  - \cos \alpha } }} \\ 
  &\times \sum\limits_{m = 0}^\infty  {\sum\limits_{n = m}^\infty  {(A_{n,m} \sinh (n + \frac{1}{2})\beta  + B_{n,m} \cosh (n + \frac{1}{2})\beta )\;} } P_n^m (\cos \alpha )\;\cos (m\phi  + \gamma _m ) \\ 
  &+ \sqrt {\cosh \beta  - \cos \alpha }  \\ 
 & \times \sum\limits_{m = 0}^\infty  {\sum\limits_{n = m}^\infty  {(C_{n,m} \sinh (n + \frac{1}{2})\beta  + D_{n,m} \cosh (n + \frac{1}{2})\beta )} } \;P_n^m (\cos \alpha )\;\cos (m\phi  + \gamma _m ), 
\label{vel3}  \\ 
\end{split}
\end{eqnarray}
where ${A_{n,m} }$, $B_{n,m}$, ..., ${H_{n,m} }$ are 8 sets of unknown coefficients that have to be determined by applying the relevant boundary conditions and the equation of continuity. The detailed analysis for problems ($a$) - ($d$) are all given in the Appendices B-E. 
\subsection{Calculation of hydrodynamic drag forces and torques}
To determine the swimming translational and angular velocities of the colloid, we first take advantage of the the fact that the net force and torque exerted on the particle is zero. The drag forces and torques associate with each one of the problems ($a$) - ($d$) of Fig. \ref{sch} are summed and set equal to zero. Having determined the unknown coefficients of Stokes equation, we proceed by calculating the force and torque exerted on the colloid due to the diffusiophoretic motion of fluid around stationary colloid where, by definition we have,
\begin{eqnarray}
{\bf{F}} = \mathop{{\int\!\!\!\!\!\int}\mkern-21mu \bigcirc}\limits_{S_p} 
 {\;({\bf{\sigma}}.\;{\bf{e}}_{\bf{n}} )} \;dS,
\end{eqnarray}

where ${\bf{e}}_{\bf{n}}$ is a unit vector normal to the surface of colloid $S_p$, and $\sigma$ is the stress tensor
\begin{eqnarray}
\sigma  =  - p{\bf{I}} + \mu \;[\nabla {\bf{v}} + (\nabla {\bf{v}})^T ].
 \end{eqnarray}
\indent In the axisymmetric problem where the tilt angle $\Xi  = 0^{\circ}$ or $180^{\circ}$, it is obvious by symmetry that the fluid exerts a force on the colloid  solely in $z$-direction and the vertical force exerted on the colloid in terms of coefficients of Stokes solution governed by \cite{JFM2}
\begin{eqnarray}
F^{P} _z  = -2\sqrt {2\;} \pi \;\epsilon \;\mu \;V_{\infty}  \;\sum\limits_{n = 0}^\infty  {(C_{n,0}  + } D_{n,0} ) + (n + \frac{1}{2})\;(A_{n,0}  + B_{n,0} ), \label{force1}
\end{eqnarray}
where superscript $P$ stands for propulsion. We note here for a Janus colloid in an arbitrary configuration, the vertical force on the particle in the $z$-direction is only a contribution from $m=0$ in the previous formula (\ref{aux1}-\ref{vel3}).  
The colloidal particle also experiences a force in $x$-direction which is a contribution from $m=1$  component of coefficients of equations (\ref{aux1}-\ref{vel3}). In this case,  it can be shown that \cite{Keh},
\begin{eqnarray}
F^{P} _x  =  - \sqrt {2\;} \pi \;\epsilon \;\mu \;V_{\infty} \;\sum\limits_{n = 0}^\infty  {(G_{n,1}  + } H_{n,1} ) - n(n + 1)\;(A_{n,1}  + B_{n,1} ),
\end{eqnarray}
and lastly we need to evaluate the torque that has been exerted on the colloid due to the fluid slip. The general expression for torque is:
\begin{eqnarray}
{\bf{T}} = \mathop{{\int\!\!\!\!\!\int}\mkern-21mu \bigcirc}\limits_{S_p} 
 {({\bf{r}-\bf{r}}_p) }  \times (\sigma .\;{\bf{e}}_{\bf{n}} )\;dS.
\end{eqnarray}
 Here, the only non-zero component of the torque is in $y$-direction, and it is solely a  contribution of $m=1$ terms in coefficients of pressure and velocity field equations (\ref{aux1}-\ref{vel3}) and can be written as  \cite{Keh}:
\begin{eqnarray}
\begin{split}
 &T^{P} _y  = \sqrt {2\;} \pi \;\epsilon^2 \;\mu \;V_{\infty}  \\ 
 &\sum\limits_{n = 0}^\infty{ - n(n + 1)({2(C_{n,1}  + D_{n,1}) + \coth \beta _0 (A_{n,1}  + B_{n,1} )}) - (2n + 1 - \coth } \beta _0 )(G_{n,1}  + H_{n,1} ). \\ 
\end{split}
\end{eqnarray}
\indent To calculate the swimming velocity of the particle we use the fact that the colloid is force and torque free. In this case, the swimming velocity in $z$ direction can be obtained from
\begin{eqnarray}
F^N _z  + F^{P} _z  = 0,\label{fz}
\end{eqnarray}
where the superscripts $N,P$ refer to normal motion and propulsion problems respectively as noted before.  Similarly, the translational velocity in the $x$-direction, $V_x$, and the angular velocity in the $y$ direction, $\Omega_y$, satisfy
\begin{eqnarray}
\left\{ \begin{array}{l}
 F^{T} _x  + F^{R} _x  + F^{P} _x  = 0, \\
  T_y ^{T}  +  T^{R} _y  +  T^{P} _y  = 0 ,\label{fxty} 
 \end{array} \right.
\end{eqnarray}
where $T,R$ refer to translation and rotation problems respectively.
Expressions for the various forces and torques related to problems ($b$)-($d$) can be found in the Appendices C-E.
\subsection{Reynolds Reciprocal Theorem}
In this section we utilize an alternative approach to determine the diffusiophoretic force and torque on the colloidal particle associated with the first problem shown in Fig.~\ref{sch} . This approach is most useful for cases where the details of the flow field are not important, and is based on the Reynolds Reciprocal Theorem (RRT) \cite{hapbren83}. 
Suppose (${\bf{v}}^{\prime},\sigma^{\prime}$) and (${{\bf{v''}}},\sigma''$) are the solutions of Stokes flow for incompressible fluid within an arbitrary fluid volume $\Sigma$. The RRT states that:
 \begin{eqnarray}
 \mathop{{\int\!\!\!\!\!\int}\mkern-21mu \bigcirc}\limits_{\partial \Sigma } 
 {{{\bf{n}}_{\Sigma}} \cdot \sigma ' \cdot {\bf{v''}}dS}  = \mathop{{\int\!\!\!\!\!\int}\mkern-21mu \bigcirc}\limits_{\partial \Sigma } 
 {{{\bf{n}}_{\Sigma}} \cdot \sigma '' \cdot {\bf{v'}}dS}.\label{rrt}
 \end{eqnarray}
 Here, the boundary $\partial \Sigma$ consists of the particle surface, the wall $z=0$, and a hemisphere at large distances from the particle.
Since for a colloidal swimmer the flow field decays as a dipole ${\bf{v}}\sim r^{-2}$ and ${\sigma}\sim r^{-3}$ at large distances, the contribution from the outer surface integral is always zero and because of the no-slip condition the wall contribution vanishes, and hence we are left with an integral on the colloid surface.\\
\indent For a general configuration where the colloid moves in an arbitrary direction with respect to the wall, similar to previous section we need to obtain the propulsion forces, $F_z^P$ and $F_x^P$ in $z$ (normal) and $x$ (parallel) directions respectively as well as the torque $T_y^P$, around the $y$ axis. To evaluate these unknowns, we take advantage the detailed solution of the three classical problems explained extensively in literature \cite{ONeal,DO,Maude,Brennerwall,JFM2} and are recapitulated in the Appendices C-E. The propulsion force in the $z$ direction can be found by letting ($\bf{v}''$, $\sigma''$) be the solution of translational motion of a colloid normal to the planar wall governed by Brenner and Maude (discussed in Appendix D) as \cite{Brennerwall,Maude},
 \begin{align}
&F_z^{P} = -\mathop{{\int\!\!\!\!\!\int}\mkern-21mu \bigcirc}\limits_{{S_p}} 
 {{{\bf{e}}_\beta } \cdot {{\sigma }_{\alpha \beta }}~{{\bf{e}}_\beta }{{\bf{e}}_\alpha } \cdot \frac{{\cosh \beta  - \cos \alpha }}{\epsilon}\frac{{\partial C}}{{\partial \alpha }}{{\bf{e}}_\alpha }dS}  = \nonumber \\
&~~~~-\int_0^{2\pi } {d\phi  \int_0^\pi  {{{\sigma }_{\alpha \beta }}} \frac{{\epsilon\sin \alpha }}{{(\cosh {\beta _0} - \cos \alpha )}}{{\left. {\frac{{\partial C}}{{\partial \alpha }}} \right|}_{\beta  = {\beta _0}}}d\alpha },\label{fzn}
 \end{align}
where $\bf{e}_{\alpha}$ and $\bf{e}_{\beta}$ are the unit normals in bispherical coordinate and $\sigma_{\alpha \beta}$ is the stress tensor associate with the problem discussed in Appendix D. Inasmuch as ${\sigma}_{\alpha \beta}$ is related to an axisymmetric problem, it is independent of azimuthal angle and  the only term inside the integrand which is a function of $\phi$ is $\frac{{\partial C}}{{\partial \alpha }}$. Utilizing the orthogonality, we realize that the only surviving term is for the case where $m=0$ and therefore the double sum in \ref{p22} reduces to one sum.
\begin{eqnarray}
\int_0^{2\pi } {{{\left. {\frac{{\partial C}}{{\partial \alpha }}} \right|}_{\beta  = {\beta _0}}}} d\phi  = \frac{{\sin \alpha }}{{2\sqrt {\cosh {\beta _0} - \cos \alpha } }}\sum\limits_{n = 0}^\infty  {\cosh (n + 0.5)\beta ({\widetilde{A}_{n0}}\cos m\phi  + {\widetilde{B}_{n0}}\sin m\phi )} P_n^0(\cos \alpha ), \nonumber
\end{eqnarray}
\begin{eqnarray}
 - \sin \alpha \sqrt {\cosh {\beta _0} - \cos \alpha } \sum\limits_{n = 0}^\infty  {\cosh (n + 0.5)\beta ({\widetilde{A}_{n0}}\cos m\phi  + {\widetilde{B}_{n0}}\sin m\phi )} \frac{{dP_n^0(\cos \alpha )}}{{d\cos \alpha }},
\end{eqnarray}
at this point, we can no longer proceed with analytical calculation and the integral in Eq.~\ref{fzn} needed to be evaluated numerically. To do so, we first calculate non-dimensional concentration field and its derivatives as well as surface stress tensor component according to appropriate relations and thereafter we undertake integration utilizing trapezoidal rule to have the propulsion force in $z$ direction.\\
\indent To determine the propulsion force parallel to the wall (in $x$ direction), we let ($\bf{v}''$, $\sigma''$) in Eq.~\ref{rrt} be the solution of translational motion of a non-rotating colloid parallel to a planar wall which satisfy the no slip boundary condition on particle surface and the wall. This problem was governed first by O'Neill \cite{ONeal}. Having used RRT, the propulsion force in the $x$ direction can be written as,\begin{eqnarray}
F_x^{P} =  -\mathop{{\int\!\!\!\!\!\int}\mkern-21mu \bigcirc}\limits_{{S_p}} 
 {{{\bf{e}}_n} \cdot \sigma_{2} \cdot \nabla CdS =  - ({I_1} + {I_2} + {I_3}} ),
\end{eqnarray}
where,
\begin{align}
&{I_1} = \pi \int_0^\pi  {(\sigma_2  \cdot {{\bf{e}}_n} \cdot {{\bf{e}}_r})\frac{{\epsilon \sin \alpha (\cosh \beta_0 \cos \alpha  - 1)}}{{{{(\cosh \beta_0  - \cos \alpha )}^2}}}}   \nonumber \\
&\times[\frac{{\sin \alpha }}{{2\sqrt {\cosh \beta_0  - \cos \alpha } }}\sum\limits_{n = 1}^\infty  {\cosh (n + 0.5)\beta_0 {\widetilde{B}_{n1}}P_n^1(\cos \alpha )}\nonumber\\
&-\sin \alpha \sqrt {\cosh \beta_0  - \cos \alpha } \sum\limits_{n = 1}^\infty  {\cosh (n + 0.5)\beta_0 {\widetilde{B}_{n1}}\frac{{dP_n^1(\cos \alpha )}}{{d\cos \alpha }}} ]d\alpha,\label{ie11} 
\end{align},
\begin{eqnarray}
{I_2} =   \pi \int_0^\pi  {\frac{{\epsilon~(\sigma_2  \cdot {{\bf{e}}_n} \cdot {{\bf{e}}_\phi })}}{{(\cosh \beta_0  - \cos \alpha )}}[-\sqrt {\cosh \beta_0  - \cos \alpha } \sum\limits_{n = 1}^\infty  {\cosh (n + 0.5)\beta_0 {\widetilde{B}_{n1}}P_n^1(\cos \alpha )} ]d\alpha },\nonumber\\
\end{eqnarray}
\begin{align}
&{I_3} =  - \pi \int_0^\pi  {(\sigma_2  \cdot {{\bf{e}}_n} \cdot {{\bf{e}}_z})\frac{{\epsilon\sinh \beta_0 {{\sin }^2}\alpha }}{{{{(\cosh \beta_0  - \cos \alpha )}^2}}}} [\frac{{\sin \alpha }}{{2\sqrt {\cosh \beta_0  - \cos \alpha } }}\sum\limits_{n = 1}^\infty  {\cosh (n + 0.5)\beta_0 {\widetilde{B}_{n1}}P_n^1(\cos \alpha )} \nonumber \\
&-\sin \alpha \sqrt {\cosh \beta_0  - \cos \alpha } \sum\limits_{n = 1}^\infty  {\cosh (n + 0.5)\beta_0 {\widetilde{B}_{n1}}\frac{{dP_n^1(\cos \alpha )}}{{d\cos \alpha }}} ]d\alpha,\label{ie33} 
\end{align}
where $\sigma_2$ accounts for the stress tensor associate with the O'Neill's analysis \cite{ONeal} (see Appendix C) and $\widetilde{B}_{n1}$ can be found according to the recursive relations given in Eq.~\ref{p26} and \ref{p27}.\\
\indent Lastly, to obtain the propulsion torque around the $y$ axis, we let  ($\bf{v}''$, $\sigma''$) in Eq.~\ref{rrt} be the solution of rotation of a non-translating colloid around the $y$ axis in a proximity of a wall. This problem was derived in detail by Dean and O'Neill \cite{DO} (see Appendix E). Having used RRT, the diffusiophoretic torque around the $y$ axis can be written likewise as,
\begin{eqnarray}
T_y^{P} = -\epsilon ({\tilde{I}_1} + {\tilde{I}_2} + {\tilde{I}_3} ),
\end{eqnarray}   
where $\tilde{I}_1$, $\tilde{I}_2$ and $\tilde{I}_3$ are given in \ref{ie11} - \ref{ie33} except the stress tensor $\sigma_3$ associate with the Dean and O'Neill's solution should be replaced \cite{DO}.\\
\indent Having determined the propulsion force and torque, we again utilize the fact that the total force and torque on colloid should be zero and via a force/torque balance in three directions, one can obtain the swimming velocity and angular velocity of the colloid by using Eq.~\ref{fz} and \ref{fxty}. Our results from RRT analysis are in complete agreement (variations are less than $1\%$ in all separation distances and coverages) with the direct solution of Stokes flow given in previous section.\\

\section{Result and discussion}\label{R&D}
\subsection{Concentration distribution}
\begin{figure}[h]
\centering
\linespread{0.5}
\subfigure{\label{d20}\includegraphics[width=1.0\textwidth]{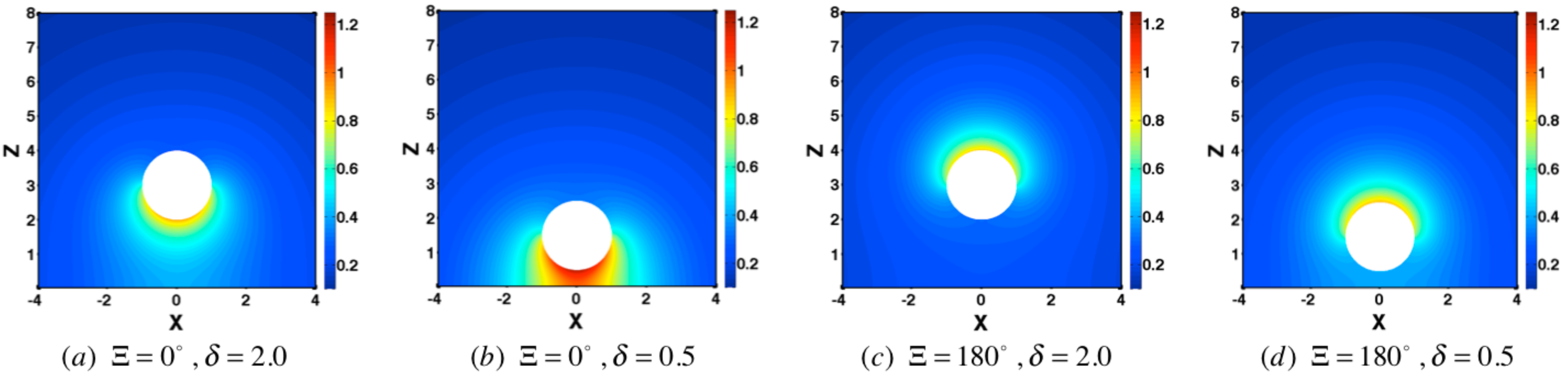}}
\caption{Non-dimensional concentration field for half active Janus particle $(\theta _{cap}  = 90)$ in the $x$-$z$ plane for axisymmetric orientation of colloids $\Xi  = 0^ \circ ,\Xi  = 180^ \circ$ at two different non-dimensional distances from wall $(\delta)$.}
\label{conce1}
 \end{figure}

We begin with the case of an inclination angle of $0^ \circ$ or $180^ \circ$, where the concentration field around the colloid is independent of azimuthal angle $\phi$. When the active section of the colloid faces away from the wall, $\Xi  = 180^ \circ$, the solute distribution does not change drastically as the colloid approaches the wall since the solute can diffuse freely away from the colloid to distant regions far from it. On the contrary, if the active side of the colloid faces the wall, $\Xi  = 0^ \circ$, as the particle approaches the wall the concentration field is highly distorted by the presence of an impenetrable barrier resulting in a very high solute concentration in a thin gap region adjacent to the wall (see Fig.~\ref{conce1} for the case $\theta _{cap}  = 90$ ). 
 
\begin{figure}[h]
\centering
\subfigure{\label{d26}\includegraphics[width=1.0\textwidth]{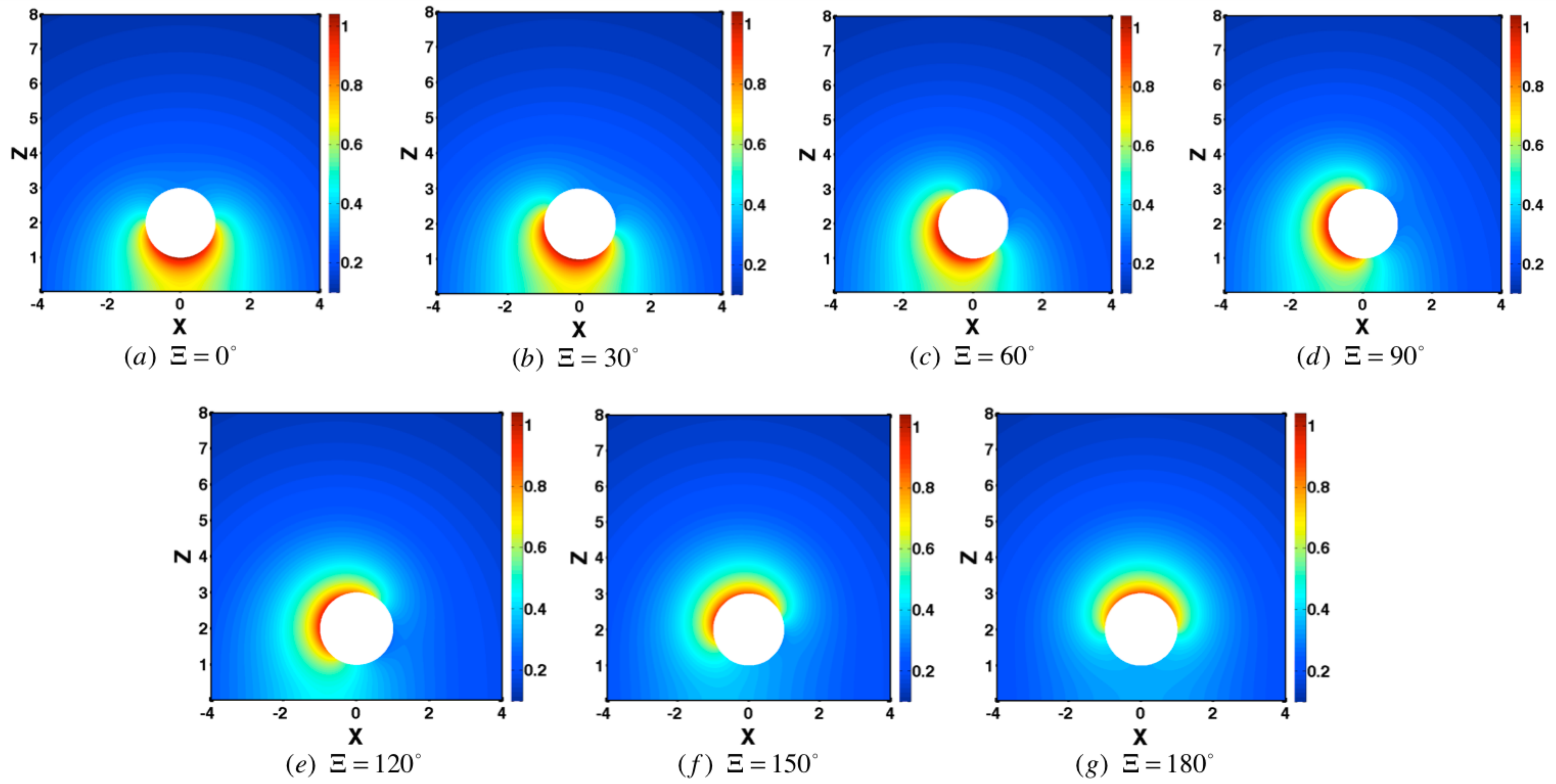}}
\caption{Non-dimensional concentration field for a Janus particle $(\theta _{cap}  = 90)$ in the $x$-$z$ plane for different inclination angles $(\Xi)$ when the particle is one radius away from the wall.}
\label{conce2}
\end{figure}

For all other colloid orientations, $0^ \circ < \Xi  < 180^ \circ$, the concentration field is fully three dimensional in bispherical coordinates $C\left( {\alpha ,\beta ,\phi } \right)$. Representative concentration fields in $x$-$z$ plane around a Janus particle $(\theta _{cap}  = 90)$, for various inclination angles and a fixed separation distance $\delta=1$, are shown in Fig. \ref{conce2}. The effect of the wall is again more pronounced when the active sections faces the wall and as the inclination angle approaches  zero, where the solute concentration is augmented in the thin gap. In general, the concentration fields are no longer symmetric around the colloid axis (see Fig.~\ref{Janusnearwalls}) which results in a rotation about it which  we will discuss in detail presently. The asymmetric distribution around the colloid axis is more pronounced as the inclination angle approaches to $\Xi  = 90^ \circ$.\\

\begin{figure}[h]
\centering
\subfigure{\label{d26}\includegraphics[width=0.8\textwidth]{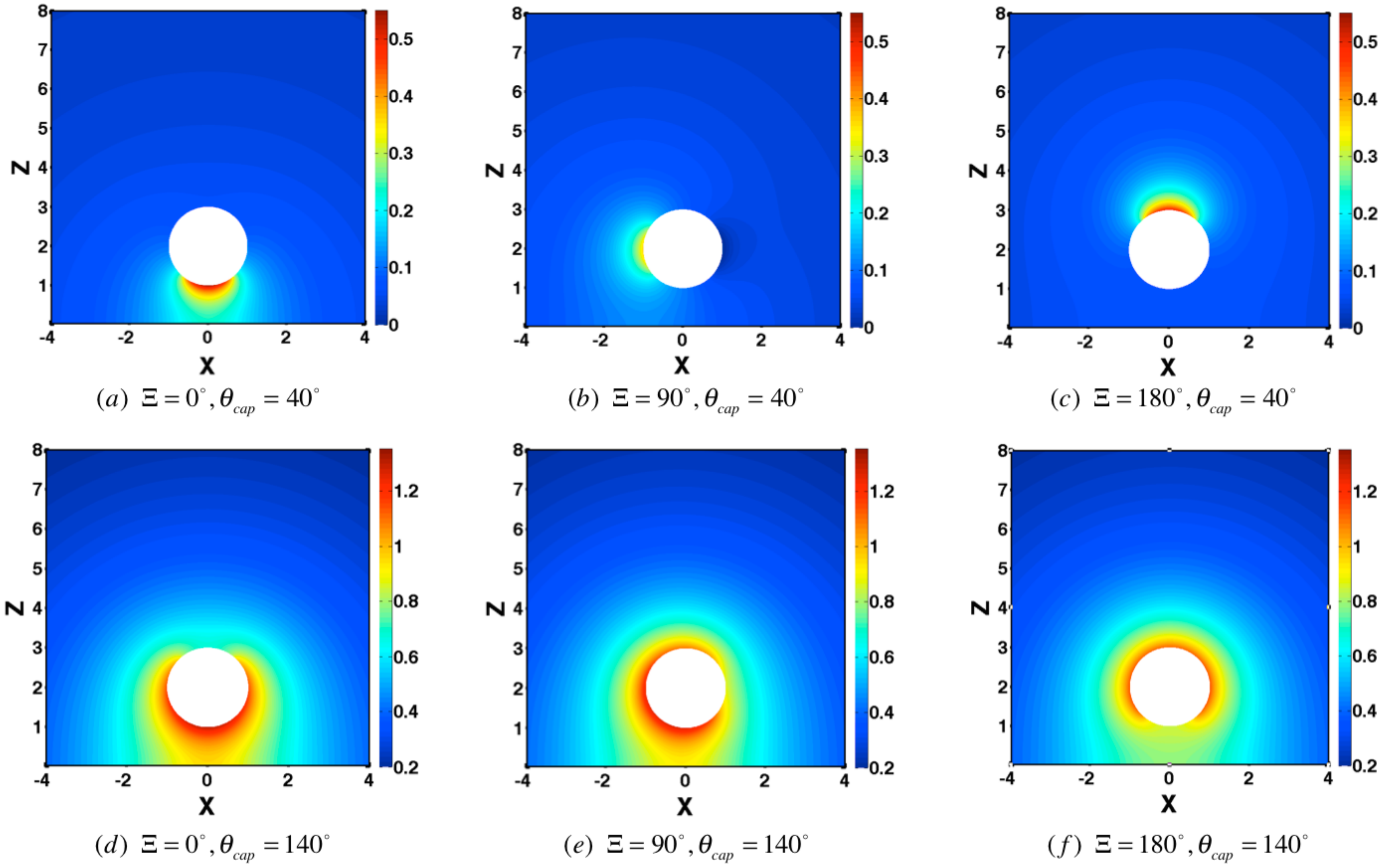}}
\caption{Non-dimensional concentration field for low active coverage $(\theta _{cap}  = 40)$ and high active coverage $(\theta _{cap}  = 140)$ in the $x$-$z$ plane for different tilt angles $(\Xi)$ at fixed distance from wall $\delta=1$.}
\label{conce3}
\end{figure}

\indent The degree of concentration asymmetry about the colloid axis can be altered as well by varying the extent of the active section of the colloid surface. Fig.~\ref{conce3} illustrates the concentration fields in the $x-z$ plane for the case of small active coverage, $\theta _{cap}  = 40^ \circ$, as well as for a greater active section, $\theta _{cap}  = 140^ \circ$. Note that for $\theta _{cap}  = 140^ \circ$, there is a substantial increase in solute concentration in the gap relative to the case where $\theta _{cap}  = 90^ \circ$ for similar inclination angles. Thus, we see that the effect of solid wall for colloids with high active coverage is more significant compared to the ones with smaller coverage.
\subsection{Swimming velocities and trajectories}
\subsubsection{ Swimming normal to the solid wall}
 Fig.~\ref{velo12} illustrates normalized swimming velocities (non-dimensionalized with the swimming velocity ${V_\infty}$ for an infinite medium) as a function of non-dimensional separation distance $\delta$ (non-dimensionalized with the colloid particle radius $R$), for different inclination angles for a half catalytically active Janus particle $(\theta _{cap}  = 90^ \circ)$ . When the initial orientation is axisymmetric, $\Xi  = 0^ \circ$ or $\Xi  = 180^ \circ$, the particle translates in the $z$-direction without rotation, but the behavior differs in the two cases. When the active side of the colloid faces the wall the solute concentration builds up in the gap region, resulting in high gradient and a large slip velocity, leading to a large diffusiophoretic propulsive force driving the particle away at an enhanced velocity $\displaystyle{\frac{{V_z }}{{V_\infty  }} > 1}$. Of course at large separations the wall lose relevance and the velocity relaxes to an infinite medium value $\displaystyle{\frac{{V_z }}{{V_\infty  }} = 1}$. In the opposite case when the active side faces away from the wall, the concentration filed and diffusiophoretic force is little altered by the boundary but the hydrodynamic lubrication force \cite{Brennerwall}, growing as $\displaystyle{\sim \frac{1}{\delta }}$, dominates at small gaps and brings the particle nearly to rest. Again, at large distance from the wall, the infinite medium result is recovered, $\displaystyle{\frac{{V_z }}{{V_\infty  }} = -1}$ in this case. These general remarks still apply for other values of $\theta _{cap}$ except for the case of very high coverage which will be discussed later.  

\subsubsection{Repulsion and reflection of a colloid from wall}
When the orientation of particle is no longer axisymmetric, the concentration field is fully three dimensional, so that the particle moves in the $x$-$z$ plane and rotates about the $y$-axis. The wall is effective only at separations  $\delta  < 5$ and at larger distances the infinite medium result applies with ($\displaystyle{\frac{{V_x }}{{V_\infty  }}}$, $\displaystyle{\frac{{V_z }}{{V_\infty  }}}$) $\to$ ($\sin \Xi$, $\cos \Xi $). 
 
\begin{figure}[!]
\centering
\subfigure[]{\label{velo12}\includegraphics[width=0.35\textwidth]{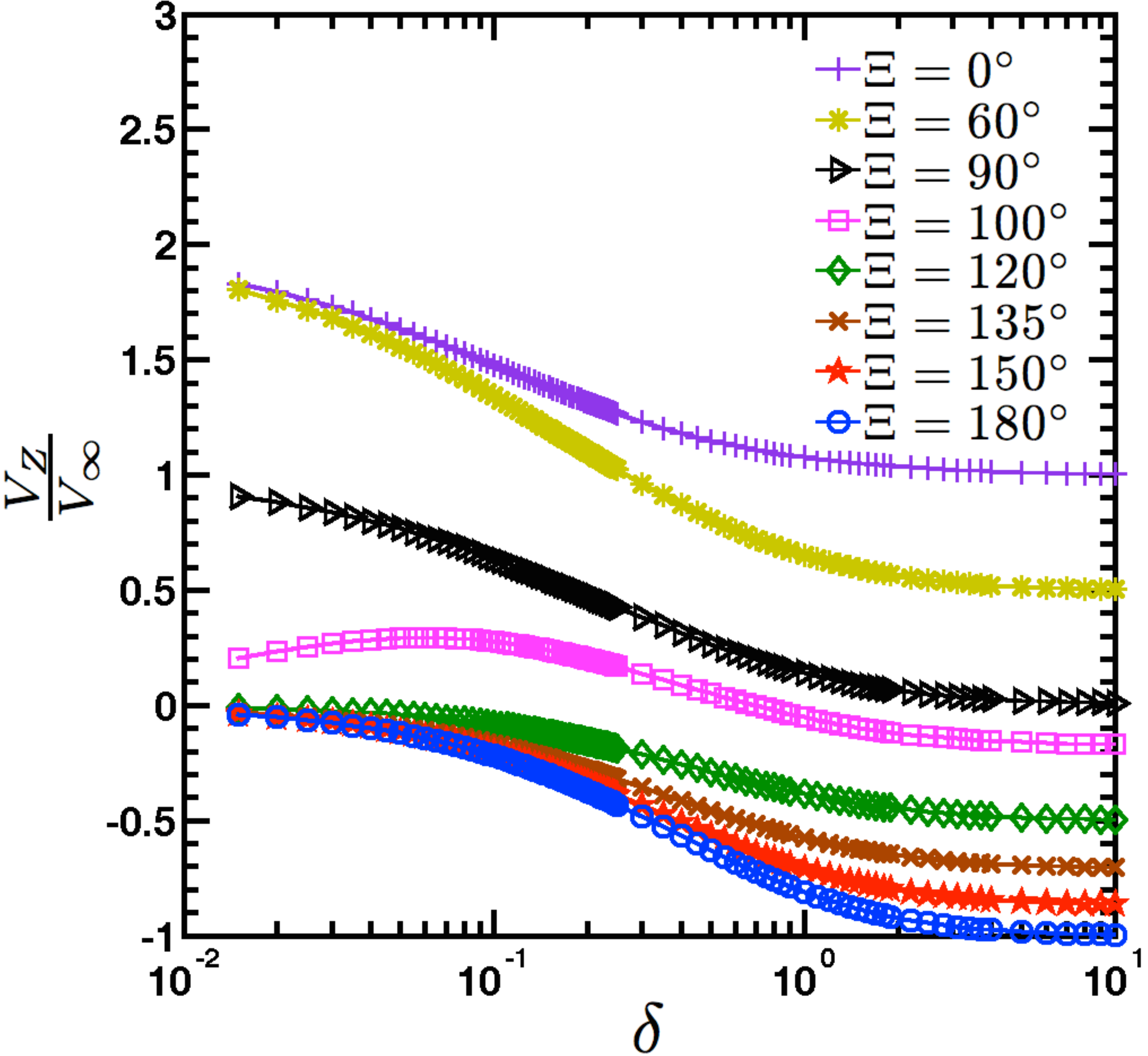}}
\subfigure[]{\label{velo13}\includegraphics[width=0.35\textwidth]{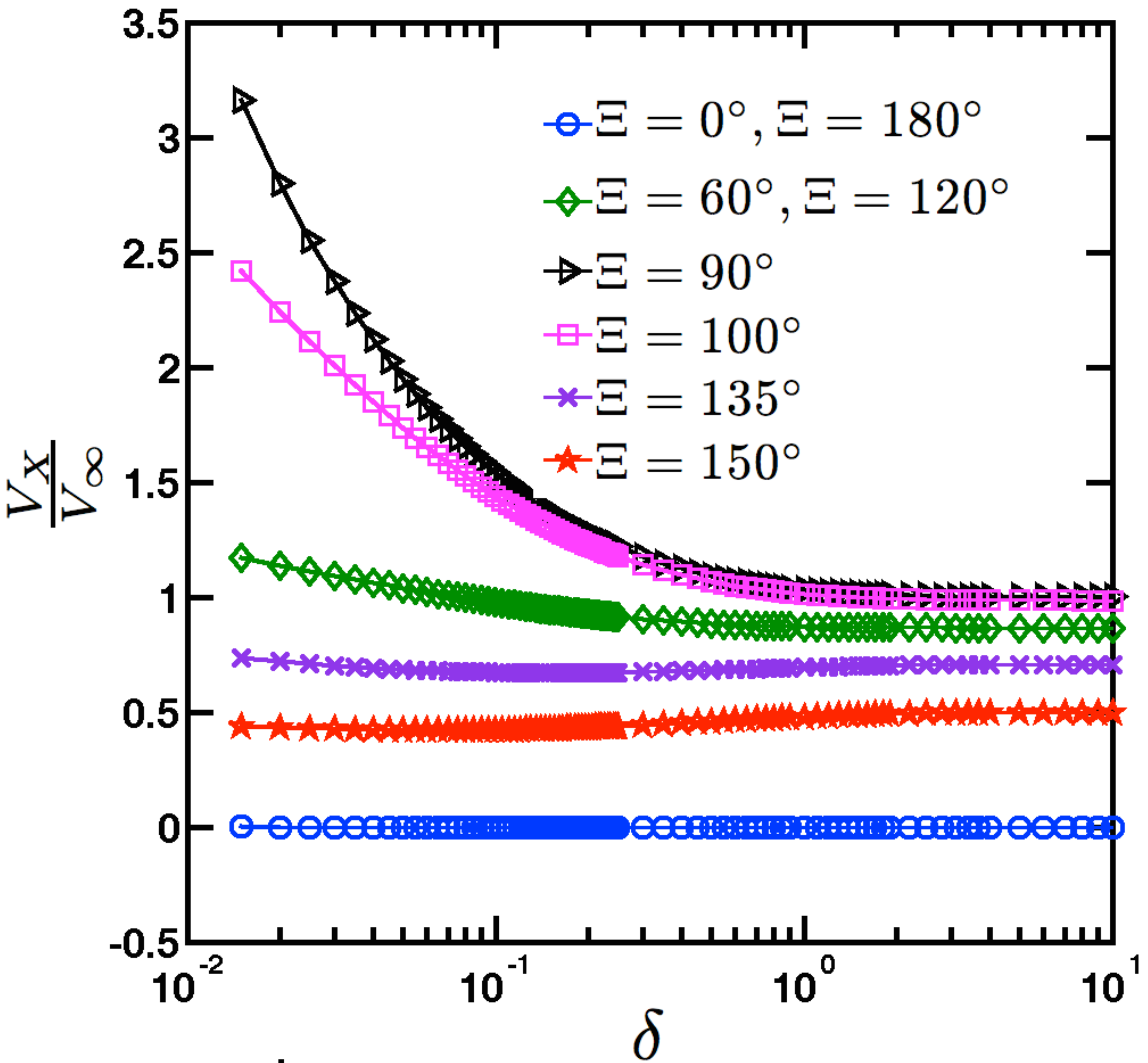}}
\subfigure[]{\label{velo14}\includegraphics[width=0.35\textwidth]{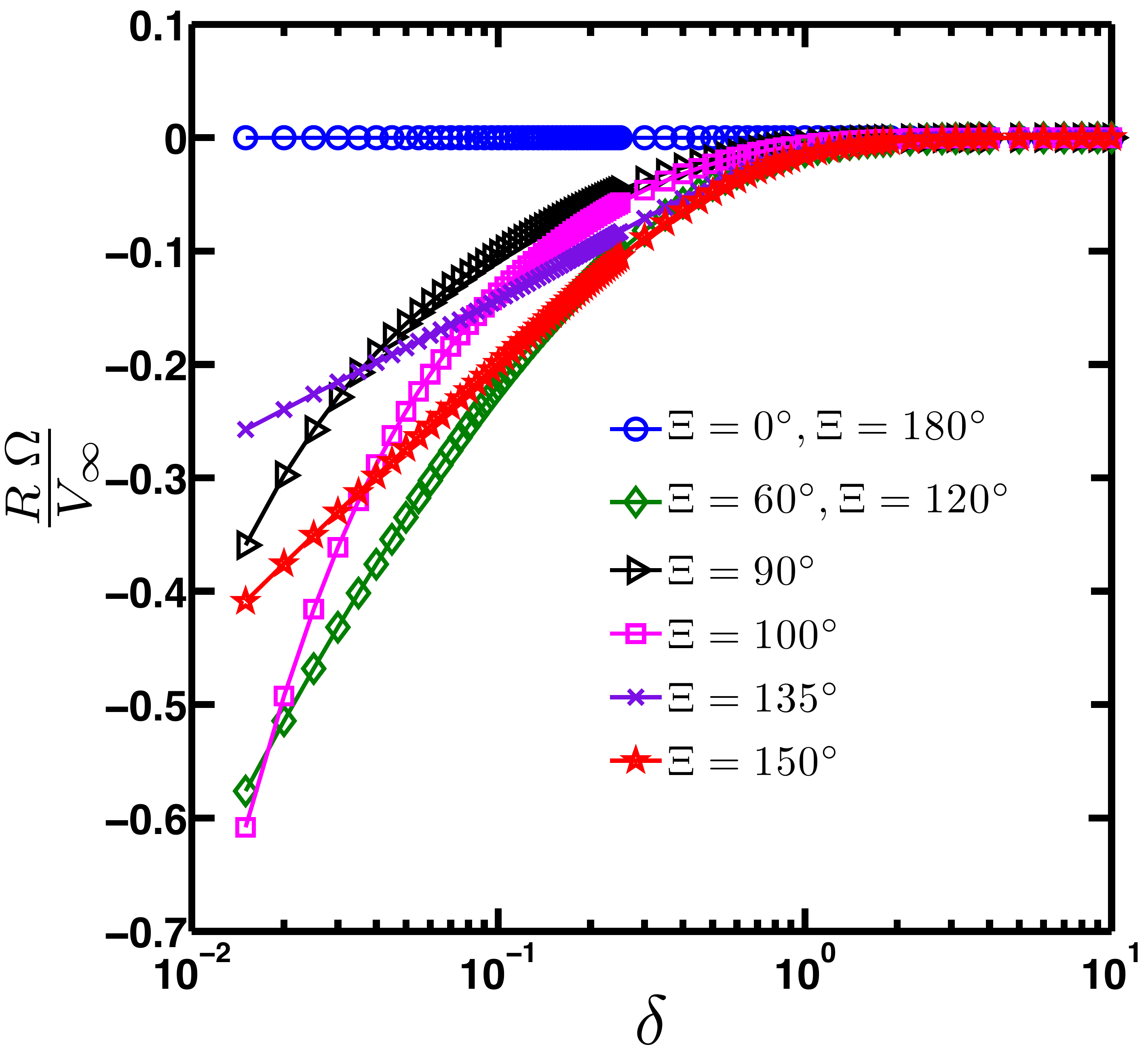}}
\subfigure[]{\label{traj12}\includegraphics[width=0.46\textwidth]{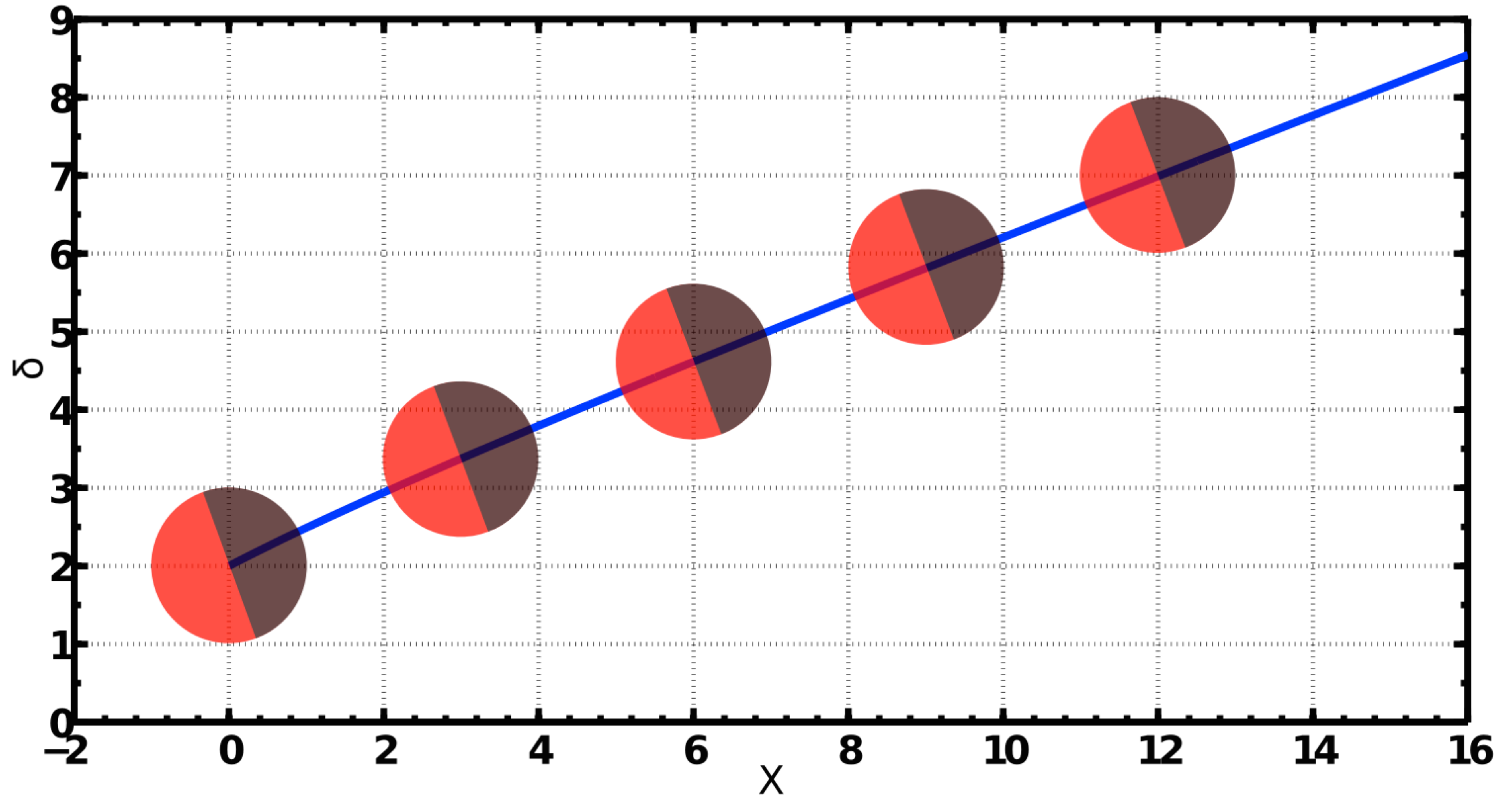}}
\subfigure[]{\label{traj13}\includegraphics[width=0.70\textwidth]{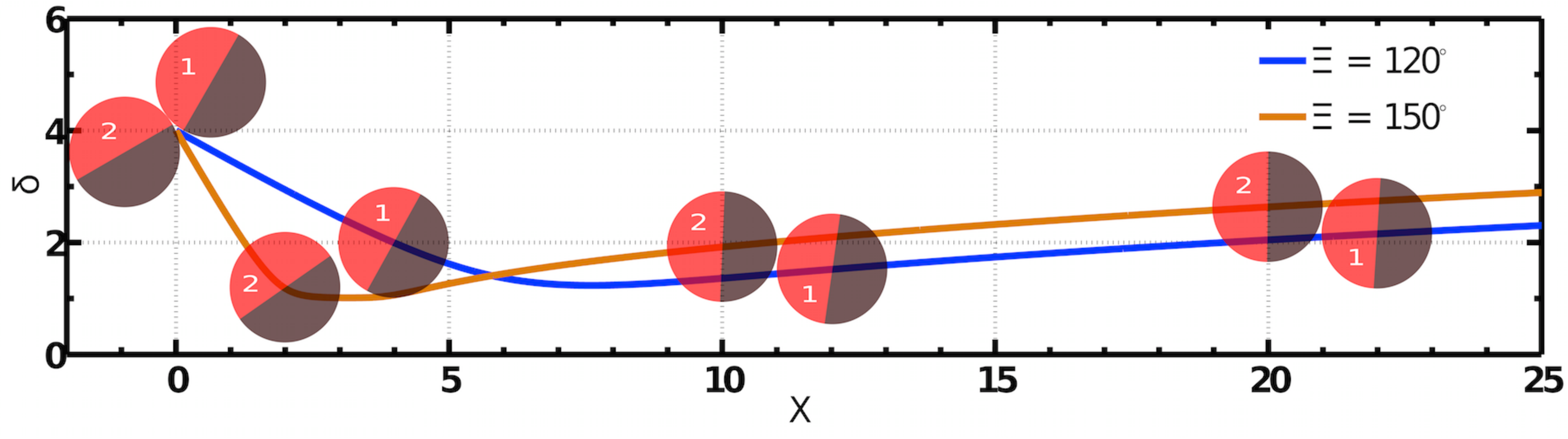}}
\caption{\ref{velo12} Non-dimensional swimming velocity in the $z$ direction, \ref{velo13} non-dimensional swimming velocity in the $x$ direction, \ref{velo13} Non-dimensional angular velocity as a function of separation distance $\delta$ for half covered Janus particle $(\theta _{cap}  = 90^ \circ )$. Sample trajectories for half covered Janus particle $(\theta _{cap}  = 90^ \circ )$ \ref{traj12} initial configuration: $\delta=1.0$, $\Xi=70^ \circ$ and \ref{traj13} initial configuration: $\delta=3.0$, $\Xi=120^ \circ, \Xi=150^ \circ $.}
\label{velo1} 
\end{figure}
\begin{figure}[!]
\centering
\subfigure[]{\label{velo22}\includegraphics[width=0.35\textwidth]{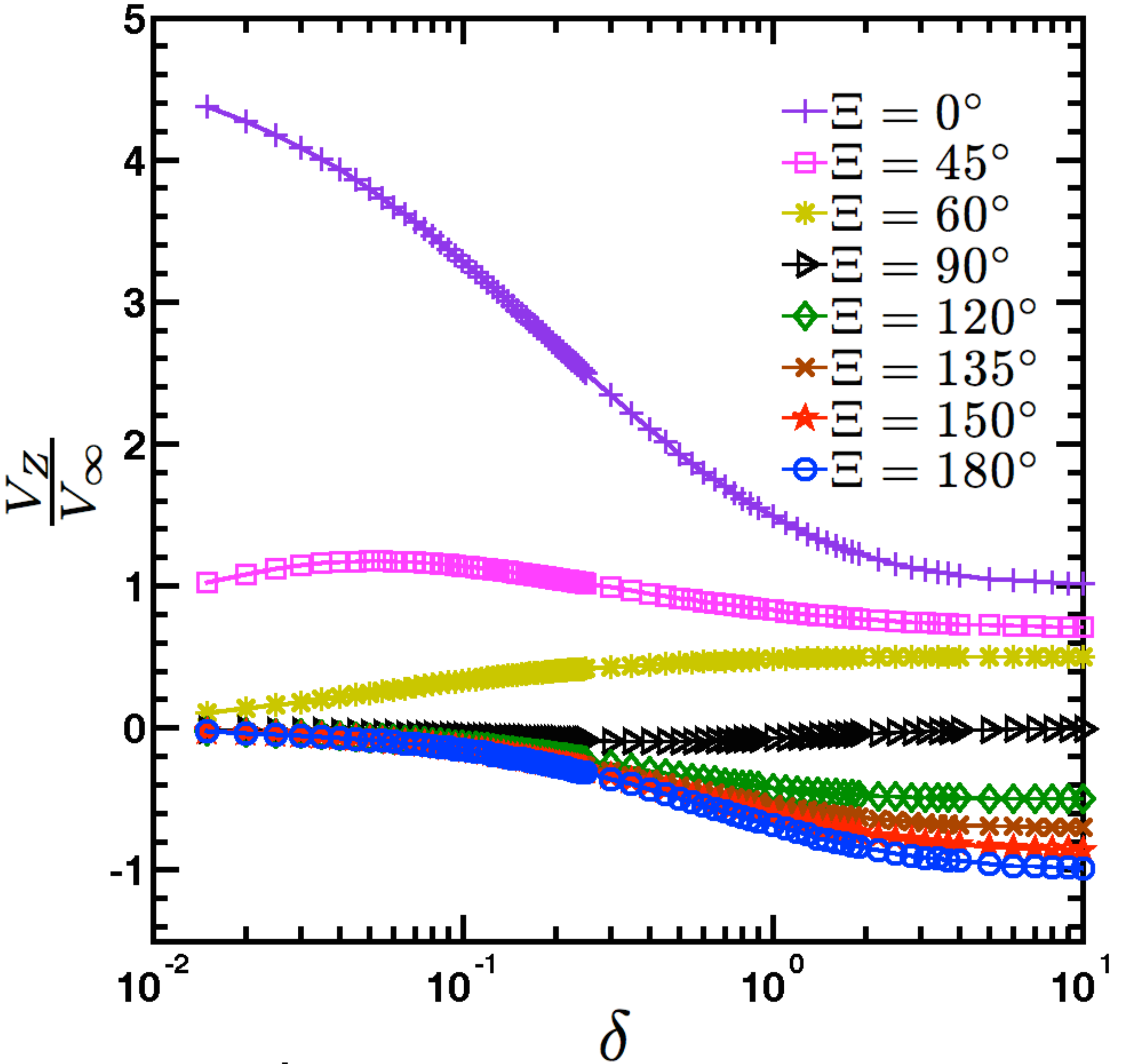}}
\subfigure[]{\label{velo23}\includegraphics[width=0.35\textwidth]{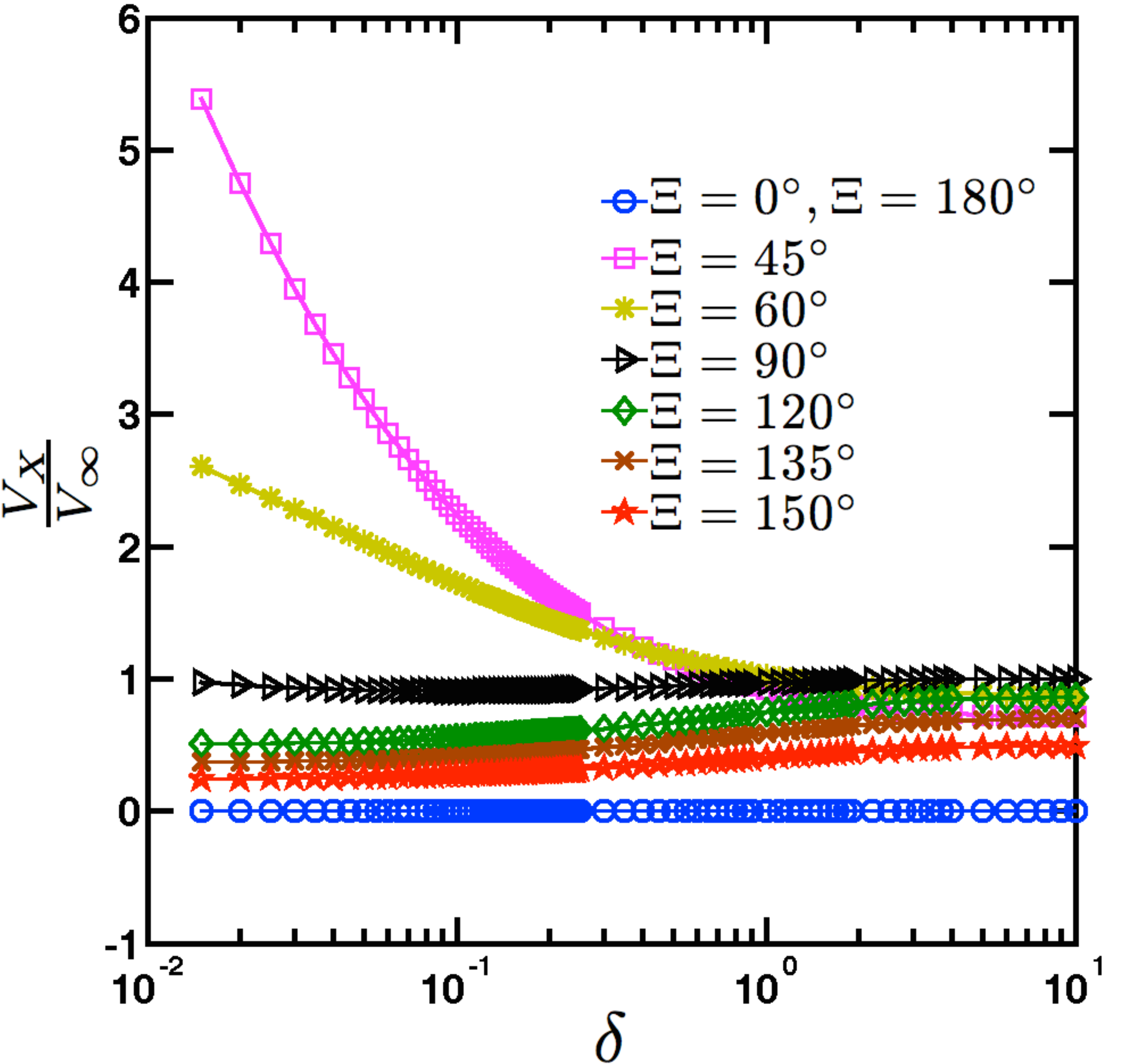}}
\subfigure[]{\label{velo24}\includegraphics[width=0.35\textwidth]{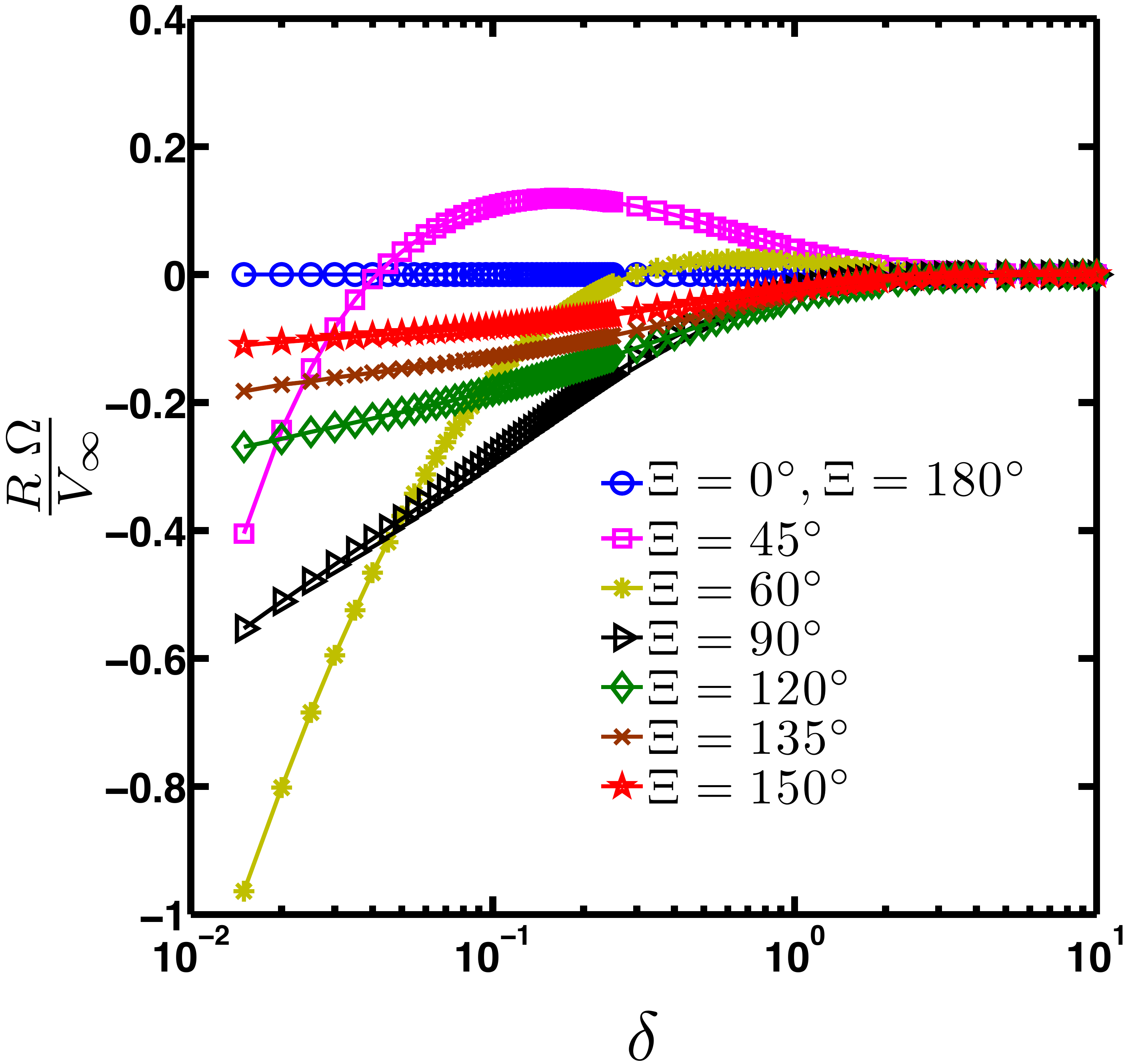}}
\subfigure[]{\label{traj2}\includegraphics[width=0.58\textwidth]{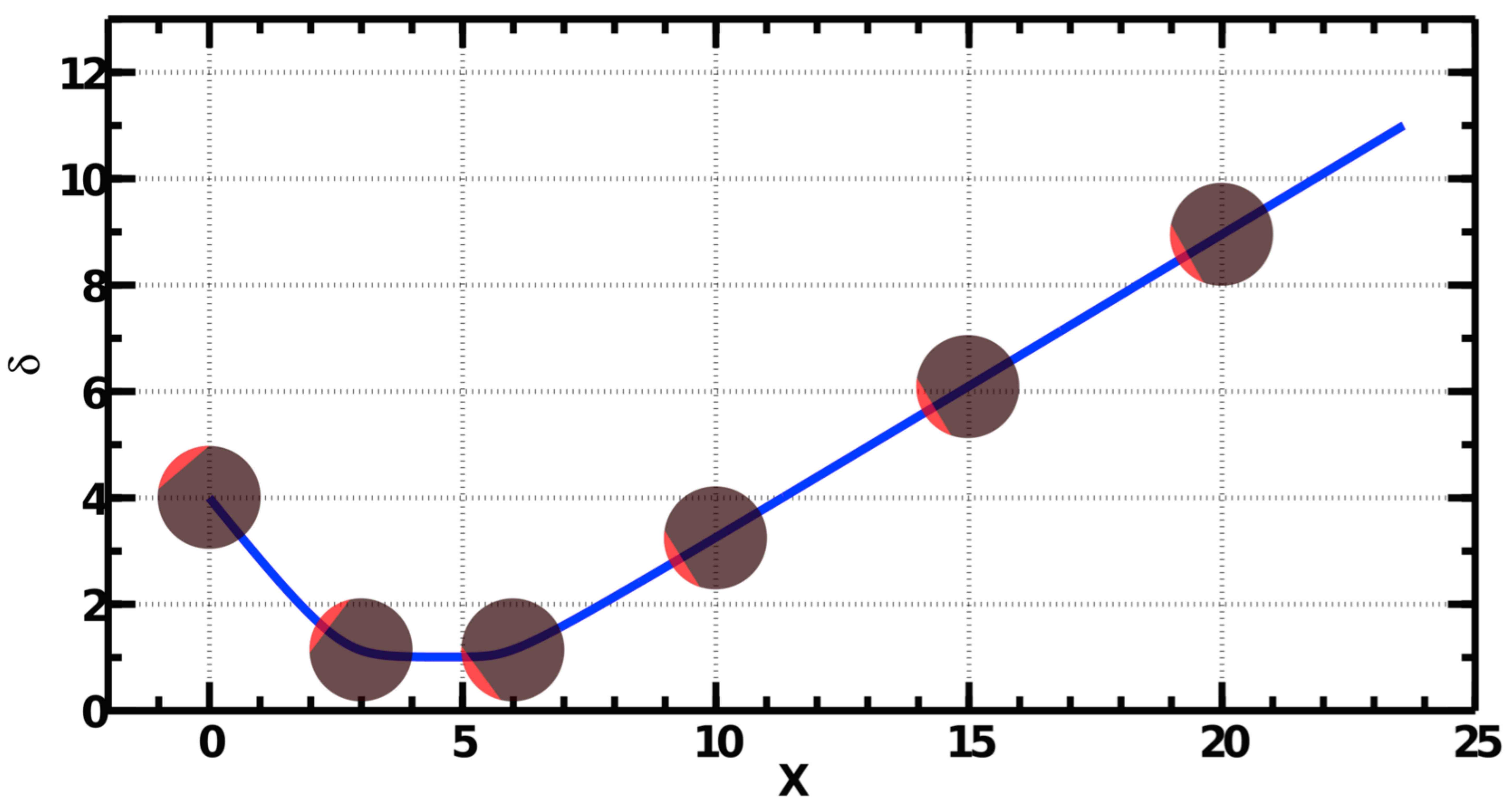}}
\caption{\ref{velo22} Non-dimensional swimming velocity in the $z$ direction, \ref{velo23} non-dimensional swimming velocity in the $x$ direction, \ref{velo23} Non-dimensional angular velocity  as a function of separation distance $\delta$ for low coverage colloid $(\theta _{cap}  = 40^ \circ )$ and \ref{traj2} sample trajectories for colloid with coverage $(\theta _{cap}  = 40^ \circ )$ and initial configuration: $\delta=3.0$, $\Xi=140^ \circ$.}
\label{velo2} 
\end{figure}

Consider first the case of a hemispherical active area ($\theta _{cap}  = 90^ \circ$). The trajectory of a Janus colloid initially located at $\delta=1.0$ with an inclination angle of $\Xi=70^\circ$ is plotted in figure \ref{traj12} where the colloid is repelled from the wall without considerable rotation. This trajectory is a typical of all partially active catalytically colloids with different surface coverage and initial separations comparable to the particle radius with initial orientation $\Xi  < 90^\circ$. This behavior can be easily understood from the results given in Fig.~\ref{velo1}, where we see that for  $\Xi  < 90^\circ$, $V_z  > 0$ for all separation distances and therefore the colloid would never approach the wall.\\
\indent For initial orientation $\Xi  > 90^\circ$, the diffusiophoresis force is towards the wall and the behavior is more interesting. Figure \ref{traj13} exhibits two trajectories of a Janus colloids $\theta _{cap}  = 90^ \circ$ with a same initial separation from the wall, $\delta=3$, but different inclination angles $\Xi=120^ \circ$ and $\Xi=150^ \circ$. In both cases the colloid  approaches the wall with a negligible counterclockwise rotation, followed by a fast counterclockwise rotation in the vicinity of wall which causes the active side of the colloid to face the wall  causing it to be eventually repelled from the wall. As the initial inclination angle approaches $180^\circ$, the colloid experiences reflection from the wall at smaller separations. Note that the angular velocity grows noticeably as the colloid approaches the wall (see Fig.~\ref{velo14}). The final configuration of the particle has inclination angles less than but close to $90^\circ$ which causes the colloid to move away very slowly in the positive $z$ direction. Finally we ask whether there is any other trajectory beside reflection from the wall for a Janus particle $(\theta _{cap}  = 90^ \circ)$. We first note that at each separation distance from the wall, there is a specific inclination angle for which $V_z=0$. Furthermore, since $\Omega<0$ at all separation distances, the colloid rotates counterclockwise, which decreases $\Xi$, which leaves the colloid at an orientation where $V_z>0$ and it tends to move away.\\ 
To explore the other possible scenarios for the swimmer against planar wall, we vary the active surface coverage of the colloid. Fig.~\ref{velo2} illustrates non-dimensional swimming velocities as a function of normalized separation distance for various inclination angles for a colloid which is mostly passive; $\theta _{cap}  = 40^ \circ$. From these curves, it can be concluded again that repulsive interaction of solute and particle eventually brings about a net repulsion of the colloid from the wall. A typical trajectory of a particle with low active coverage $\theta _{cap}  = 40^ \circ$, starting from separation distance $\delta=3.0$ and inclination angle $\Xi=140^ \circ$, is shown in Fig.~\ref{traj2}. We see that the colloid approaches the wall with a small counterclockwise rotation, but due to the strong hydrodynamic repulsion in the vicinity of the wall, the colloid rotates into a configuration where $V_z$ becomes positive and the particle ultimately is reflected form the wall. According to the angular velocity curves in Fig.~\ref{velo24}, an orientation with positive rotation is possible but since at that orientation the velocity in the $z$ direction is positive the colloid moves away from the wall and subsequently its rotation decays to zero. Our analysis suggests that repulsion from the wall is typical for colloids with initial orientations of $\Xi  < 90^\circ$ for any active coverage section $\theta_{cap}$, and reflection from the wall is obtained for orientation angle $\Xi  > 90^\circ$ for active coverage of $0^ \circ   < \theta _{cap}  < 115^ \circ $.
\subsubsection{Skimming of a colloid along planar wall}
\begin{figure}[!]
\centering
\subfigure[]{\label{velo31}\includegraphics[width=0.35\textwidth]{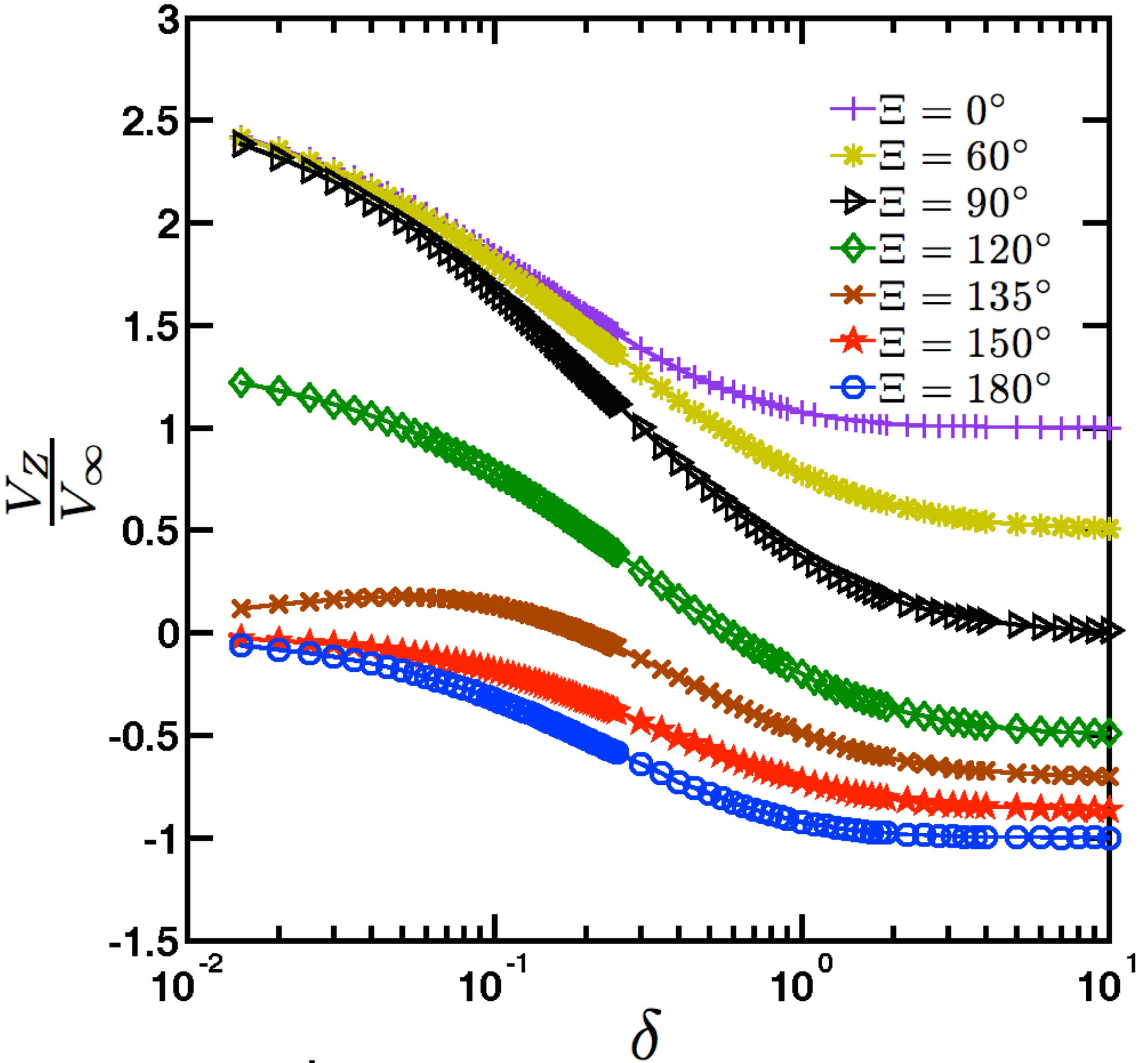}}
\subfigure[]{\label{velo32}\includegraphics[width=0.35\textwidth]{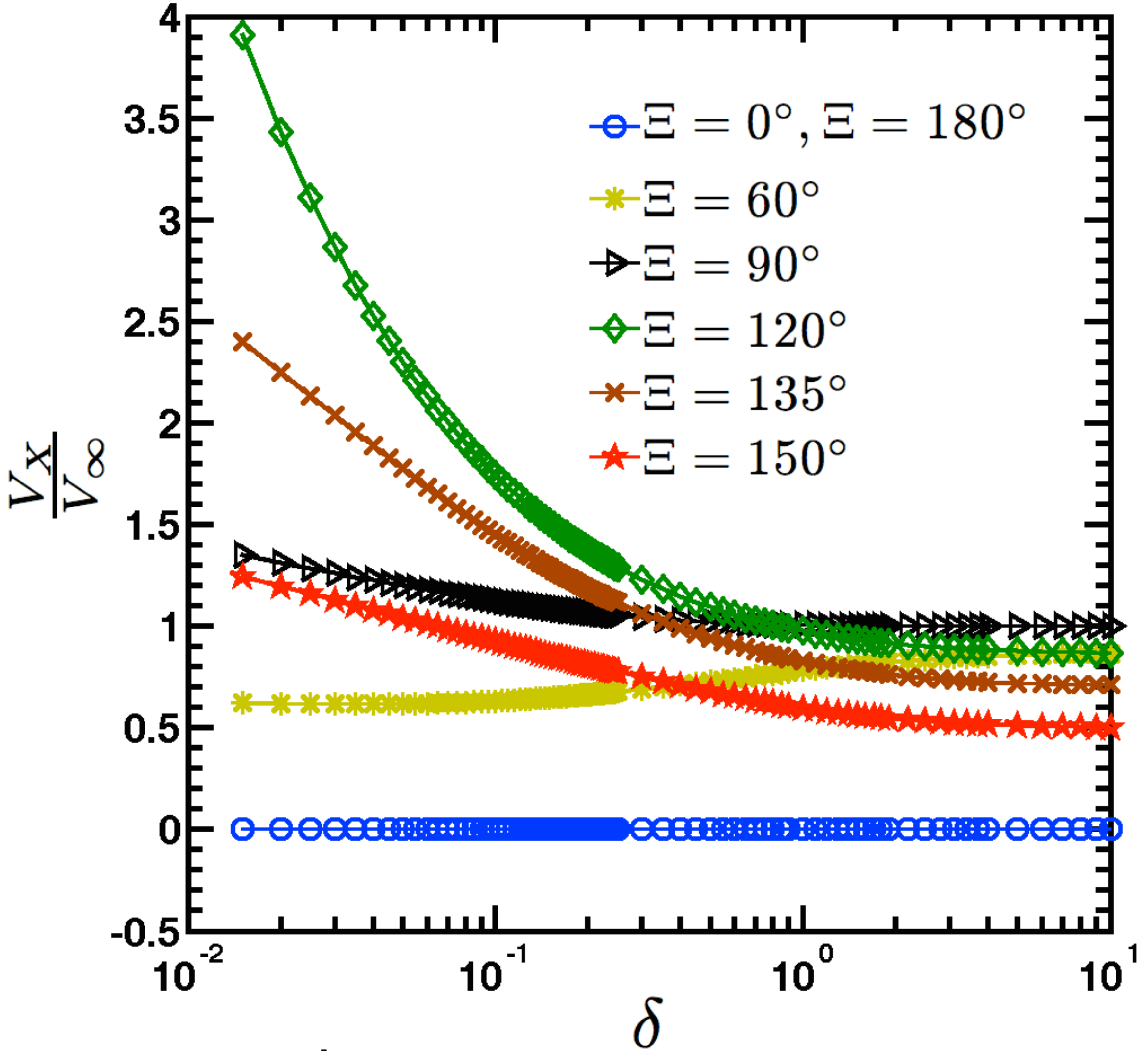}}
\subfigure[]{\label{velo33}\includegraphics[width=0.34\textwidth]{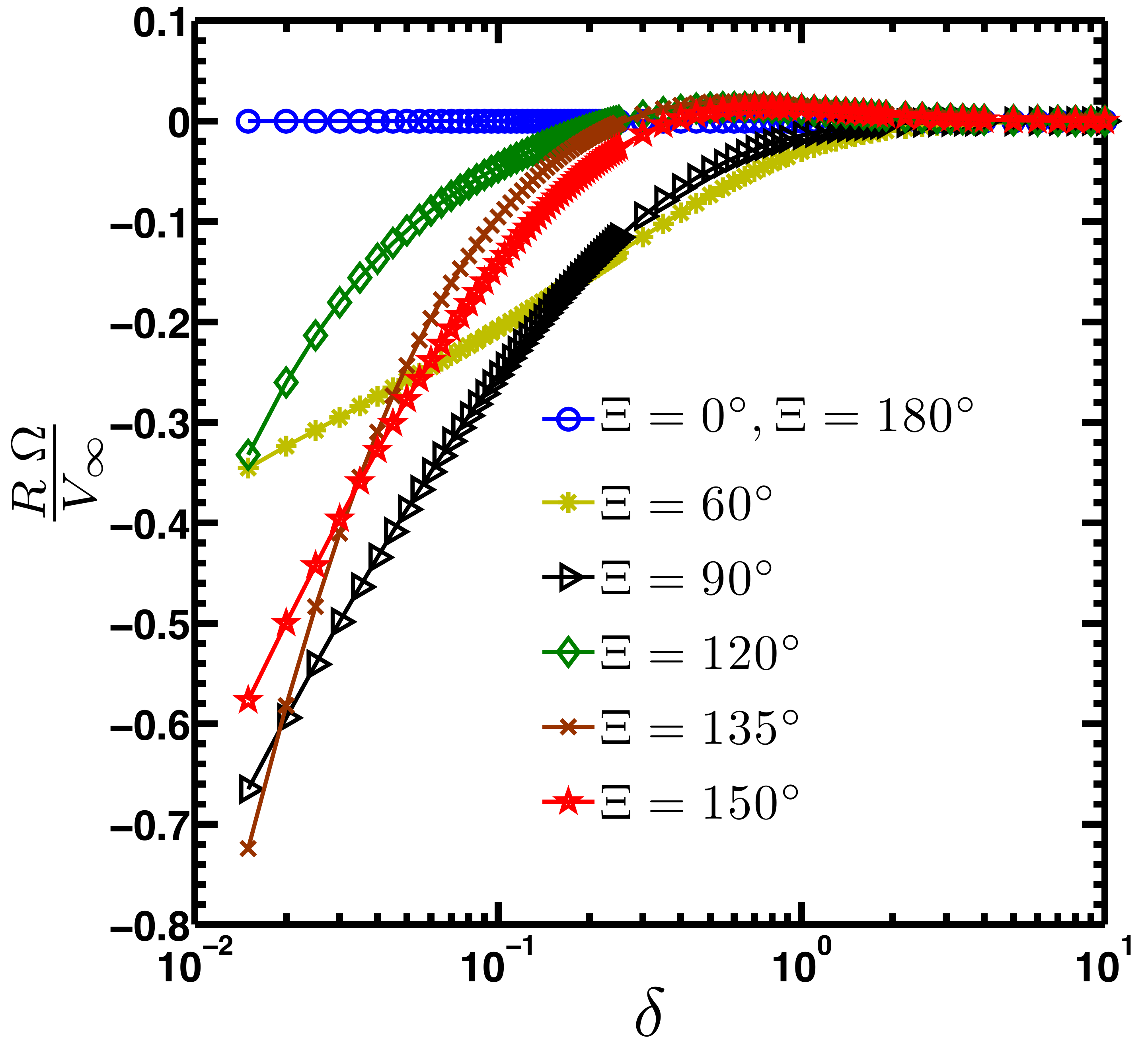}}
\subfigure[]{\label{traj3}\includegraphics[width=0.64\textwidth]{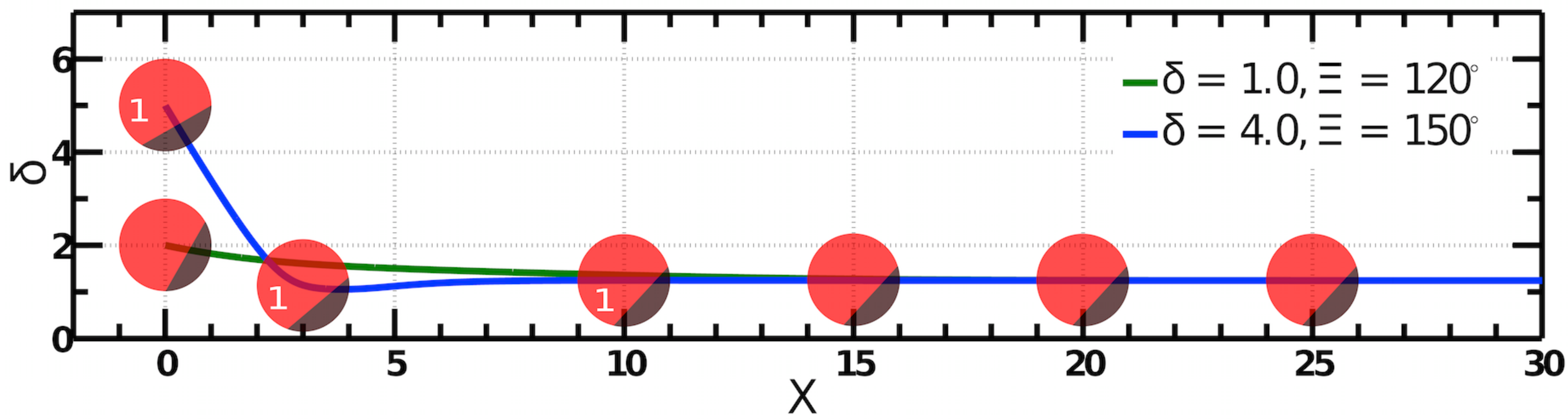}}
\subfigure[]{\label{velo34}\includegraphics[width=0.33\textwidth]{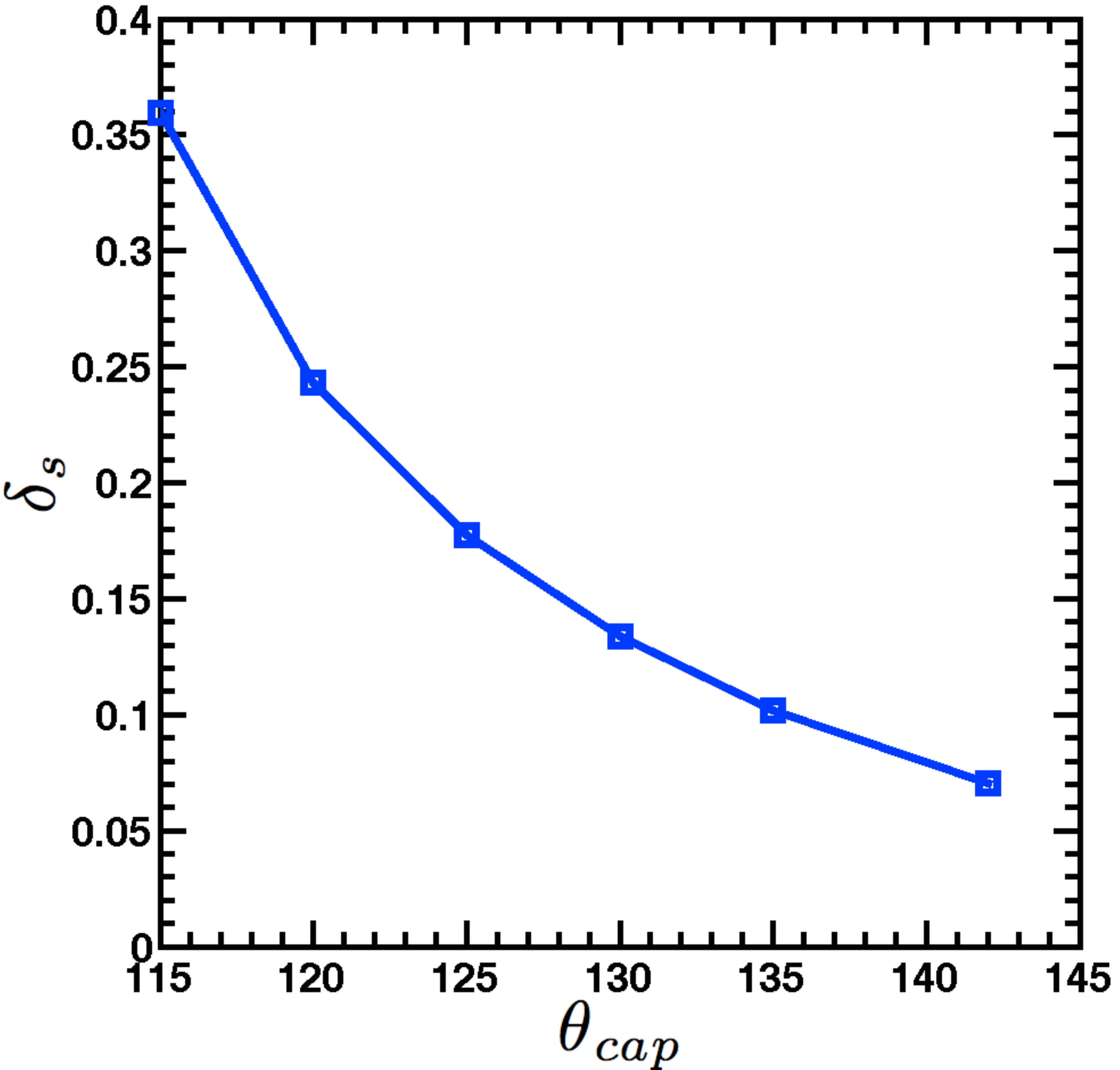}}
\subfigure[]{\label{velo35}\includegraphics[width=0.33\textwidth]{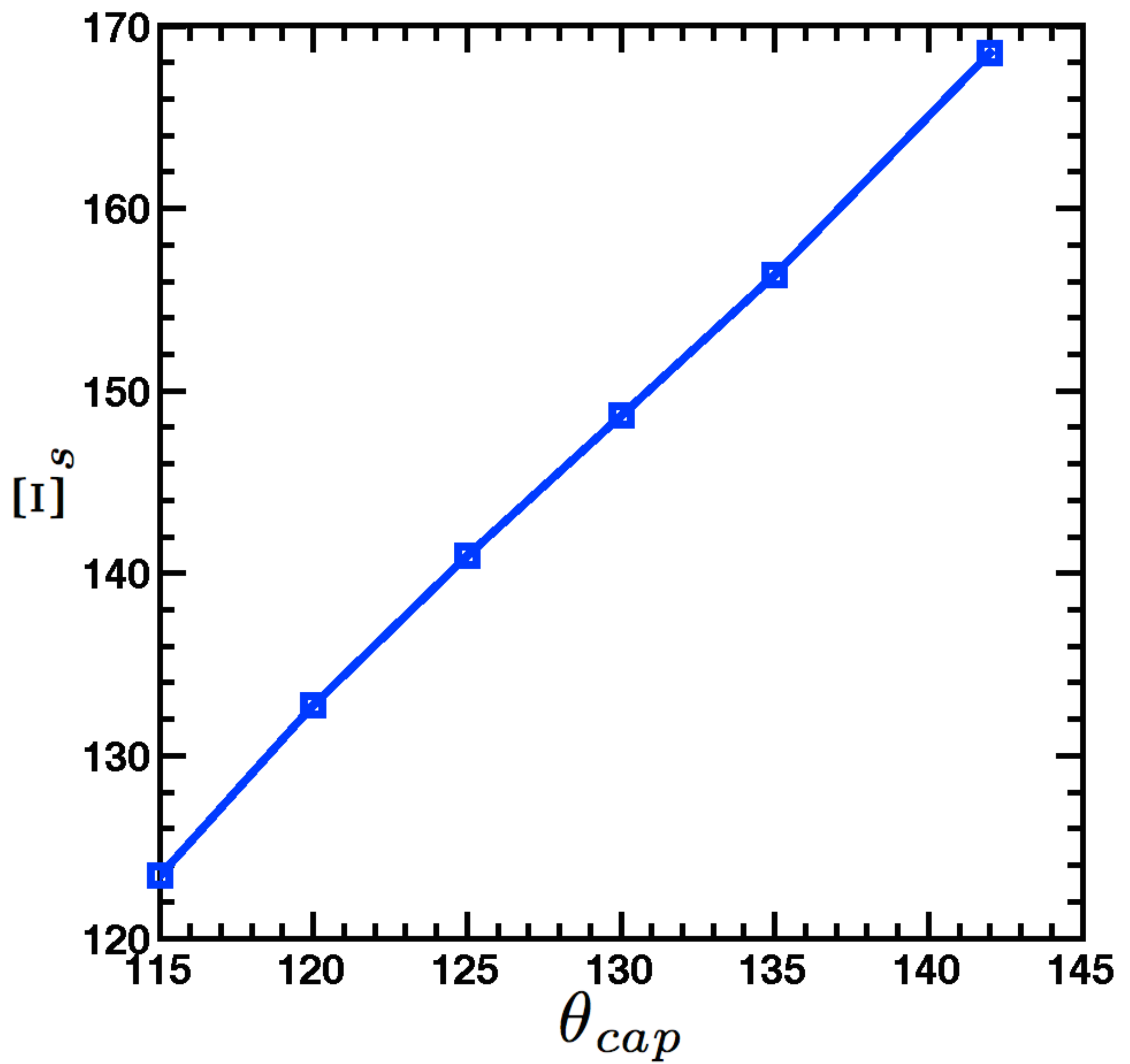}}
\caption{\ref{velo31} Non-dimensional swimming velocity in the $z$-direction, \ref{velo32} non-dimensional swimming velocity in the $X$-direction, \ref{velo33} Non-dimensional angular velocity  as a function of separation distance $\delta$ for high coverage colloid $(\theta _{cap}  = 120^ \circ )$, \ref{traj3} sample trajectories for colloid with coverage $(\theta _{cap}  = 120^ \circ )$ and initial configuration: $\delta=1.0$, $\Xi=120^ \circ$ and $\delta=4.0$, $\Xi=150^ \circ$, \ref{velo34}-\ref{velo35} equilibrium skimming distance and tilt angles respectively as a function of surface coverage.}
\label{velo3} 
\end{figure}

As the coverage increases above $\theta _{cap}  = 115^ \circ $ other types of trajectory appear for the colloids with the active side facing away from the wall. In Fig.~\ref{velo31}-\ref{velo33} we show the velocities for the case $\theta _{cap}  = 120^ \circ$, indicating that there is always a specific configuration of colloid (separation distance and inclination angle, $\delta _s$ and $\Xi _s$) in which $V_z$ and $\Omega_y$ are close to zero, so that the colloid slides along the wall while maintaining its specific orientation and separation from the wall. If the particle initially has an inclination angle greater than $90^ \circ $, eventually it slides along the wall with trajectories similar to the ones shown in Fig.~\ref{traj3}. This figure indicates different initial configurations leading to identical sliding behavior. Generally, having known the initial configuration of colloid, the final orientation of the colloid with respect to the wall can be extracted from the curves in Fig.~\ref{velo34}-\ref{velo35}, which give the variation of skimming distance and tilt angle with surface coverage. In this steady condition, the inclination angle and separation distance remains constant while propulsion force in the $z$ direction and propulsion torque are close to zero. As the area of active cap increases, particle skims in the smaller separation distance and inclination angle increases and approaches $180^\circ$.

\subsubsection{Stationary Particle}
\begin{figure}[!]
\centering
\subfigure[]{\label{confix}\includegraphics[width=0.34\textwidth]{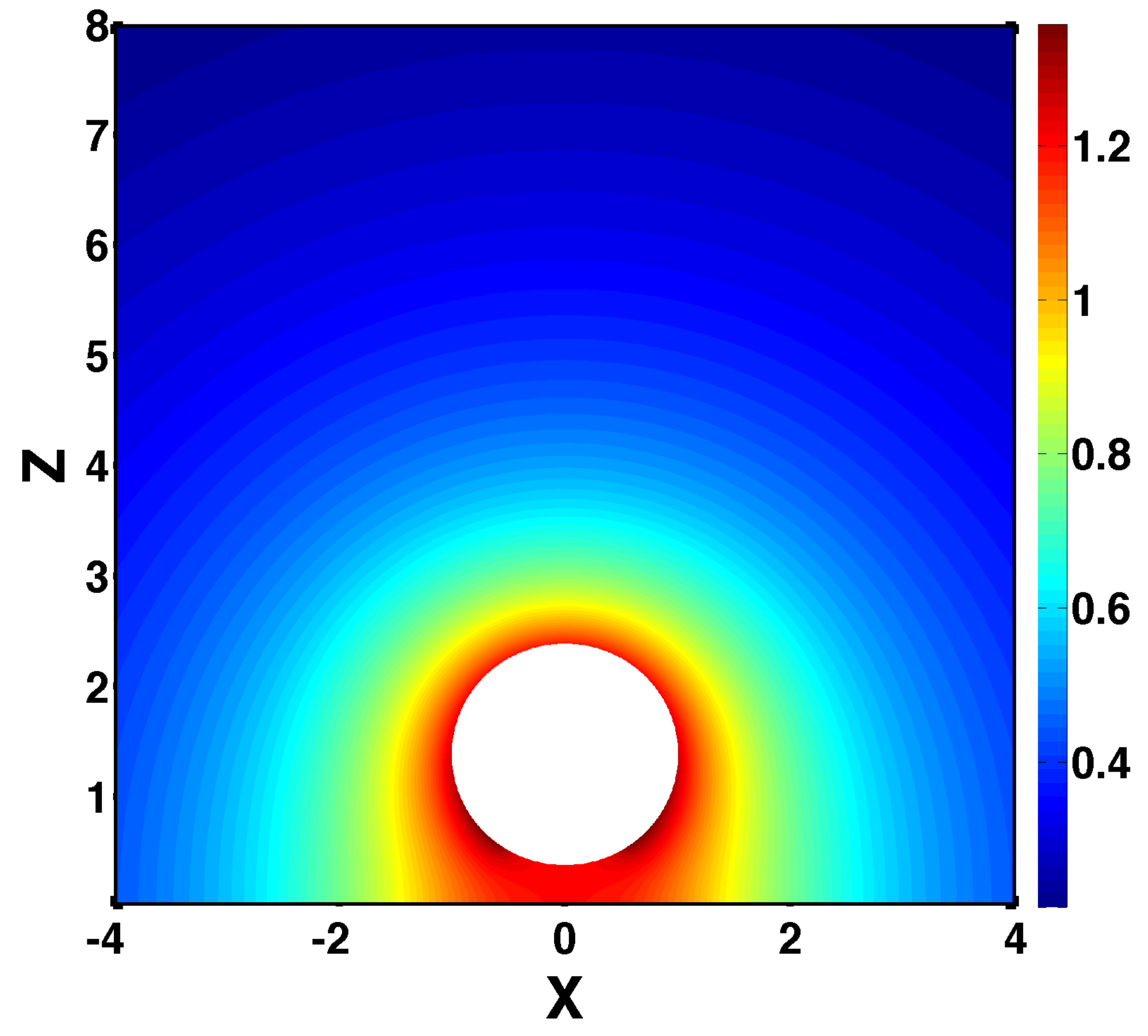}}
\subfigure[]{\label{velo41}\includegraphics[width=0.35\textwidth]{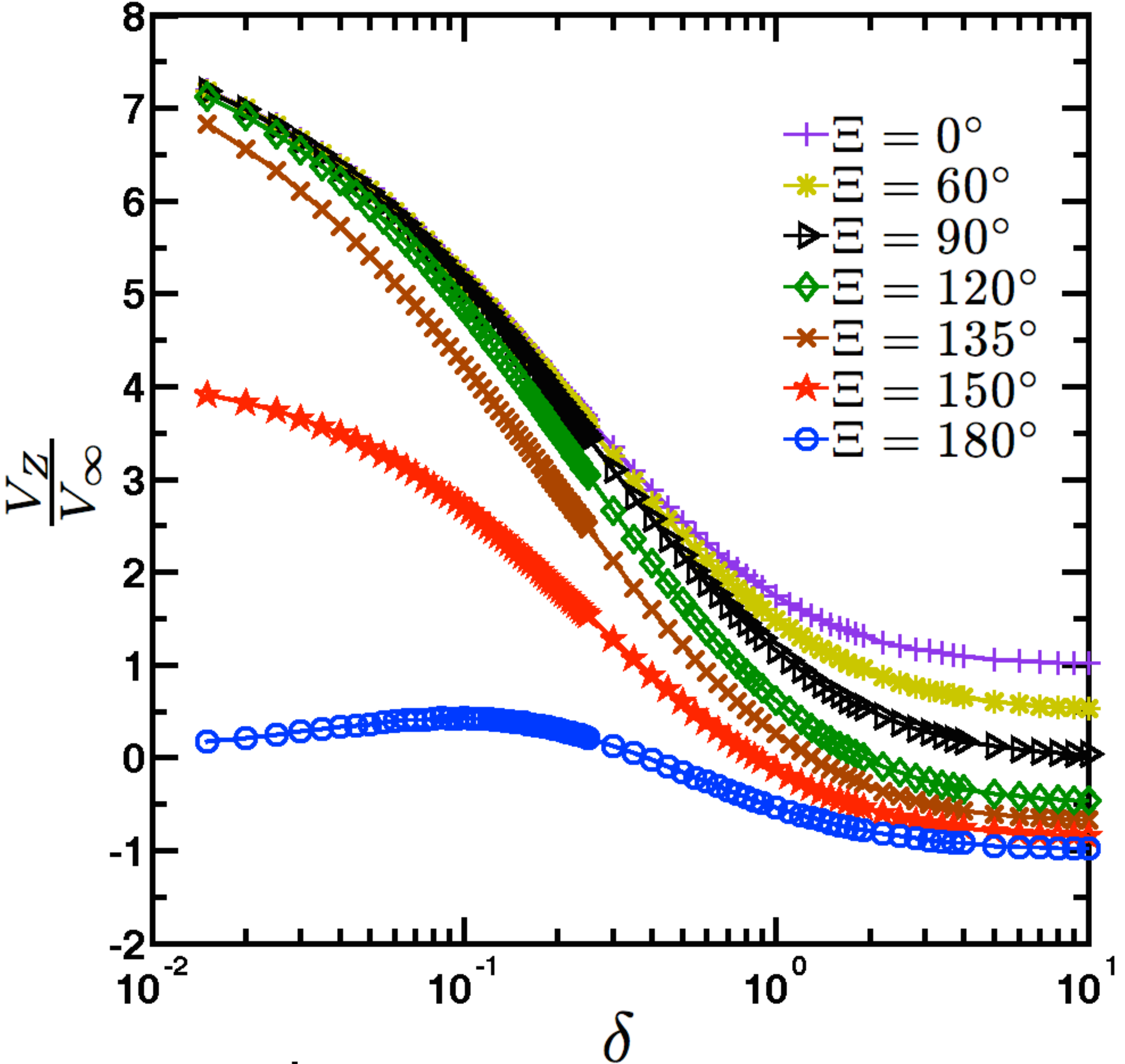}}
\subfigure[]{\label{velo42}\includegraphics[width=0.35\textwidth]{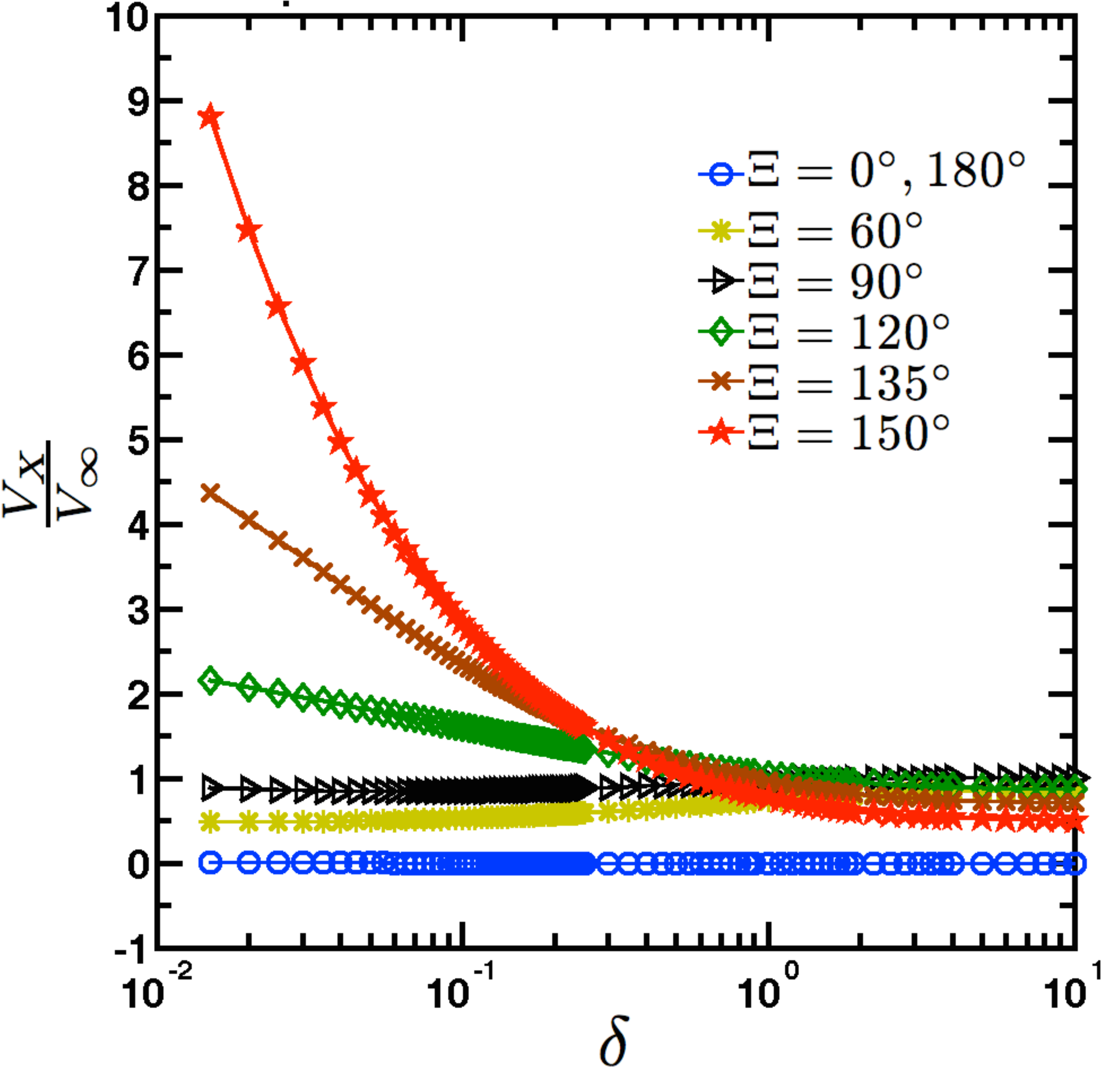}}
\subfigure[]{\label{velo43}\includegraphics[width=0.37\textwidth]{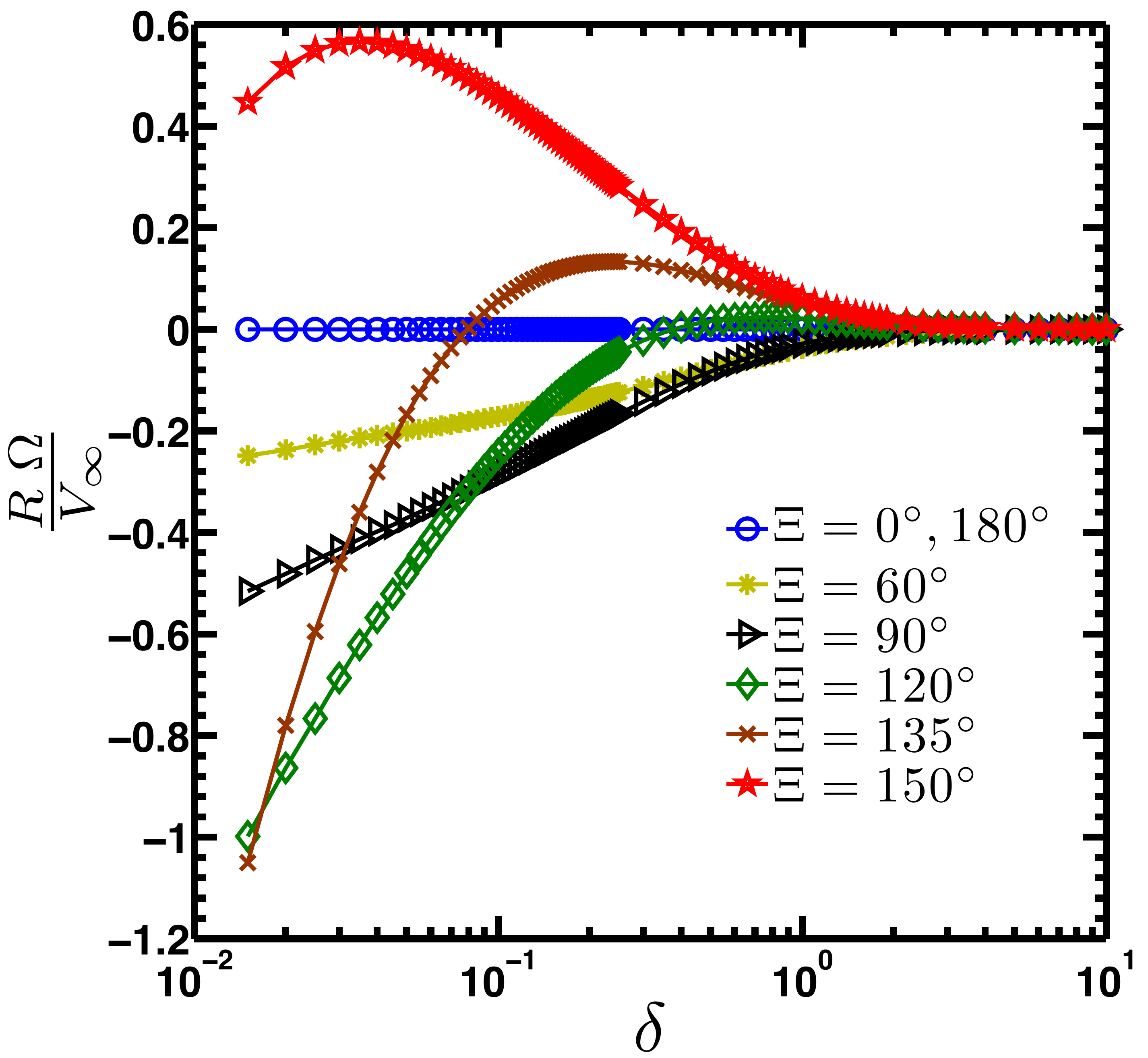}}
\subfigure[]{\label{traj4}\includegraphics[width=0.65\textwidth]{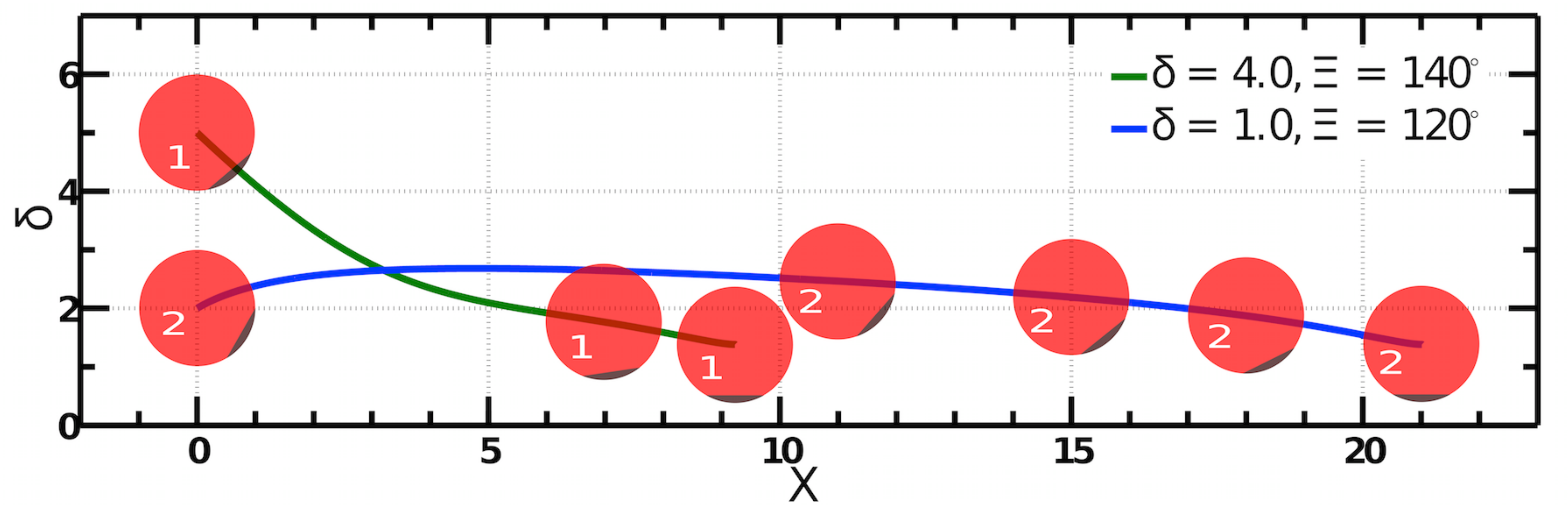}}
\subfigure[]{\label{velo45}\includegraphics[width=0.31\textwidth]{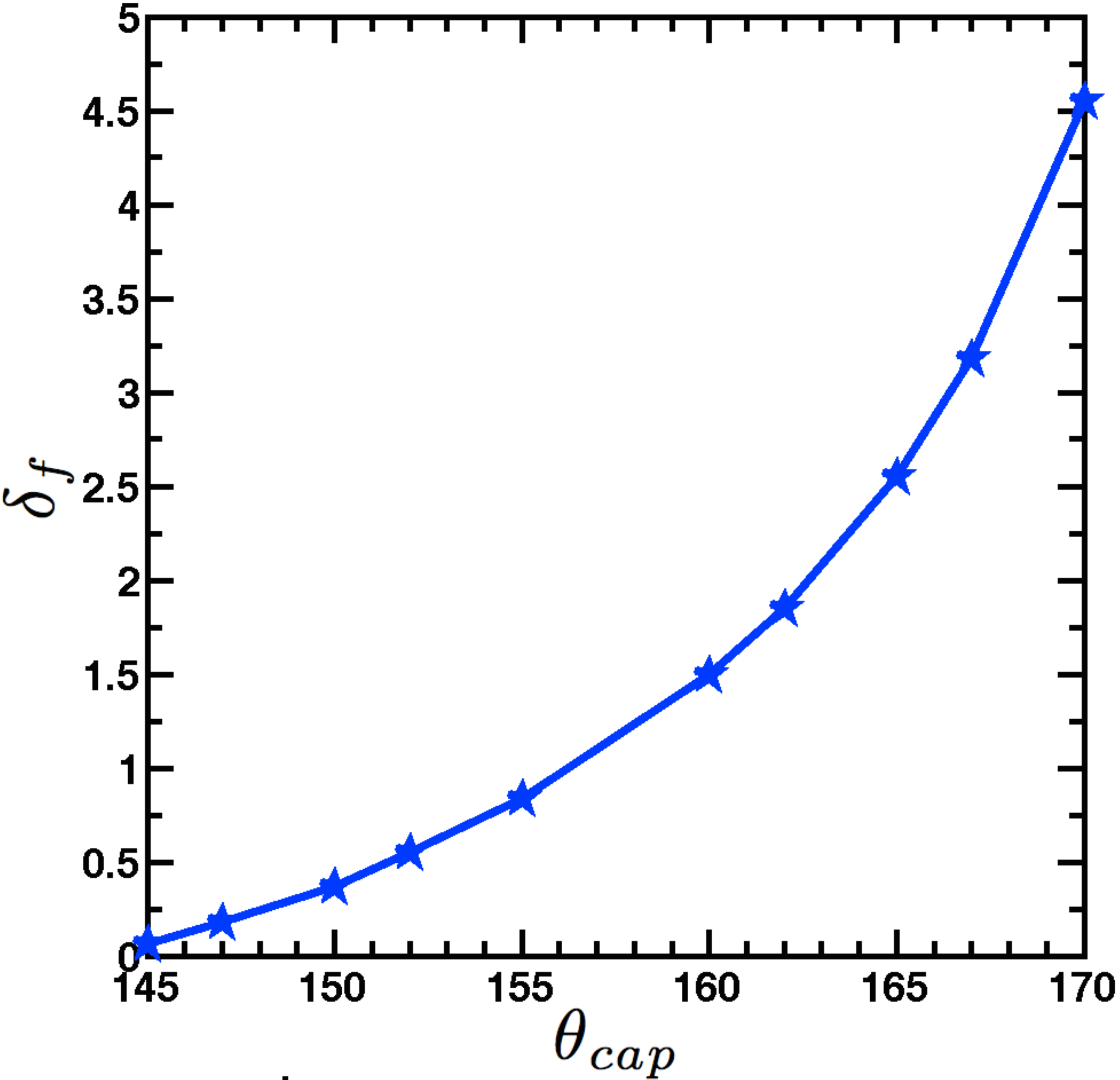}}
\caption{ \ref{confix}, concentration field for a final configuration of particle with $(\theta _{cap}  = 150^ \circ )$, \ref{velo41} Non-dimensional swimming velocity in the $z$-direction, \ref{velo42} non-dimensional swimming velocity in the $X$-direction, \ref{velo43} Non-dimensional angular velocity as a function of separation distance $\delta$ for very high surface coverage of particle $(\theta _{cap}  = 150^ \circ )$, \ref{traj3} sample trajectories for colloid with coverage $(\theta _{cap}  = 150^ \circ )$ and initial configurations: $\delta=1.0$ , $\Xi=120^ \circ$ and $\delta=4.0$ , $\Xi=140^ \circ$ and \ref{velo45} equilibrium fix position distance of particle to the wall as a function of coverage.}
\label{velo4} 
\end{figure}

We have seen that when the surface coverage is not too high, $\theta _{cap}  < 145^ \circ$ and the inclination angle is close to $180^ \circ$, i.e. the active side facing away from the wall, the solute repulsion drives the particle towards the wall at all separation distances. Surprisingly, when the coverage exceeds this value of $\theta _{cap}$ the particle is instead repelled at small separations! The reason is that if the particle is near the wall the active surface region extends into the thin gap region and produces a local repulsion which reduces the total diffusiophoretic force to values comparable to the lubrication resistance. The concentration field in such case is given in Fig.~\ref{confix}. When a particle of high coverage approaches a wall at a large value of $\Xi$, this effect produces a positive rotation (Fig.~\ref{velo43}) which drives the orientation to $\Xi = 180^\circ$, where the force in the $x$-direction vanishes and the particle moves to rest (Fig.~\ref{velo41}-Fig.~\ref{velo43}). Two examples of such stopping trajectories are given in Fig.~\ref{traj4}. We see in these figures at $V_x=0$ since $\Xi=180^\circ$, but $V_z=0$ only at a certain value of $\delta$, so the final separation distance $\delta_f$ depends on $\theta_{cap}$, as indicated in Fig.~\ref{velo45}, but as the trajectory plot indicates $\delta_f$ is insensitive to initial orientation. 

\section{Summary and Conclusion}\label{concl}
\begin{figure} 
\centering
\includegraphics[width=0.6 \textwidth]{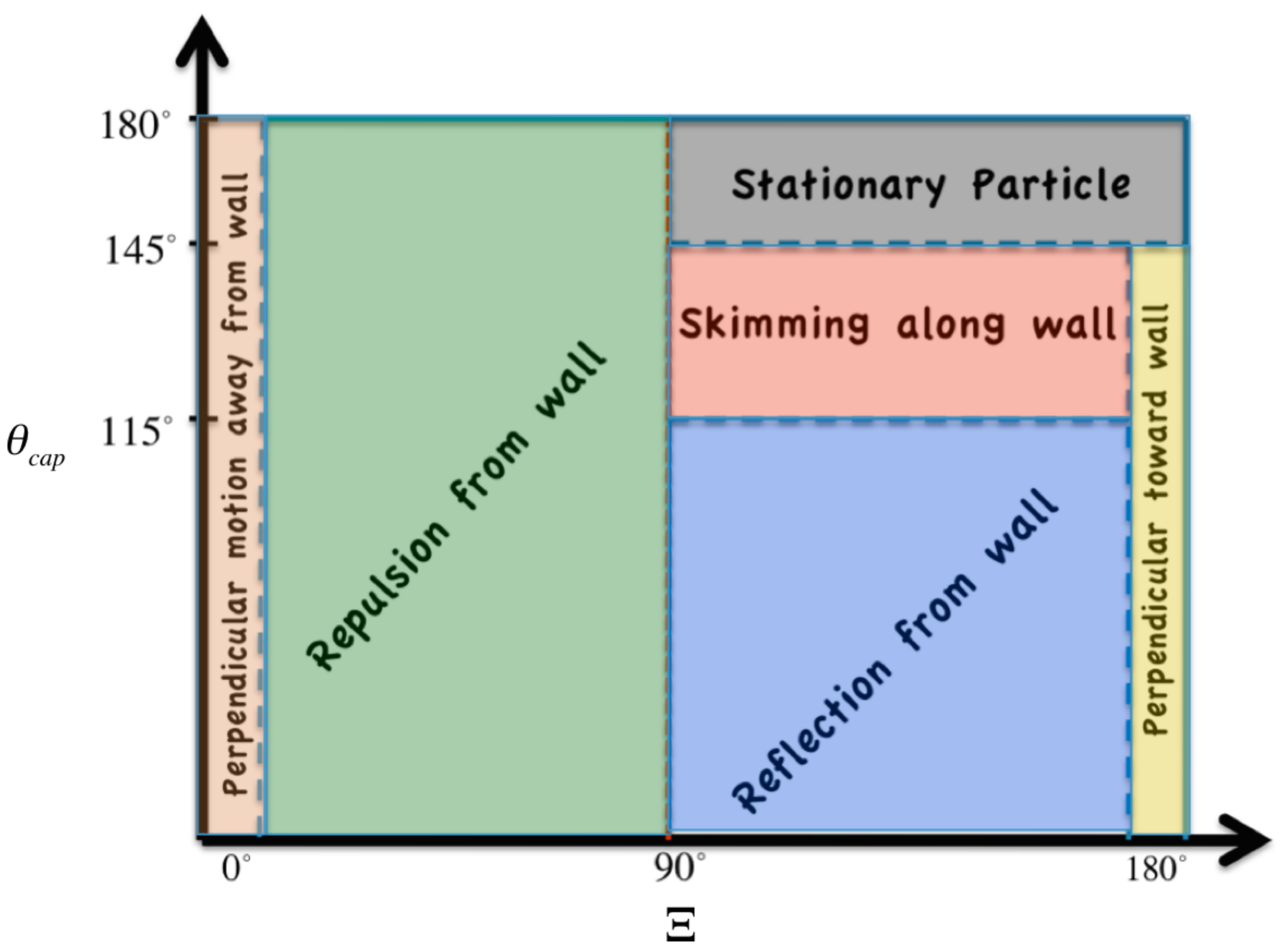}	     	    
 \caption{\footnotesize{Types of particle trajectory as a function of tilt angle and active cap size.} } 
\label{phase}
 \end{figure}

This study has aimed to study the self-diffusiophoretic motion of a catalytically active swimmers near a solid planar boundary. In self-diffusiophoresis, the colloid creates and sustains a solute gradient as a means of autonomous motion, and the colloid functions as an engine or motor. The hydrodynamic and mass transfer equations which describe the self-phoretic motion of the colloid are formulated utilizing slip velocity approach \cite{a89} in the small Damk\"{o}hler $Da\ll1 $ and P\'eclet $Pe\ll1$ number regimes. The hydrodynamic problem was solved exactly, both by directly solution of the boundary value problem and via Reynolds Reciprocal Theorem. \\ 
 \indent We have shown the complex dynamics of catalytically active swimmers near a solid boundary is controlled by the orientation of the colloid relative to the solid surface, $\Xi$ as well as the active coverage of the catalytic cap, $\theta_{cap}$. For situations where the tilt angle equals to $0^{\circ}$ or $180^{\circ}$, the motion is axisymmetric and simple rectilinear motion in the $z$ direction results, where symmetry insures that the colloid does not rotate and its trajectories are perpendicular to the wall. When the active area faces the wall, the solute concentration rises in the gap between the wall and the colloid due to the zero flux condition at the wall surface, and the particle experiences a boost in diffusiophoresis relative to the propulsion in an infinite medium. This effect is opposed by viscous drag, as modified by the presence of the wall, and the balance between these two effects determines the swimming velocity. \\
 \indent When the active area of the swimmer is not aligned perfectly towards or away from the wall, the inclination generates a diffusiophoretic propulsion along the wall,  perpendicular  to the wall as well as a torque perpendicular to the plane of symmetry passing through the particle.   
For $0^\circ<\Xi<90^\circ$ and for all coverages, the net diffusiophoretic propulsive force is in the positive $z$ direction, since  the orientation is such  that solute accumulates between the wall and the particle, pushing the particle away from the wall with small rotation that vanishes as the particle moves further from the wall. 
For  $\Xi > 90^{\circ}$, the diffusiophoretic propulsion in the z direction may vanish at a certain critical tilt. In this case, the reactive cap is obliquely oriented to the surface so that the accumulation in the gap is reduced and the increased concentration above the colloid balances the reduced upward propulsion from the bottom hemisphere. 
  In this state the particle, at least instantly, skims along the surface. For a given $\theta_{cap}$, the critical tilt will be a function of the separation distance, and this relationship will vary with the cap size.  We can anticipate that as the cap size increases, the  critical angle for a fixed $\delta$ will increase as more of the cap is required to face away from the wall to reduce the diffusiophoretic boost due to the accumulation in the gap between the wall and the particle. For particles with low to medium surface coverages at that critical tilt angles which is a function of separation distance at all values of $\delta$ rotation is counterclockwise at the critical tilt angles, which leads to repulsion of the particle from the wall, along with a reduction of the inclination angle. For particles with high coverage there is a unique configuration at each $\theta_{cap}$ at which the normal force and torque are close to zero, so that the particle skims at that steady configuration. Finally particles with very high $\theta_{cap}$ attain their fixed position with an axisymmetric orientation and the particle will halt.\\
\indent Based on the results we have obtained, we can construct a phase diagram which gives the long-time behavior of a partially active catalytic particle near a solid planar wall: Fig.~\ref{phase}. We believe the phase diagram can have an important implications on designing the next generation of micro-nano engines for various biological applications. For example, the skimming regime is an area of great interest since it offers the possibility of steering a particle along a solid wall from one place to another. Likewise, the stationary regime has obvious applications to drug delivery at particular positions. 

One important issue not considered here is the effects of thermal (Brownian) fluctuations on colloidal particle motion. In general, one would anticipate that large colloidal particles would be resistive to thermal fluctuations and the current results would apply, while smaller particles might behave differently, in particular the existence of skimming or stationary states may be problematic for sufficiently smaller colloids. We intend to return to this issue in a future publication.

\newpage
\section{Appendix}\label{app}
We first discuss the concentration field calculations, and then the strategy used to obtain solutions to the problems ($a$)-($d$) of Fig.~\ref{sch}. As discussed before, due to the linearity of Stokes equation and boundary conditions, it is possible to decompose the general problem into some subproblems and by superposing the solutions to the problem we can construct the flow associate with the general problem. The general solution for the flow field around a spherical particle close to a wall was proposed by Leal \cite{JFM2}, given in Eq.~\ref{aux1} - \ref{vel3}. In the following, we review the boundary conditions associated with each problem shown in Fig.~\ref{sch} and provide the analysis and related recursive relations.
\subsection{Concentration field around the particle} 
The boundary condition (\ref{p23}) is satisfied trivially when 
\begin{equation}
\widetilde{A}_{n,m}  = 0 \;\;\;  \;n,m \ge 0.
\end{equation}
\indent Furthermore, we choose the axis in such a way that the active cap is completely symmetric about the $x$-$z$ plane and therefore $\gamma _m =0$.  
By imposing boundary conditions (\ref{p24} -\ref{p25}) at the colloid surface, and utilizing the appropriate orthogonality and recursion relations among the associated Legendre polynomials, we have
\begin{equation}
\begin{split}
 &[ -n\sinh (n-\frac{1}{2})\beta _0 \;\widetilde{B}_{n - 1,0}  + (n\;\sinh (n - \frac{1}{2})\beta _0  + (n + 1)\;\sinh (n + \frac{3}{2})\beta _0 )\widetilde{B}_{n,0}  \\ 
 &- (n + 1)\;\sinh (n + \frac{3}{2})\beta _0 \;\widetilde{B}_{n + 1,0} ] \\ 
 &= \frac{{(2n + 1)}}{{2\pi }}\;\int_0^{2\pi } {\int_0^\pi  {\frac{{\epsilon \;f(\alpha ,\phi )}}{{\sqrt {\cosh \beta_0  - \cos \alpha } }}} } \;\;\sin \alpha \;P_n^0 (\cos \alpha )\;d\alpha \;d\phi   \;\;\; (m=0),\label{p26}
\end{split}
\end{equation}
\begin{equation}
\begin{split}
 &[ - (n - m)\;\sinh (n - \frac{1}{2})\beta _0 \;\widetilde{B}_{n - 1,m}  + (n\;\sinh (n - \frac{1}{2})\beta _0  + (n + 1)\;\sinh (n + \frac{3}{2})\beta _0 )\widetilde{B}_{n,m}  \\ 
 &- (n + m + 1)\;\sinh (n + \frac{3}{2})\beta _0 \;\widetilde{B}_{n + 1,m} ] = \frac{{(2n + 1)(n - m)!}}{{\pi (n + m)!}} \\ 
 &\times \;\int_0^{2\pi } {\int_0^\pi  {\frac{{\epsilon \;f(\alpha ,\phi )}}{{\sqrt {\cosh \beta_0  - \cos \alpha } }}} } \;\cos m\phi \;\sin \alpha \;P_n^m (\cos \alpha )\;d\alpha \;d\phi \;\;\;(m \ge 1). \label{p27}
\end{split}
\end{equation}
 The special cases $ \Xi  = 0^0 $ and $\Xi  = 180^0 $ are two axisymmetric situations (where the active side faces toward and away from the wall, respectively) where the concentration is independent of azimuthal angle $\phi$, and the only relevant term is $m=0$.\\
\indent The infinite series solution is approximated by truncating the series for large values of $n$ and $m$, based on the fact that the coefficients vanishes for sufficiently large number of terms, and the remaining unknown coefficients are obtained by solving equations \ref{p26} - \ref{p27}. Lastly, we note that more terms in the truncated series are required for small separation distances to reach numerically accurate and consistent result for the concentration field. \\

\subsection{Stationary sphere near a wall with slip on the sphere surface} 
A spherical particle is located in a vicinity of an infinite plane wall $z=0$, where the sphere is stationary but the fluid slips with velocity ${\bf{v}}_s$ on its surface. 
In this case non-dimensional boundary conditions are,
\begin{align}
&\left. {\bf{v}} \right|_{\beta  = 0}  = 0,\label{bond1}\\ 
&\left. {\bf{v}} \right|_{\beta  = \beta _0 }  = {\bf{v}}_s  =  - \nabla _s C,\label{bond2}\\ 
&{\bf{v}} \to 0\;\;, as\; \sqrt {r^2  + z^2 }  \to \infty.
\end{align}
\indent The no-slip boundary condition at the wall can be expressed in cylindrical coordinate as,
\begin{equation}
\left. {v_r } \right|_{\beta  = 0}  = \left. {v_\phi  } \right|_{\beta  = 0}  = \left. {v_z } \right|_{\beta  = 0}  = 0. \label{bon1}
\end{equation}
We proceed first by determining the velocity field for $m=0$, which corresponds to the axisymmetric problem, by evaluating ${A_{n,0} }$, ${B_{n,0} }$, ..., ${H_{n,0} }$. 
The slip velocity at the particle surface is proportional to surface gradient of concentration $\nabla _s C$, where $\nabla_s={\bf{I}}-{\bf{nn}}$. Using the relations between tangent unit vectors to the surface of sphere in bispherical coordinate and their counterparts in cylindrical coordinates one can show that
\begin{equation}
\begin{split}
& \left. {v_r } \right|_{\beta  = \beta _0 }  =  - \;\nabla _s C \cdot {\bf{e}}_{\bf{r}}  =  - [\frac{{(\cosh \beta _0 \;\cos \alpha  - 1)\;\sin \alpha }}{{2\epsilon \;\sqrt {\cosh \beta _0  - \cos \alpha } }}\sum\limits_{n = 0}^\infty  {\widetilde{B}_{n,0} \cosh (n + \frac{1}{2})\beta _0 \;} P_n (\cos \alpha ) \\ 
 & + \frac{{(\cosh \beta _0 \;\cos \alpha  - 1)\;\sqrt {\cosh \beta _0  - \cos \alpha } }}{{\epsilon \;}}\sum\limits_{n = 1}^\infty  {\widetilde{B}_{n,0} \cosh (n + \frac{1}{2})\beta _0 \;} P^1 _n (\cos \alpha )] , \label{grad4}
\end{split}
\end{equation}
\begin{equation}
\left. {v_\phi  } \right|_{\beta  = \beta _0 }  =  - \nabla _s C \cdot {\bf{e}}_\phi   = 0,\label{grad5}
\end{equation}
\begin{equation}
\begin{split}
 &\left. {v_z } \right|_{\beta _0 }  =  - \nabla _s C \cdot {\bf{e}}_{\bf{z}}  = [\frac{{\sinh \beta \;\sin ^2 \alpha }}{{2\epsilon \;\sqrt {\cosh \beta _0  - \cos \alpha } }}\sum\limits_{n = 0}^\infty  {\widetilde{B}_{n,0} \cosh (n + \frac{1}{2})\beta _0 \;} P_n (\cos \alpha ) \\ 
  &+ \frac{{\sinh \beta _0 \;\sin \alpha \;\sqrt {\cosh \beta _0  - \cos \alpha } }}{{\epsilon \;}}\sum\limits_{n = 1}^\infty  {\widetilde{B}_{n,0} \cosh (n + \frac{1}{2})\beta _0 \;} P^1 _n (\cos \alpha )]. \label{grad6} 
 \end{split}
\end{equation}
Axisymmetry implies that $\gamma _0=0$, and 
\begin{equation}
G_{n,0}  = H_{n,0}  = 0. \label{recu1}
\end{equation}
The no slip condition at the wall (\ref{bon1}) yields the following recursive relation
\begin{eqnarray}
- \frac{1}{{2(2n - 1)}}B_{n - 1,0}  + \frac{1}{{2(2n + 3)}}B_{n + 1,0}  - \frac{{\left( {n - 1} \right)}}{{(2n - 1)}}F_{n - 1,0}  + F_{n,0}- \frac{{\left( {n + 2} \right)}}{{(2n + 3)}}F_{n + 1,0}  = 0\nonumber\\n \ge 1,  \label{recu2}
\end{eqnarray}
and
\begin{eqnarray}
&D_{n,0}  = 0\;\;\;;n \ge 0.  \label{recu3}
\end{eqnarray}
Applying the slip velocity condition at the particle surface, Eqs.~\ref{grad4} and \ref{grad6}, to the solution of equation of motion for $m=0$, Eq.~\ref{vel1} and \ref{vel3}, we obtain
\begin{equation}
\begin{split}
 & - \frac{1}{{2(2n - 1)}}[A_{n - 1,0} \;\sinh (n - \frac{1}{2})\beta _0  + B_{n - 1,0} \cosh (n - \frac{1}{2})\beta _0 ] \\ 
 &+ \frac{1}{{2(2n + 3)}}[A_{n + 1,0} \;\sinh (n + \frac{3}{2})\beta _0  + B_{n + 1,0} \cosh (n + \frac{3}{2})\beta _0 ] \\ 
 &- \frac{{(n - 1)}}{{(2n - 1)}}[E_{n - 1,0} \;\sinh (n - \frac{1}{2})\beta _0  + F_{n - 1,0} \cosh (n - \frac{1}{2})\beta _0 ] \\ 
 & + \cosh \beta _0 [E_{n,0} \;\sinh (n + \frac{1}{2})\beta _0  + F_{n,0} \cosh (n + \frac{1}{2})\beta _0 ] \\ 
 &- \frac{{(n + 2)}}{{(2n + 3)}}[E_{n + 1,0} \;\sinh (n + \frac{3}{2})\beta _0  + F_{n + 1,0} \cosh (n + \frac{3}{2})\beta _0 ] \\ 
 &=  - \frac{1}{{2\epsilon }}[\cosh \beta _0 \;\cosh (n - \frac{3}{2})\beta _0 \;(\frac{{ - (n - 1)}}{{2n - 1}})\;\widetilde{B}_{n - 2,0}  \\ 
 &+ \cosh (n - \frac{1}{2})\beta _0 \;(\frac{1}{{2n - 1}}(1 + 2(n - 1)(1 + \cosh ^2 \beta _0 )))\;\widetilde{B}_{n - 1,0}  \\ 
 &- \cosh \beta _0 \;\cosh (n + \frac{1}{2})\beta _0 \;(\frac{{(n - 1)}}{{2n - 1}} + \frac{{(n + 2)}}{{2n + 3}} + 2)\;\widetilde{B}_{n,0}  \\ 
 &- \cosh (n + \frac{3}{2})\beta _0 \;(\frac{1}{{2n + 3}}(1 - 2(n + 2)(1 + \cosh ^2 \beta _0 )))\widetilde{B}_{n + 1,0}  \\ 
 &+ \cosh \beta _0 \;\cosh (n + \frac{5}{2})\beta _0 \;(\frac{{ - (n + 2)}}{{2n + 3}})\widetilde{B}_{n + 2,0} ] \;\;\;(n \ge 1), \label{recu4} 
\end{split}
\end{equation}
\begin{equation}
\begin{split}
&\frac{{\sinh \beta _0 }}{2}[A_{n,0} \;\sinh (n + \frac{1}{2})\beta _0  + B_{n,0} \cosh (n + \frac{1}{2})\beta _0 ] \\ 
&- \frac{n}{{(2n - 1)}}[C_{n - 1,0} \;\sinh (n - \frac{1}{2})\beta _0  + D_{n - 1,0} \cosh (n - \frac{1}{2})\beta _0 ] \\ 
& + \cosh \beta _0 [C_{n,0} \;\sinh (n + \frac{1}{2})\beta _0  + D_{n,0} \cosh (n + \frac{1}{2})\beta _0 ] \\ 
&- \frac{{(n + 1)}}{{(2n + 3)}}[C_{n + 1,0} \;\sinh (n + \frac{3}{2})\beta _0  + D_{n + 1,0} \cosh (n + \frac{3}{2})\beta _0 ] \\ 
&=  - \frac{{\;\sinh \beta _0 }}{{2\epsilon }}[\cosh (n - \frac{3}{2})\beta _0\;(\frac{{n\;(n - 1)}}{{2n - 1}})\;\widetilde{B}_{n - 2,0}  \\ 
&- \cosh (n - \frac{1}{2})\beta _0 \;\cosh \beta _0 \;(\frac{{2n\;(n - 1)}}{{2n - 1}})\;\widetilde{B}_{n - 1,0}  \\ 
&- \cosh (n + \frac{1}{2})\beta _0 \;(\frac{1}{{(2n + 1)}}(\frac{{n\;(2n^2  + 3n - 1)}}{{2n - 1}} + \frac{{(n + 1)\;( - 2n^2  - n + 2)}}{{2n + 3}}))\;\widetilde{B}_{n,0}  \\ 
& + \cosh (n + \frac{3}{2})\beta _0 \;\cosh \beta _0 \;(\frac{{2(n + 1)\;(n + 2)}}{{2n + 3}})\;\widetilde{B}_{n + 1,0}  \\ 
&- \cosh (n + \frac{5}{2})\beta _0 \;(\frac{{(n + 1)\;(n + 2)}}{{2n + 3}})\;\widetilde{B}_{n + 2,0} ] \;\;\;(n \ge 0).  \label{recu5} 
\end{split}
\end{equation}
\indent Lastly, the continuity equation gives two more recursive relations
\begin{eqnarray}
\begin{split}
 &\frac{{ - 1}}{2}n\;A_{n - 1,0}  + \frac{5}{2}\;A_{n,0}  + \frac{1}{2}(n + 1)\;A_{n + 1,0}  - n\;D_{n - 1,0}  + (2n + 1)D_{n,0}   \\ 
 &- (n + 1)D_{n + 1,0} - n(n - 1)E_{n - 1,0}  + 2n(n + 1)E_{n,0}  - (n + 1)(n + 2)E_{n + 1,0}  = 0 , \label{recu6}\\ 
\end{split}
\end{eqnarray}
\begin{eqnarray}
\begin{split}
 &\frac{{ - 1}}{2}n\;B_{n - 1,0}  + \frac{5}{2}\;B_{n,0}  + \frac{1}{2}(n + 1)\;B_{n + 1,0}  - n\;C_{n - 1,0}  + (2n + 1)C_{n,0}  \\ 
 & - (n + 1)C_{n + 1,0} - n(n - 1)F_{n - 1,0}  + 2n(n + 1)F_{n,0}  - (n + 1)(n + 2)F_{n + 1,0}  = 0 , \label{recu7}\\ 
\end{split}
\end{eqnarray}
\\

\indent For $m \ge 1$, no-slip boundary condition at the wall gives two sets of recursion relations
\begin{eqnarray}
\begin{split}
&- \frac{1}{{2(2n - 1)}}B_{n - 1,m}  + \frac{1}{{2(2n + 3)}}B_{n + 1,m}  - \frac{{\left( {n - m - 1} \right)}}{{(2n - 1)}}F_{n - 1,m}  + F_{n,m} \\ 
&- \frac{{\left( {n + m + 2} \right)}}{{(2n + 3)}}F_{n + 1,m}  = 0\;\;\;\;;\;m \ge 1\;,\;n \ge m + 1,
\end{split}
\end{eqnarray}
\begin{eqnarray}
\begin{split}
&\frac{{(n - m)(n - m + 1)}}{{2(2n - 1)}}B_{n - 1,m}  - \frac{{(n + m)(n + m + 1)}}{{2(2n + 3)}}B_{n + 1,m}  \\ 
&- \frac{{\left( {n - m + 1} \right)}}{{(2n - 1)}}H_{n - 1,m}  + H_{n,m}  - \frac{{\left( {n + m} \right)}}{{(2n + 3)}}H_{n + 1,m}  = 0\;\;\;\;;\;m \ge 1\;,\;n \ge m - 1.
\end{split}
\end{eqnarray}
The no-penetration condition at the solid surface of wall, $\left. {v_z } \right|_{\beta  = 0}  = 0$, is trivially satisfied if
\begin{eqnarray}
D_{n,m}  = 0\;\;\;\;;\;m \ge 1\;,\;n \ge m.
\end{eqnarray}
Next, we impose the slip boundary condition at the colloid surface (Eq.~\ref{bond2}). Transforming the unit vector from bipolar spherical to cylindrical coordinates, 
\begin{eqnarray}
\left. {v_r } \right|_{\beta=\beta _0 }  =  - \;\nabla _s C \cdot {\bf{e}}_r  = \; - {\frac{{(\cosh \beta _0 \cos \alpha  - 1)}}{{(\cosh \beta _0  - \cos \alpha )\;}}}\sum\limits_{m = 1}^\infty  {\left( {\vartheta _m  + \zeta _m  + \kappa _m } \right)} \;\cos (m\phi  + \gamma _m ), \label{gr1}
\end{eqnarray}
\begin{eqnarray}
\left. {v_\phi  } \right|_{\beta=\beta _0 }  =  - \;\nabla _s C \cdot {\bf{e}}_\phi   =  - \sum\limits_{m = 1}^\infty  {\nu _m \sin (m\phi  + \gamma _m )}, \label{gr2}
\end{eqnarray}
\begin{eqnarray}
\left. {v_z } \right|_{\beta=\beta _0 }  =  - \;\nabla _s C \cdot {\bf{e}}_z  =  { \frac{{\sinh \beta _0 \;\sin \alpha }}{{(\cosh \beta _0  - \cos \alpha )}}} \sum\limits_{m = 1}^\infty  {\left( {\vartheta _m  + \zeta _m  + \kappa _m } \right)} \;\cos (m\phi  + \gamma _m ). \label{gr3}
\end{eqnarray}
where
\begin{eqnarray}
\vartheta _m  = \frac{{\;\sin \alpha \;\sqrt {\cosh \beta _0  - \cos \alpha } }}{{2\epsilon \;}}\;\sum\limits_{n = m}^\infty  {\widetilde{B}_{nm} \cosh (n + \frac{1}{2})\beta _0 } \;P_n^m (\cos \alpha ),
\end{eqnarray}
\begin{eqnarray}
\zeta _m  = \frac{{(\cosh \beta _0  - \cos \alpha )^{\frac{3}{2}} }}{{ - 2\epsilon }}\sum\limits_{n = m - 1}^\infty  {\widetilde{B}_{nm} (n + m)(n - m + 1)\cosh (n + \frac{1}{2})\beta _0 } \;P_n^{m - 1} (\cos \alpha ),
\end{eqnarray}
\begin{eqnarray}
\kappa _m  = \frac{{(\cosh \beta _0  - \cos \alpha )^{\frac{3}{2}} }}{{2\epsilon }}\sum\limits_{n = m + 1}^\infty  {\widetilde{B}_{nm} \;\cosh (n + \frac{1}{2})\beta _0 } \;P_n^{m + 1} (\cos \alpha ),
\end{eqnarray}
\begin{eqnarray}
\nu _m  = \frac{{(\cosh \beta _0  - \cos \alpha )^{\frac{3}{2}} }}{{ - \epsilon \;\sin \alpha }}\sum\limits_{n = m}^\infty  {\widetilde{B}_{nm} \;m\;\cosh (n + \frac{1}{2})\beta _0 } \;P_n^m (\cos \alpha ).
\end{eqnarray}
\indent Inserting those boundary values from Eqs.~\ref{gr1}-\ref{gr2} into the solution of the Stokes equations in cylindrical coordinate (Eqs.~\ref{vel1}-\ref{vel2}) and using orthogonality along with the recursion relations between associated Legendre polynomials, two sets of recursion relations among unknown coefficients found. For $m \ge 1$ and $l \ge m + 1$,
\begin{eqnarray}
\begin{split}
&  - \frac{1}{{2(2l - 1)}}[{A_{l - 1,m} \;\sinh (l - \frac{1}{2})\beta _0  + B_{l - 1,m} \cosh (l - \frac{1}{2})\beta _0 }] \\ 
&+ \frac{1}{{2(2l + 3)}}[ {A_{l + 1,m} \;\sinh (l + \frac{3}{2})\beta _0  + B_{l + 1,m} \cosh (l + \frac{3}{2})\beta _0 }] \\ 
&- \frac{{(l - m - 1)}}{{(2l - 1)}}[ {E_{l - 1,m} \;\sinh (l - \frac{1}{2})\beta _0  + F_{l - 1,m} \cosh (l - \frac{1}{2})\beta _0 }] \\ 
&+ \cosh \beta _0 [ {E_{l,m} \;\sinh (l + \frac{1}{2})\beta _0  + F_{l,m} \cosh (l + \frac{1}{2})\beta _0 } ] \\ 
& - \frac{{(l + m + 2)}}{{(2l + 3)}} [ {E_{l + 1,m} \;\sinh (l + \frac{3}{2})\beta _0  + F_{l + 1,m} \cosh (l + \frac{3}{2})\beta _0 }] =  \frac{{(2l + 1)(l - m-1)!}}{{2(l + m+1)!}} \\ 
&\int\limits_0^\pi  { - ( {( {\frac{{\cosh \beta _0 \;\cos \alpha  - 1}}{{\sqrt {\cosh \beta _0  - \cos \alpha } }}}) ( {\vartheta _m  + \zeta _m  + \kappa _m }) + \sqrt {\cosh \beta _0  - \cos \alpha } \;\nu _m })} \;P_l^{m + 1} ( {\cos \alpha } )\;\sin \alpha \;d\alpha,  \\ 
\end{split}
\end{eqnarray}
while for $m \ge 1$ and $l \ge m -1 $,
\begin{eqnarray}
\begin{split}
 &\frac{{(l - m)(l - m + 1)}}{{2(2l - 1)}} [ {A_{l - 1,m} \;\sinh (l - \frac{1}{2})\beta _0  + B_{l - 1,m} \cosh (l - \frac{1}{2})\beta _0 } ] \\ 
  &+ \frac{{(l + m)(l + m + 1)}}{{2(2l + 3)}} [ {A_{l + 1,m} \;\sinh (l + \frac{3}{2})\beta _0  + B_{l + 1,m} \cosh (l + \frac{3}{2})\beta _0 } ] \\ 
&  - \frac{{(l - m + 1)}}{{(2l - 1)}} [ {G_{l - 1,m} \;\sinh (l - \frac{1}{2})\beta _0  + H_{l - 1,m} \cosh (l - \frac{1}{2})\beta _0 } ] \\ 
 & + \cosh \beta _0 [ {G_{l,m} \;\sinh (l + \frac{1}{2})\beta _0  + H_{l,m} \cosh (l + \frac{1}{2})\beta _0 }] \\ 
&  - \frac{{(l + m)}}{{(2l + 3)}} [ {G_{l + 1,m} \;\sinh (l + \frac{3}{2})\beta _0  + H_{l + 1,m} \cosh (l + \frac{3}{2})\beta _0 } ]= \frac{{(2l + 1)(l - m + 1)!}}{{2(l + m - 1)!}}\\ 
&\int\limits_0^\pi  { - ( {( {\frac{{\cosh \beta _0 \;\cos \alpha  - 1}}{{\sqrt {\cosh \beta _0  - \cos \alpha } }}} ) ( {\vartheta _m  + \zeta _m  + \kappa _m } ) - \sqrt {\cosh \beta _0  - \cos \alpha } \;\nu _m } )} \;P_l^{m - 1} ( {\cos \alpha } )\;\sin \alpha \;d\alpha.
\end{split}
\end{eqnarray}
Inserting the boundary condition for $V_z$, Eq.~\ref{gr3}, into the $z$ component of Stokes equation (Eq.~\ref{vel3}) we have, 
\begin{eqnarray}
\begin{split}
& \frac{{\sinh \beta _0 }}{2}\;[A_{n,m} \;\sinh (n + \frac{1}{2})\beta _0  + B_{n,m} \cosh (n + \frac{1}{2})\beta _0 ] \\ 
&  - \frac{{(n - m)}}{{(2n - 1)}}\;[C_{n - 1,m} \;\sinh (n - \frac{1}{2})\beta _0  + D_{n - 1,m} \cosh (n - \frac{1}{2})\beta _0 ] \\ 
&  + \cosh \beta _0 \;[C_{n,m} \;\sinh (n + \frac{1}{2})\beta _0  + D_{n,m} \cosh (n + \frac{1}{2})\beta _0 ] \\ 
&  - \frac{{(n + m + 1)}}{{(2n + 3)}}\;[C_{n + 1,m} \;\sinh (n + \frac{3}{2})\beta _0  + D_{n + 1,m} \cosh (n + \frac{3}{2})\beta _0 ] \\ 
&  =  - \frac{{\;\sinh \beta _0 }}{{2\epsilon }}[\cosh (n - \frac{3}{2})\beta _0 \frac{{(n - m - 1)\;(n - m)}}{{2n - 1}}\;\widetilde{B}_{n - 2,0}  \\ 
&  - \;\cosh (n - \frac{1}{2})\beta _0 \;\cosh \beta _0 \;\frac{{2(n - m)\;(n - 1)}}{{2n - 1}}\;\widetilde{B}_{n - 1,0}  \\ 
&  - \cosh (n + \frac{1}{2})\beta _0 ((\frac{{(n + m)\;((n + m - 1) + 2(n + 1)(n - m))}}{{2n - 1}} \\ 
&  + \frac{{(n - m + 1)((n - m + 2) - 2n(n + m + 1))\;}}{{2n + 3}})\frac{1}{{(2n + 1)}})\;\widetilde{B}_{n,0}  \\ 
&  + \;\cosh (n + \frac{3}{2})\beta _0 \;\cosh \beta _0 \;\frac{{2(n + m + 1)\;(n + 2)}}{{2n + 3}}\;\widetilde{B}_{n + 1,0}  \\ 
&  - \cosh (n + \frac{5}{2})\beta _0 \frac{{(n + m + 1)\;(n + m + 2)}}{{2n + 3}}\;\widetilde{B}_{n + 2,0} ]\;\;\;\;;m \ge 1\;,\;n \ge m.\\ 
\end{split}
\end{eqnarray}
\indent To complete the solution two more relations are needed, and these are obtained from the continuity equation. Using  Eqs.~\ref{vel1} - \ref{vel3} the continuity equation (Eq.~\ref{hyd2}) can be expressed in the form of a double series of terms of the form $\cosh (n + \frac{1}{2})\beta \;P_n^m (\cos \alpha )\cos (m\phi  + \gamma _m )$ and $\sinh (n + \frac{1}{2})\beta \;P_n^m (\cos \alpha )\cos (m\phi  + \gamma _m )$, setting the coefficients to zero we find for $m \ge 1$ and $n \ge m$ 
\begin{eqnarray}
\begin{split}
 &\frac{{ - 1}}{2}(n - m)\;A_{n - 1,m}  + \frac{5}{2}\;A_{n,m}  + \frac{1}{2}(n + m + 1)\;A_{n + 1,m}  - (n - m)\;D_{n - 1,m}  + (2n + 1)D_{n,m}  \\ 
 &- (n + m + 1)D_{n + 1,m}  - \frac{1}{2}(n - m - 1)(n - m)E_{n - 1,m}  + (n + m + 1)(n - m)E_{n,m}  \\ 
 &- \frac{1}{2}(n + m + 1)(n + m + 2)E_{n + 1,m}  + \frac{1}{2}G_{n - 1,m}  - G_{n,m}  + \frac{1}{2}G_{n + 1,m}  = 0, \\ \label{contin1}
\end{split}
\end{eqnarray}
\begin{eqnarray}
\begin{split}
 &\frac{{ - 1}}{2}(n - m)\;B_{n - 1,m}  + \frac{5}{2}\;B_{n,m}  + \frac{1}{2}(n + m + 1)\;B_{n + 1,m}  - (n - m)\;C_{n - 1,m}  + (2n + 1)C_{n,m}  \\ 
 &- (n + m + 1)C_{n + 1,m}  - \frac{1}{2}(n - m - 1)(n - m)F_{n - 1,m}  + (n + m + 1)(n - m)F_{n,m}  \\ 
 &- \frac{1}{2}(n + m + 1)(n + m + 2)F_{n + 1,m}  + \frac{1}{2}H_{n - 1,m}  - H_{n,m}  + \frac{1}{2}H_{n + 1,m}  = 0, \\ \label{contin2}
\end{split}
\end{eqnarray}
We have obtained 8 independent recursion relations for $\phi$-independent and 8 more for $\phi$-dependent parts of the field, which are solved simultaneously to find the unknown coefficients ${A_{n,0} }$, ${B_{n,0} }$,..., ${H_{n,0} }$ and ${A_{n,m} }$, ${B_{n,m} }$,..., ${H_{n,m} }$. 
\subsection{Translation of a non-rotating sphere parallel to the wall}\label{vparallel}
Problem ($b$) concerns the translational motion of a non-rotating solid sphere parallel to a solid wall, which has been solved by O'Neill \cite{ONeal}. In this case, the boundary conditions on the wall are
\begin{eqnarray}
\left. {V_r } \right|_{\beta  = 0}  = \left. {V_\phi  } \right|_{\beta  = 0}  = \left. {V_z } \right|_{\beta  = 0}  = 0,
\end{eqnarray}
and the no-slip condition on the particle surface is 
\begin{align}
&\left. {V_r } \right|_{\beta _0 }  = \cos \phi \label{sor1}, \\ 
&\left. {V_\phi  } \right|_{\beta _0 }  =  - \sin \phi \label{sor2}, \\ 
&\left. {V_z } \right|_{\beta _0 }  = 0 \label{sor3}.
\end{align}
One may easily show that in the general solution of the Stokes equation just the $m=1$ term survives and furthermore ${\gamma _0 }=0$. Using Eqs.~\ref{sor1} - \ref{sor2} we have two recursion relations:
\begin{equation}
\begin{split}
 & - \frac{1}{{2(2n - 1)}}\left[ {A_{n - 1,1} \;\sinh (n - \frac{1}{2})\beta _0  + B_{n - 1,1} \cosh (n - \frac{1}{2})\beta _0 } \right] \\ 
  &+ \frac{1}{{2(2n + 3)}}\left[ {A_{n + 1,1} \;\sinh (n + \frac{3}{2})\beta _0  + B_{n + 1,1} \cosh (n + \frac{3}{2})\beta _0 } \right] \\ 
  &- \frac{{(n - 2)}}{{(2n - 1)}}\left[ {E_{n - 1,1} \;\sinh (n - \frac{1}{2})\beta _0  + F_{n - 1,1} \cosh (n - \frac{1}{2})\beta _0 } \right] \\ 
  &+ \cosh \beta _0 \left[ {E_{n,1} \;\sinh (n + \frac{1}{2})\beta _0  + F_{n,1} \cosh (n + \frac{1}{2})\beta _0 } \right] \\ 
  &- \frac{{(n + 3)}}{{(2n + 3)}}\left[ {E_{n + 1,1} \;\sinh (n + \frac{3}{2})\beta _0  + F_{n + 1,1} \cosh (n + \frac{3}{2})\beta _0 } \right] = 0,\;\;\;(n \ge 2), \label{recurs1} \\ 
\end{split}
\end{equation}
\begin{equation}
\begin{split}
 &\frac{{(n - 1)\;n}}{{2(2n - 1)}}\left[ {A_{n - 1,1} \;\sinh (n - \frac{1}{2})\beta _0  + B_{n - 1,1} \cosh (n - \frac{1}{2})\beta _0 } \right] \\ 
  &+ \frac{{(n + 1)(n + 2)}}{{2(2n + 3)}}\left[ {A_{n + 1,1} \;\sinh (n + \frac{3}{2})\beta _0  + B_{n + 1,1} \cosh (n + \frac{3}{2})\beta _0 } \right] \\ 
  &- \frac{n}{{(2n - 1)}}\left[ {G_{n - 1,1} \;\sinh (n - \frac{1}{2})\beta _0  + H_{n - 1,1} \cosh (n - \frac{1}{2})\beta _0 } \right] \\ 
  &+ \cosh \beta _0 \left[ {G_{n,1} \;\sinh (n + \frac{1}{2})\beta _0  + H_{n,1} \cosh (n + \frac{1}{2})\beta _0 } \right] \\ 
  &- \frac{{(n + 1)}}{{(2n + 3)}}\left[ {G_{n + 1,1} \;\sinh (n + \frac{3}{2})\beta _0  + H_{n + 1,1} \cosh (n + \frac{3}{2})\beta _0 } \right] =  \\ 
&2\sqrt 2 {\rm{ }}\left[ {\cosh {\beta _0}\;{e^{ - \left( {n + \frac{1}{2}} \right){\beta _0}}} - \left( {\frac{n}{{2n - 1}}} \right){e^{ - \left( {n - \frac{1}{2}} \right){\beta _0}}} - \left( {\frac{{n + 1}}{{2n + 3}}} \right){e^{ - \left( {n + \frac{3}{2}} \right){\beta _0}}}} \right],~~(n \ge 0).\label{recurs2}\\ 
\end{split}
\end{equation}
From the boundary condition Eq.~\ref{sor3}, the no-slip boundary condition and the no-penetration condition at the wall surface, we obtain
\begin{equation}
\begin{split}
 &\frac{{\sinh \beta _0 }}{2}\left[ {A_{n,1} \;\sinh (n + \frac{1}{2})\beta _0  + B_{n,1} \cosh (n + \frac{1}{2})\beta _0 } \right] \\ 
  &- \frac{{(n - 1)}}{{(2n - 1)}}\left[ {C_{n - 1,1} \;\sinh (n - \frac{1}{2})\beta _0  + D_{n - 1,1} \cosh (n - \frac{1}{2})\beta _0 } \right] \\ 
  &+ \cosh \beta _0 \left[ {C_{n,1} \;\sinh (n + \frac{1}{2})\beta _0  + D_{n,1} \cosh (n + \frac{1}{2})\beta _0 } \right] \\ 
  &- \frac{{(n + 2)}}{{(2n + 3)}}\left[ {C_{n + 1,1} \;\sinh (n + \frac{3}{2})\beta _0  + D_{n + 1,1} \cosh (n + \frac{3}{2})\beta _0 } \right] = 0,\;\;\;\;(n \ge 1). \label{recurs3} \\ 
\end{split}
\end{equation}
\begin{eqnarray}
  - \frac{1}{{2(2n - 1)}}B_{n - 1,1}  + \frac{1}{{2(2n + 3)}}B_{n + 1,1}  - \frac{{\left( {n - 2} \right)}}{{(2n - 1)}}F_{n - 1,1}  \nonumber  \\
  + F_{n,1}  - \frac{{\left( {n + 3} \right)}}{{(2n + 3)}}F_{n + 1,1}  = 0\;\;\;(n \ge 2), \label{recurs4}\\ 
 \frac{{(n - 1)\;n}}{{2(2n - 1)}}B_{n - 1,1}  - \frac{{(n + 1)(n + 2)}}{{2(2n + 3)}}B_{n + 1,1}  - \frac{n}{{(2n - 1)}}H_{n - 1,1} \nonumber \\ 
  + H_{n,1}  - \frac{{\left( {n + 1} \right)}}{{(2n + 3)}}H_{n + 1,1}  = 0\;\;\;\; \;\;(n \ge 0), \label{recurs5}
\end{eqnarray}
\begin{equation}
D_{n,1}  = 0\quad \quad \;(n \ge 1).\label{recurs6}
\end{equation}
To complete the solution we must satisfy the equation of continuity, which from Eqs.~\ref{contin1} -\ref{contin2} with $m=1$ can be written as:
\begin{equation}
\begin{split}
 &\frac{{ - 1}}{2}(n - 1)\;A_{n - 1,1}  + \frac{5}{2}\;A_{n,1}  + \frac{1}{2}(n + 2)\;A_{n + 1,1}  - (n - 1)\;D_{n - 1,1}  \\ 
  &+ (2n + 1)D_{n,1}  - (n + 2)D_{n + 1,1}  - \frac{1}{2}(n - 2)(n - 1)E_{n - 1,1}  + (n + 2)(n - 1)E_{n,1}  \\ 
  &- \frac{1}{2}(n + 2)(n + 3)E_{n + 1,1}  + \frac{1}{2}G_{n - 1,1}  - G_{n,1}  + \frac{1}{2}G_{n + 1,1}  = 0, \label{recurs7}\\ 
\end{split}
\end{equation}
\begin{equation}
\begin{split}
 &\frac{{ - 1}}{2}(n - 1)\;B_{n - 1,1}  + \frac{5}{2}\;B_{n,1}  + \frac{1}{2}(n + 2)\;B_{n + 1,1}  - (n - 1)\;C_{n - 1,1}  \\ 
  &+ (2n + 1)C_{n,1}  - (n + 2)C_{n + 1,1}  - \frac{1}{2}(n - 2)(n - 1)F_{n - 1,1}  + (n + 2)(n - 1)F_{n,1}  \\ 
  &- \frac{1}{2}(n + 2)(n + 3)F_{n + 1,1}  + \frac{1}{2}H_{n - 1,1}  - H_{n,1}  + \frac{1}{2}H_{n + 1,1}  = 0. \label{recurs8}\\ 
\end{split}
\end{equation}
We obtain the unknown coefficients from simultaneous solution of the above equations. The fluid exerts a force on the particle, given by \cite{ONeal}:
\begin{equation}
F^{T} _x  =  - \sqrt {2\;} \pi \;\epsilon \;\mu \;U_x \;\sum\limits_{n = 0}^\infty  {(G_{n,1}  + } H_{n,1} ) - n(n + 1)\;(A_{n,1}  + B_{n,1} ) ,\\
\end{equation}
where $U_x$ is the particle velocity 
and similarly the torque exerted on the colloid is in the $y$ direction, given by
\begin{align}
 &{\rm T}^{T} _y  = \frac{2}{{3\sqrt 2 }}\pi \epsilon ^2 \mu \;U_x \sum\limits_{n = 0}^\infty  {\left[ {\left( {2 + e^{ - \left( {2n + 1} \right)\beta _0 } } \right)} \right.} \left( {n\left( {n + 1} \right)\left( {2C_{n,1}  + \coth \beta _0 A_{n,1} } \right) + (2n + 1 - \coth \beta _0 G_{n,1} } \right) \nonumber \\ 
 &\left. {\left( {2 - e^{ - \left( {2n + 1} \right)\beta _0 } } \right)\left( {n\left( {n + 1} \right)\left( {2D_{n,1}  + \coth \beta _0 B_{n,1} } \right) + (2n + 1 - \coth \beta _0 H_{n,1} } \right)} \right]. 
\end{align}
\subsection{Translation of a sphere normal to the solid wall}\label{vnormal}
This problem has been solve by Brenner and Maude separately from stream function solution because of axisymmetric al nature of problem \cite{Brennerwall,Maude}. Here we address this problem by directly solving the Stokes equation in cylindrical coordinates, using the expressions in Eqs.~\ref{vel1} - \ref{vel3}. The no-slip boundary conditions at the wall and on the particle are
\begin{align}
\left. {V_r } \right|_{\beta  = 0}  = \left. {V_\phi  } \right|_{\beta  = 0}  = \left. {V_z } \right|_{\beta  = 0}  = 0,\\
 \left. {V_r } \right|_{\beta _0 }  = \left. {V_\phi  } \right|_{\beta _0 }  = 0, \\ 
 \left. {V_z } \right|_{\beta _0 }  = 1,
\end{align}
and from the above boundary conditions and by symmetry we see that the solution only requires $m=0$ and ${\gamma _0 }=0$. Using the expression for the generating function of Legendre polynomials, we can expand boundary condition at the colloid surface, and after some algebraic manipulation we find 
\begin{eqnarray}
\begin{split}
 & - \frac{1}{{2(2n - 1)}}\left[ {A_{n - 1,0} \;\sinh (n - \frac{1}{2})\beta _0  + B_{n - 1,0} \cosh (n - \frac{1}{2})\beta _0 } \right] \\ 
  &+ \frac{1}{{2(2n + 3)}}\left[ {A_{n + 1,0} \;\sinh (n + \frac{3}{2})\beta _0  + B_{n + 1,0} \cosh (n + \frac{3}{2})\beta _0 } \right] \\ 
  &- \frac{{(n - 1)}}{{(2n - 1)}}\left[ {E_{n - 1,0} \;\sinh (n - \frac{1}{2})\beta _0  + F_{n - 1,0} \cosh (n - \frac{1}{2})\beta _0 } \right] \\ 
  &+ \cosh \beta _0 \left[ {E_{n,0} \;\sinh (n + \frac{1}{2})\beta _0  + F_{n,0} \cosh (n + \frac{1}{2})\beta _0 } \right] \\ 
  &- \frac{{(n + 2)}}{{(2n + 3)}}\left[ {E_{n + 1,0} \;\sinh (n + \frac{3}{2})\beta _0  + F_{n + 1,0} \cosh (n + \frac{3}{2})\beta _0 } \right] = 0,~~(n \ge 1), \label{bren1}
\end{split}
\end{eqnarray}

and 
\begin{eqnarray}
\begin{split}
 &\frac{{\sinh \beta _0 }}{2}\left[ {A_{n,0} \;\sinh (n + \frac{1}{2})\beta _0  + B_{n,0} \cosh (n + \frac{1}{2})\beta _0 } \right] \\ 
  &- \frac{n}{{(2n - 1)}}\left[ {C_{n - 1,0} \;\sinh (n - \frac{1}{2})\beta _0  + D_{n - 1,0} \cosh (n - \frac{1}{2})\beta _0 } \right] \\ 
  &+ \cosh \beta _0 \left[ {C_{n,0} \;\sinh (n + \frac{1}{2})\beta _0  + D_{n,0} \cosh (n + \frac{1}{2})\beta _0 } \right] \\ 
  &- \frac{{(n + 1)}}{{(2n + 3)}}\left[ {C_{n + 1,0} \;\sinh (n + \frac{3}{2})\beta _0  + D_{n + 1,0} \cosh (n + \frac{3}{2})\beta _0 } \right] =  \\ 
 &\sqrt 2 \left[ {\cosh \beta _0 \;e^{ - \left( {n + \frac{1}{2}} \right)\beta _0 }  - \left( {\frac{n}{{2n - 1}}} \right)e^{ - \left( {n - \frac{1}{2}} \right)\beta _0 }  - \left( {\frac{{n + 1}}{{2n + 3}}} \right)e^{ - \left( {n + \frac{3}{2}} \right)\beta _0 } } \right].~~~(n \ge 0) \label{bren2}
\end{split} 
\end{eqnarray}
Furthermore, Eqs.~\ref{recu1}-\ref{recu3} and \ref{recu6}-\ref{recu7} can be used again in order to find the unknown coefficients, ${A_{n,0} }$, ${B_{n,0} }$, ${C_{n,0} }$, ${E_{n,0} }$ and ${F_{n,0} }$.

By symmetry the force on the particle is in the $z$-direction and there is no torque; explicitly 
\begin{eqnarray}
F^{N}_z  = 2\sqrt {2\;} \pi \;\epsilon \;\mu \;U_z \;\sum\limits_{n = 0}^\infty  {(C_{n,0}  + } D_{n,0} ) + (n + \frac{1}{2})\;(A_{n,0}  + B_{n,0} ).\label{forceb}
\end{eqnarray}
\subsection{Rotation of a non-translating sphere parallel to a solid wall}\label{vrotation}
This problem was first discussed by Dean and O'Neill \cite{DO}.The boundary conditions on the wall are no-slip and no-penetration,
\begin{eqnarray}
\left. {V_r } \right|_{\beta  = 0}  = \left. {V_\phi  } \right|_{\beta  = 0}  = \left. {V_z } \right|_{\beta  = 0}  = 0,
\end{eqnarray}
and the velocity at the particle surface is:
\begin{eqnarray}
 \left. {V_r } \right|_{\beta _0 }  = \left( {z - \left( {\delta  + R} \right)} \right)\cos \phi, \label{rott1} \\ 
 \left. {V_\phi  } \right|_{\beta _0 }  =  - \left( {z - \left( {\delta  + R} \right)} \right)\sin \phi,\label{rott2}  \\ 
 \left. {V_z } \right|_{\beta _0 }  =  - r\;\cos \phi.  \label{sorz}
\end{eqnarray}
From the boundary conditions at the particle surface it can be shown that  only the $m=1$ term survives in the general solution of the Stokes equation and that ${\gamma _0 }=0$.\\
\indent By applying the boundary conditions and using appropriate relations among associated Legendre polynomials, we obtain recursion relation \ref{recurs1} along with

\begin{equation}
\begin{split}
& \frac{{(n - 1)\;n}}{{2(2n - 1)}}\left[ {A_{n - 1,1} \;\sinh (n - \frac{1}{2})\beta _0  + B_{n - 1,m} \cosh (n - \frac{1}{2})\beta _0 } \right] \\ 
  &+ \frac{{(n + 1)(n + 2)}}{{2(2n + 3)}}\left[ {A_{n + 1,1} \;\sinh (n + \frac{3}{2})\beta _0  + B_{n + 1,m} \cosh (n + \frac{3}{2})\beta _0 } \right] \\ 
  &- \frac{n}{{(2n - 1)}}\left[ {G_{n - 1,1} \;\sinh (n - \frac{1}{2})\beta _0  + H_{n - 1,m} \cosh (n - \frac{1}{2})\beta _0 } \right] \\ 
  &+ \cosh \beta _0 \left[ {G_{n,1} \;\sinh (n + \frac{1}{2})\beta _0  + H_{n,m} \cosh (n + \frac{1}{2})\beta _0 } \right] \\ 
  &- \frac{{(n + 1)}}{{(2n + 3)}}\left[ {G_{n + 1,1} \;\sinh (n + \frac{3}{2})\beta _0  + H_{n + 1,m} \cosh (n + \frac{3}{2})\beta _0 } \right] =  \\ 
 &2\sqrt 2 \;\epsilon \left[ { - \left( {\frac{1}{{\sinh \beta _0 }}} \right)\;e^{ - \left( {n + \frac{1}{2}} \right)\beta _0 }  + \coth \beta _0 \left( {\frac{n}{{2n - 1}}} \right)e^{ - \left( {n - \frac{1}{2}} \right)\beta _0 }  + \coth \beta _0 \left( {\frac{{n + 1}}{{2n + 3}}} \right)e^{ - \left( {n + \frac{3}{2}} \right)\beta _0 } } \right],
\end{split}
\end{equation}
 for $n \ge 0$. Applying the boundary condition ( \ref{sorz}) into the expression (\ref{vel3}), we have:
\begin{equation}
\begin{split}
& \frac{{\sinh \beta _0 }}{2}\left[ {A_{n,1} \;\sinh (n + \frac{1}{2})\beta _0  + B_{n,1} \cosh (n + \frac{1}{2})\beta _0 } \right] \\ 
 & - \frac{{(n - 1)}}{{(2n - 1)}}\left[ {C_{n - 1,1} \;\sinh (n - \frac{1}{2})\beta _0  + D_{n - 1,1} \cosh (n - \frac{1}{2})\beta _0 } \right] \\ 
 & + \cosh \beta _0 \left[ {C_{n,1} \;\sinh (n + \frac{1}{2})\beta _0  + D_{n,1} \cosh (n + \frac{1}{2})\beta _0 } \right] \\ 
  &- \frac{{(n + 2)}}{{(2n + 3)}}\left[ {C_{n + 1,1} \;\sinh (n + \frac{3}{2})\beta _0  + D_{n + 1,1} \cosh (n + \frac{3}{2})\beta _0 } \right] =  \\ 
  &- \sqrt 2 \;\epsilon \left[ { - \left( {\frac{1}{{2n - 1}}} \right)e^{ - \left( {n - \frac{1}{2}} \right)\beta _0 }  + \left( {\frac{1}{{2n + 3}}} \right)e^{ - \left( {n + \frac{3}{2}} \right)\beta _0 } } \right], ~~~(n \ge 1) \\ 
\end{split}
\end{equation}

The no slip and no penetration condition at the wall yields Eq.~\ref{recurs4}-\ref{recurs6}. 
 Two more relations obtained form mass conservation, Eq.~\ref{recurs7}-\ref{recurs8}.

The unknown coefficients of velocity fields are found form simultaneous solution of above relations and the force and torque on the particle are \cite{DO}
\begin{equation}
F^{R} _x  =  - \sqrt {2\;} \pi \;\epsilon \;\mu \;R\;\Omega _y \;\sum\limits_{n = 0}^\infty  {(G_{n,1}  + } H_{n,1} ) - n(n + 1)\;(A_{n,1}  + B_{n,1} ),
\end{equation}
\begin{equation}
\begin{split}
&T^{R} _y  = 8\pi \;\mu \;R^3 \;\Omega _y \left[ { - \frac{1}{3}} \right. - \frac{{\sinh ^2 \beta _0 }}{{12\sqrt 2 }}\sum\limits_{n = 0}^\infty  {\left[ {\left( {2 + e^{ - \left( {2n + 1} \right)\beta _0 } } \right)} \right.}  \\ 
& \left( {n\left( {n + 1} \right)\left( {2C_{n,1}  + \coth \beta _0 A_{n,1} } \right) + (2n + 1 - \coth \beta _0 G_{n,1} } \right)) \\ 
& \left. {\left( {2 - e^{ - \left( {2n + 1} \right)\beta _0 } } \right)\left( {n\left( {n + 1} \right)\left( {2D_{n,1}  + \coth \beta _0 B_{n,1} } \right) + (2n + 1 - \coth \beta _0 H_{n,1} } \right))} \right], 
 \end{split}
\end{equation}
with all other components of force and torque vanishing. 
\section{Acknowledgments}\label{concl}
JK is supported in part by the National Science Foundation through award CBET-1264550.
\newpage
%
\end{document}